\documentclass{article}

\PassOptionsToPackage{numbers, compress}{natbib}
\usepackage[preprint]{neurips_data_2024}
\usepackage{multirow}
\usepackage[utf8]{inputenc} % allow utf-8 input
\usepackage[T1]{fontenc}    % use 8-bit T1 fonts
\usepackage{hyperref}       % hyperlinks
\usepackage{url}            % simple URL typesetting
\usepackage{booktabs}       % professional-quality tables
\usepackage{amsfonts}       % blackboard math symbols
\usepackage{nicefrac}       % compact symbols for 1/2, etc.
\usepackage{xcolor}         % colors

\usepackage{fullpage,graphicx,caption,subcaption}
\usepackage{amsmath,amsthm,amssymb}
\usepackage{appendix}
\usepackage{authblk}
\hypersetup{colorlinks=true, urlcolor=blue, linkcolor=blue, citecolor=red}

\title{A Comprehensive Evaluation of Generative Models in Calorimeter Shower Simulation}

\author[1]{Farzana Yasmin Ahmad}
\author[2]{Vanamala Venkataswamy}
\author[1,2]{Geoffrey Fox}
\affil[1]{Department of Computer Science, University of Virginia, VA 22903}
\affil[2]{Biocomplexity Institute and Initiative, University of Virginia, VA 22903}

\begin{document}

\maketitle

\begin{abstract}

The pursuit of understanding fundamental particle interactions has reached unparalleled precision levels. Particle physics detectors play a crucial role in generating low-level object signatures that encode collision physics. However, simulating these particle collisions is a demanding task in terms of memory and computation which will be exasperated with larger data volumes, more complex detectors, and a higher pileup environment in the High-Luminosity LHC. The introduction of "Fast Simulation" has been pivotal in overcoming computational bottlenecks. The use of deep-generative models has sparked a surge of interest in surrogate modeling for detector simulations, generating particle showers that closely resemble the observed data. Nonetheless, there is a pressing need for a comprehensive evaluation of their performance using a standardized set of metrics. In this study, we conducted a rigorous evaluation of three generative models using standard datasets and a diverse set of metrics derived from physics, computer vision, and statistics. Furthermore, we explored the impact of using full versus mixed precision modes during inference. Our evaluation revealed that the CaloDiffusion and CaloScore generative models demonstrate the most accurate simulation of particle showers, yet there remains substantial room for improvement. Our findings identified areas where the evaluated models fell short in accurately replicating Geant4 data.

\end{abstract}

\section{Introduction}

The calorimeter is used to understand primary particles' fundamental properties and interactions. The calorimeter acts as a detector that measures the energy of incoming particles. When high-energy particles collide with the calorimeter, it produces particle showers. A particle shower can be defined as a cascade of secondary particles produced as the result of high-energy particles interacting with the dense matter of the calorimeter. This phenomenon is crucial to identify the type and properties of particles as the energy deposition pattern varies with the type of particles like Higgs Boson~\cite{atlas_higgs} and advance the design of newer detectors. Geant4~\cite{agostinelli2003geant4} is an accurate and full detector simulator for electromagnetic and hadronic calorimeters. Geant4 is based on Monte Carlo simulations making it the most resource and compute-intensive step in the ATLAS experiment pipeline. Due to high time and resource demands~\cite{INFN_agenda}, Geant4 will become inefficient for every simulated event in the upcoming upgrade at the Large Hadron Collider (LHC)~\cite{lhc_schedule}. 

Alternatively, ~\textit{Fast Simulation} methods are employed that approximate the output of Geant4 with significantly less computation overhead. ML-based generative modeling has made significant advances in recent years serving as fast surrogate models for Geant4. The ATLAS experiment has now adopted FastCaloGAN, a fast simulation framework based on Generative Adversarial Network (GAN)~\cite{atlas2020fast}. Following CaloChallenge-2022, many variations of generative methods including GAN~\cite{atlas2020fast,de2017learning,paganini2018accelerating, paganini2018calogan,erdmann2018generating,erdmann2019precise,carminati2018three}, VAE~\cite{cresswell2022caloman,buhmann2021decoding,buhmann2021getting,diefenbacher2023new,salamani2023metahep,madulacalolatent,ernst2023normalizing} , Diffusion~\cite{amram2023denoising,mikuni2022score,mikuni2024caloscore,madulacalolatent,kobylianskii2024calograph,liu2024calo,favaro2024calodream}, Normalizing Flow ~\cite{krause2022caloflow,krause2021caloflow,krause2021caloflow_2, diefenbacher2023l2lflows,buckley2023inductive,ernst2023normalizing} models are proposed for calorimeter shower simulations. 

These methods have demonstrated remarkable performance in generating showers faster than Geant4 across various ranges of incident energy for primary particles, although less precise than Geant4. While GAN-based models can generate showers faster than other generative models, it is challenging to train GANs due to the difficulty of converging and `mode collapse'. VAE-based models can generate samples of calorimeter showers faster than Geant4 and GAN. However, they lack in quality of samples. Nevertheless, VAE-based models are used along with other GAN-based models~\cite{buhmann2021decoding,buhmann2021getting,diefenbacher2023new}, Normalizing Flow (NF)~\cite{ernst2023normalizing,cresswell2022caloman} and Diffusion~\cite{madulacalolatent} models. Among the generative methods, NF and Diffusion-based models have shown promising performance in generating shower samples with high fidelity. However, Diffusion-based models are slower in sampling, and NFs require a constraint bijective mapping making flow models restrictive. Refer to Appendix-\ref{app_related} for a more comprehensive discussion of these models.

The development of accurate surrogate models for simulating calorimeter showers is crucial except comparing the performance of existing models is challenging due to the diverse approaches and metrics used in recent research articles. We need a comprehensive approach to evaluating these models, both qualitatively and quantitatively. Without such an approach, it becomes difficult to build upon existing work or identify areas that require additional effort to continue progress.  

This study fills the gap by evaluating three different state-of-the-art calorimeter shower simulation models using a comprehensive set of metrics. The three models evaluated are CaloDiffusion~\cite{amram2023denoising}, CaloScore~\cite{mikuni2024caloscore} and CaloINN~\cite{ernst2023normalizing}. We chose CaloDiffusion and CaloScore because of their ability to generate high-quality samples and CaloINN because of its faster sample generation capability. Furthermore, these are the three open-sourced implementations that work.

The recent introduction of NVIDIA's Blackwell~\cite{nvidia_blackwell} GPU micro-architecture opens new doors for generative models. The key feature of Blackwell architecture is faster and more accurate low-precision inference for trained generative models. As an exemplar of using low-precision for inference, we explore the quality of mixed-precision inference with NVIDIA's V100 GPU. To the best of our knowledge, this is the first paper to examine mixed-precision inference for generative models to simulate calorimeter showers, comparing the generated samples with those from full-precision inference. We selected CaloDiffusion for this experiment due to its slow sample generation time and high accuracy. Our investigation aims to determine whether mixed-precision inference, in CaloDiffusion, can reduce sampling time without compromising on accuracy. 

The main \textbf{contributions} of this work are 1) Systematic and comprehensive evaluation of the three surrogate models to understand strengths and weaknesses of the models in generating the showers as accurately as possible compared to Geant4, 2) Using a common set of evaluation metrics applied consistently on all the models to benchmark their performance both using quantitative and qualitative measures and 3) Analysing the models using full and mixed precision modes on GPU to reveal any performance (speed and precision) implications of using mixed precision mode.

The rest of the paper is arranged as follows. In section~\ref{dataset}, we discuss calorimeter datasets followed by section~\ref{exp_setup} with experimental setup and metrics in ~\ref{eval_metrics}. In section~\ref{exp_findings}, we discuss the evaluation results and findings followed by related work in section~\ref{related_work} and conclusion in section~\ref{conclusion}.

\section{Datasets}\label{dataset}
We use the three datasets from CaloChallenge-2022~\cite{calochallenge_2022}, ranging in difficulty from easy to medium to hard, depending on the dimensionality of the calorimeter showers. 
%It causes a rise in the surrogate model training time. 
All datasets follow a similar layout. The detector geometry comprises concentric cylinders, and the particles travel along the z-axis. The detector is divided into discrete layers along the z-axis, and each layer has bins along the radial direction. Some layers also have bins at the angle $\alpha$. The binning XML files store the number of layers and the number of bins in r and $\alpha$. The coordinates $\Delta\varphi$ and $\Delta\eta$ correspond to the x and y-axis of the cylindrical coordinates. In dataset 1~\cite{michele_faucci_giannelli_2023_8099322}, has photons and pions as primary particles. Both dataset 2~\cite{faucci_giannelli_2022_6366271} and dataset 3~\cite{faucci_giannelli_2022_6366324} only have electrons as the primary particle. A summary of these datasets is tabulated in Table~\ref{tab:data}. 

\begin{table}[h]
\caption{Summary of CaloChallenge-2022 datasets where $N_z, N_\alpha, N_r$ represents the number of layers, angular bins and radial bins respectively.}
\label{tab:data}
\centering
\begin{tabular}{lllll}
\hline
 Dataset& $\#$ of Voxels  & $N_z$  &$N_\alpha$  &$N_r$  \\ \hline
$1$(Photon) &$368$  &$5$  &$[1,10,10,1,1]
$  &  $[8,16,19,5,5]$\\ %\hline
$1$(Pion) &  $533$&  $7$& $[1,10,10,1,10,10,1]
$ & $[8,10,10,5,15,16,10]
$ \\ %\hline
 $2$(Electron)&$6480$  &$45$  & $16$ &$9$  \\ %\hline
 $3$(Electron)&$40500$  &  $45$&  $50$& $18$ \\ \hline
\end{tabular}
\end{table}

For dataset 1, there are 15 incident energies from 256 MeV up to 4 TeV produced in powers of two. Each sample contains 10k events. Dataset 2 has two files containing 100k events for electrons each with energies sampled from a log-uniform distribution ranging from 1 GeV to 1 TeV. Dataset 3 has 4 files, each containing 50k Geant4 simulated electron showers with energies sampled from a log-uniform distribution ranging from 1 GeV to 1 TeV.

\section{Experimental setup}\label{exp_setup}
Following CaloChallenge-2022, researchers have published diverse generative techniques for calorimeter shower simulation (section~\ref{app_related}). It is difficult to evaluate all these proposed models because not all implementations are open-sourced and others lack proper documentation. We selected three models based on the following criteria: \textbf{1) Generalized}: model should be applicable to all three datasets from CaloChallenge-2022, \textbf{2) Open source and well documented}: model should be open-source and thoroughly documented, \textbf{3) Easy to execute}: model should be easy to run with minimum environment setup, \textbf{4) Reproducible}: model should exhibit consistent and reproducible results, and \textbf{5) State of the Art}: model should be the state of the art in terms of precision or speed. Based on this criteria, we selected CaloDiffusion~\cite{amram2023denoising}, CaloScore~\cite{mikuni2024caloscore}, and CaloINN~\cite{ernst2023normalizing} for comprehensive evaluation. We refer readers to the Appendix-\ref{app_a} for a detailed overview of the evaluated models, training details and total number of parameters in each of the trained models (Table~\ref{tab:params}).

\subsection{Evaluation metrics}\label{eval_metrics}
This section presents a systematic review of the evaluation metrics and their sensitivity to failure modes of generative models. The selected metrics are motivated by physics, computer vision and statistics. Some of these metrics are adopted from Calochallenge-2022 including a binary classifier trained on truth Geant4 vs. generated shower images, a binary classifier trained on a set of high-level features (like layer energies, shower shape variables),  chi-squared type measure derived from histogram differences of high-level features, and training time and calorimeter shower generation time. First, we briefly describe the metrics and then present the evaluation results. Refer to Appendix-\ref{app_b} for additional details on metrics and Appendix-\ref{app_c} for more results.

\subsubsection{Histograms of physics observables}
The histograms of the generated showers can provide insight into how well they replicate the distribution of real calorimeter shower samples. We consider them as qualitative metrics and five such physics observables (described below) are investigated in this work. 

\textbf{Layer wise energy distribution:} For datasets 2 and 3, we examine the distribution of energy as a function of the layer of the calorimeter. For most cases, the energies below 1 MeV are not significant due to the presence of noise. Therefore, the energy distributions plotted in the figures have energies above 1 MeV. The $E_{tot}/E_{inc}$ histogram represents the total deposited energy in an event, which has been normalized by the incident energy.

\textbf{Center of energy in $\eta$ and $\phi$ direction:} The center of energy along a coordinate is defined as the sum of energy deposited in each voxel times the voxel's coordinate distance from the origin, normalized by the total energy deposited. The equation to compute it is; $\Bar{x}=\frac{\langle x_i E_i \rangle}{\sum E_i}$, where $x_i$ is the cell location and $E_i$ is the energy at that location. For dataset 1, we generate the center of energy distribution for each layer. For datasets 2 and 3 we consolidate the layers by taking the average of five consecutive layers to summarize the histograms. 

\textbf{Shower width in $\eta$ and $\phi$ direction:} Another spatial property of a shower can be defined by shower width. Equation $\sqrt{\frac{\langle x_i^2 E_i \rangle}{\sum E_i}-\Bar{x}^2}$ defines the shower width.  

\textbf{Sparsity:} The sparsity is defined as the ratio of the number of voxels with nonzero deposition to the total number of voxels in each layer. 

\subsubsection{Correlation}
Identifying the correlation of energy deposition among adjacent layers in a calorimeter detector is crucial. This step is often overlooked in most proposed models for calorimeter shower simulation. To compute the correlation between layers, we first sum the deposited energy for each layer and then apply the Pearson Correlation Coefficient (PCC) across all layers.

\subsubsection{Classifier test}
One widely used metric is based on training a classifier to distinguish between the synthetic (generated) and reference samples. The closer the two samples, the closer the likelihoods, and the classifier will fail to distinguish between the two samples. Performance can be measured based on the area under the curve (AUC) of the receiver operating characteristic (ROC) curve of the classifier evaluated on a statistically independent dataset. An AUC of $1$ indicates that there is a significant difference in the generated sample. An AUC of 0.5 would indicate the classifier cannot distinguish between the two samples. To measure the similarity of the two distributions, we use Jensen-Shannon divergence (JSD) given by equation \ref{eq:JSD} (Appendix-\ref{app_c}).

\subsubsection{EMD, FPD, KPD scores}
We use the similarity measures for all of the physics observables which are sensitive to both the quality and diversity of the generated samples. We compare the total deposited energy between the models using the 1-Wasserstein distance referred to as the Earth Mover’s Distance (EMD), between the generated samples and the Geant simulation. EMD measures the dissimilarity between two distributions. Formally, the EMD between probability distributions P and Q can be defined as an infimum over joint probabilities. Refer equation \ref{eq:emd}, Appendix-\ref{app_c}. 

In~\cite{kansal2023evaluating}, authors suggest adopting Fr\'echet and Kernel Physics Distances(FPD, KPD) as two evaluation metrics to determine the similarity between real and generated images. Formally, for any two probability distributions $\mu$, $\upsilon$ over $\mathcal{R}^n$ having finite mean and variances, the Fr\'echet distance is given by equation \ref{eq:fpd}, Appendix-\ref{app_c}. When FID and KID are applied to some physical features, they are referred to as  Fr\'echet Physics Distance (FPD) and Kernel Physics Distance (KDP) respectively. Among EMD, FPD and KPD, FPD is the most efficient as it is an interpretable and highly sensitive metric for evaluating generative models in HEP. Both FPD and KPD are efficient in capturing the correlation between different features of this multidimensional feature space. 

\subsubsection{Separation power}
To quantify the similarities between two different histograms we use separation power. In information theory, it is known as triangular discrimination. To compute separation power, we use the equation $\langle S^2 \rangle=\frac{1}{2}\sum_{i=1}^{n_{\text{bins}}} \frac{(h_{1,i} - h_{2,i})^2}{h_{1,i} + h_{2,i}}$, where $h1$ and $h2$ denote two different histograms that are compared. A separation power of $0$ means that two histograms are the same and $1$ means they do not have any overlapping bins, i.e., they are dissimilar.

\subsubsection{Timing}
Timing is another widely used metric in calorimeter shower simulation due to the fact that the efficacy of the proposed generative surrogate models depends on the acceleration of the sample generation time. We investigate the sampling process from the trained models of CaloDiffusion, CaloScore, and CaloINN by considering different batch sizes 1, 10, and 100. We conduct the generation process three times for each batch size to calculate the mean and standard deviation time taken per event generation in seconds.

\section{Experimental findings}\label{exp_findings}

This section outlines various qualitative and quantitative analysis of the three surrogate models on three datasets of CaloChallenge. We benchmark the models using both full precision and mixed precision inference and share insights on performance-accuracy tradeoff.

\subsection{Histogram of physics observables}

We perform extensive analysis of various physics observables. In this section, we illustrate results for distribution of layer wise energy and sparsity metric for dataset 2. A more comprehensive analysis of these metrics and other physics observables (i.e., voxel energy distribution, $E_{ratio}$, center of energy and shower width) is available in Appendix~\ref{app_c}.

% {\color{red} mentin what are in main and what are in appendix}

% \begin{figure}  
% \subfloat[Dataset 2]
% {\includegraphics[width=0.46\textwidth]
% {images_subplot/final_dataset_2/E_layers_dataset_2_all.pdf}} 
% \subfloat[Dataset 3]
% {\includegraphics[width=0.46\textwidth]
% {images_subplot/final_dataset_3/E_layers_dataset_3_all.pdf}}
% \caption{Layer wise energy distribution considering all ranges of incident energy in GeV } \label{fig:layer_engergy_all}
% \end{figure}

\begin{figure}  
\subfloat[Layer energy distribution (GeV)  \label{fig:layer_engergy_all}]
{\includegraphics[width=0.5\textwidth]
{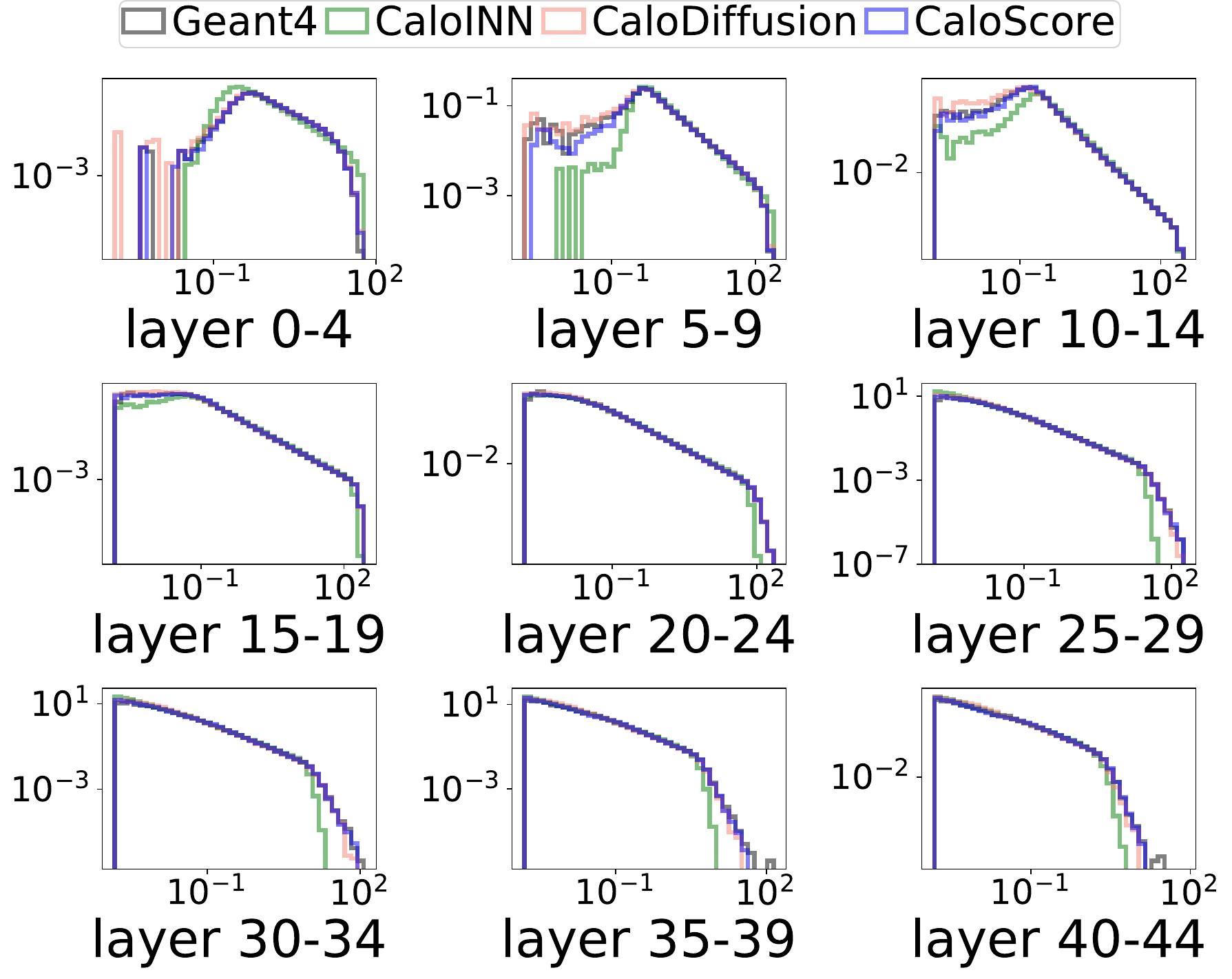}} 
\subfloat[Distribution of sparsity \label{fig:sparsity}]
{\includegraphics[width=0.5\textwidth]
{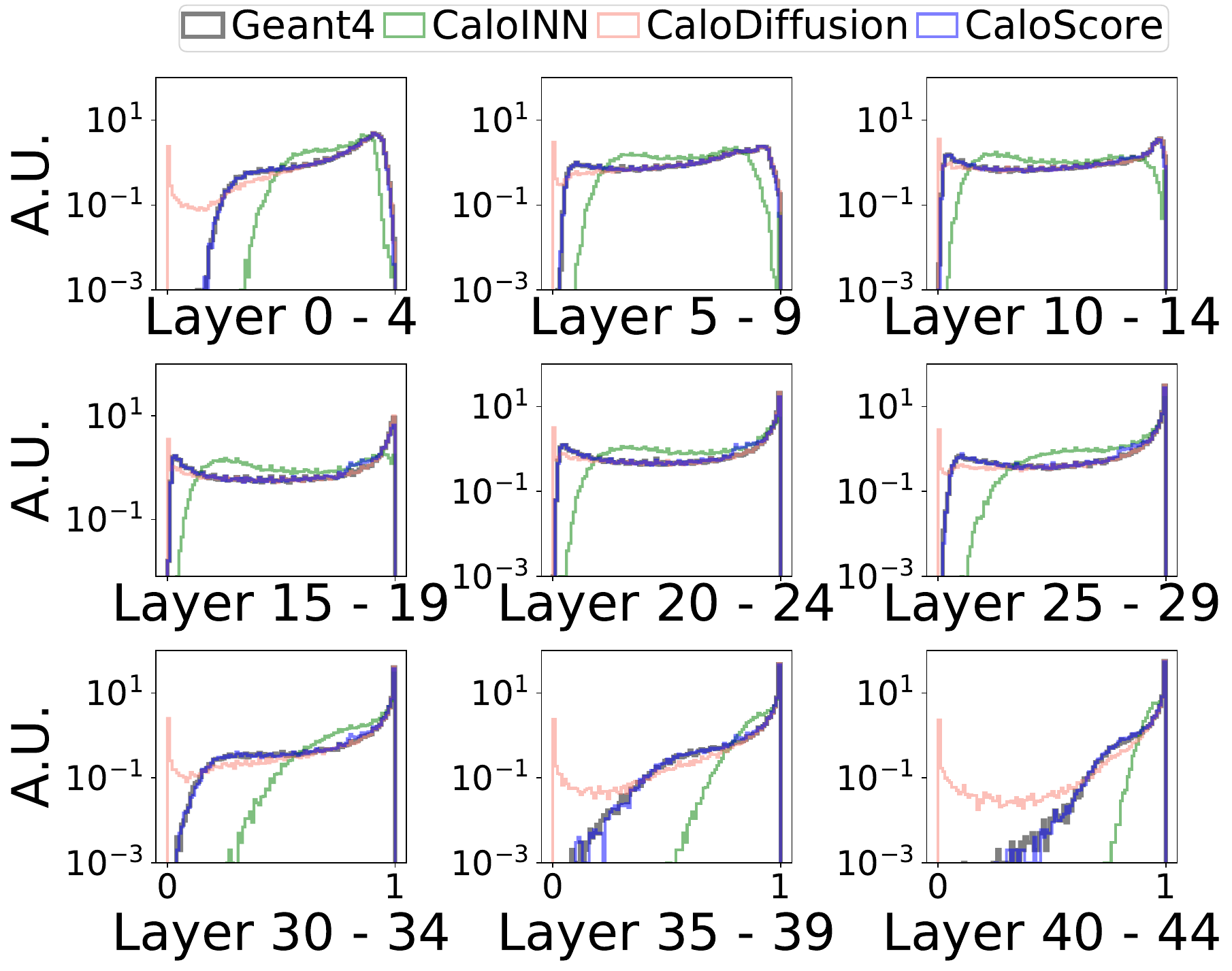}} 
\caption{Histogram of two physics observables for dataset 2.} \label{fig:histogram_observable}
\end{figure}

% \begin{figure}  
% \subfloat[Dataset 2]
% {\includegraphics[width=0.5\textwidth]
% {images_subplot/final_dataset_2/Sparsity_dataset_2.pdf}} 
% \subfloat[Dataset 3]
% {\includegraphics[width=0.5\textwidth]
% {images_subplot/final_dataset_3/Sparsity_dataset_3.pdf}}
% \caption{Distribution of sparsity} \label{fig:sparsity}
% \end{figure}

\textbf{Layer wise energy distribution:} 
% One of the most interesting observables to understand the quality of the generated shower is layer energy distribution. 
%Here we show layer wise energy distribution for dataset 2 as surrogate models usually struggle to capture the distribution when the dimension is higher.
Figure~\ref{fig:layer_engergy_all} shows layer wise energy distribution for all ranges of incident energy. Instead of showing distribution for each 45 layers individually, we compute the mean energy distribution for groups of five consecutive layers. To further isolate the performance of the surrogate models, we separated the graphs into three different incident energy ranges. Please refer to Appendix-\ref{app_c} for more comprehensive analysis (Figures~\ref{fig:energy_dist_ds1}-\ref{fig:layer_engergy_100_1000}).

% \textbf{Center of energy:}
% Figures~\ref{fig:ec_ds1_photon} through \ref{fig:ec_ds3} illustrates the center of energy deposition in each layer for all datasets. Typically, the major part of the energy deposition is concentrated near $r=0$ as incident particles are generated at the center, orthogonal to the detector plane. As the shower progresses, the interactions with the detector material leading in energy deposition spread away from the center. For instance, with layer 44, majority of the energy deposited is spread over higher values of $r$. In all cases, CaloScore is able to reproduce shower trend most accurately, followed by CaloDiffusion. CaloINN performs well on dataset 1 pion shower generation but struggles in other cases.

% \textbf{Shower width:}
% Shower width measures how far secondary particles spread perpendicular to the direction of the initial particle from the perspective of $\eta$ and $\phi$, measured in mm. Figures ~\ref{fig:sw_ds1_photon} through ~\ref{fig:sw_ds3} illustrate the angular distributions of the showers in terms of the shower width. For dataset 1 (photons), both CaloScore and CaloDiffusion perform similarly in terms of shower width. But, they show some degree of mismodeling (right end of the plots), which is especially evident in layers 25-30 for dataset 2 and 3. CaloINN shows significant mismodeling for all datasets.

\textbf{Sparsity:} Figure~\ref{fig:sparsity} illustrates the sparsity metric for dataset 2. Despite some mismodeling (e.g., layers 30-44), CaloScore generates most accurate showers. Both CaloDiffusion and CaloINN show significant deviation from Geant4. Similar observation can be made for dataset 3. Please refer to Appendix-\ref{app_c} (Figure~\ref{fig:histogram_observe_3}) for more details. %CaloScore exhibits some mismodeling in the final layers, specifically layers 35 to 44. 

\subsection{Correlation}\label{sec:pcc}

 \begin{figure}  
\subfloat[Geant4]
{\includegraphics[width=0.25\textwidth]
{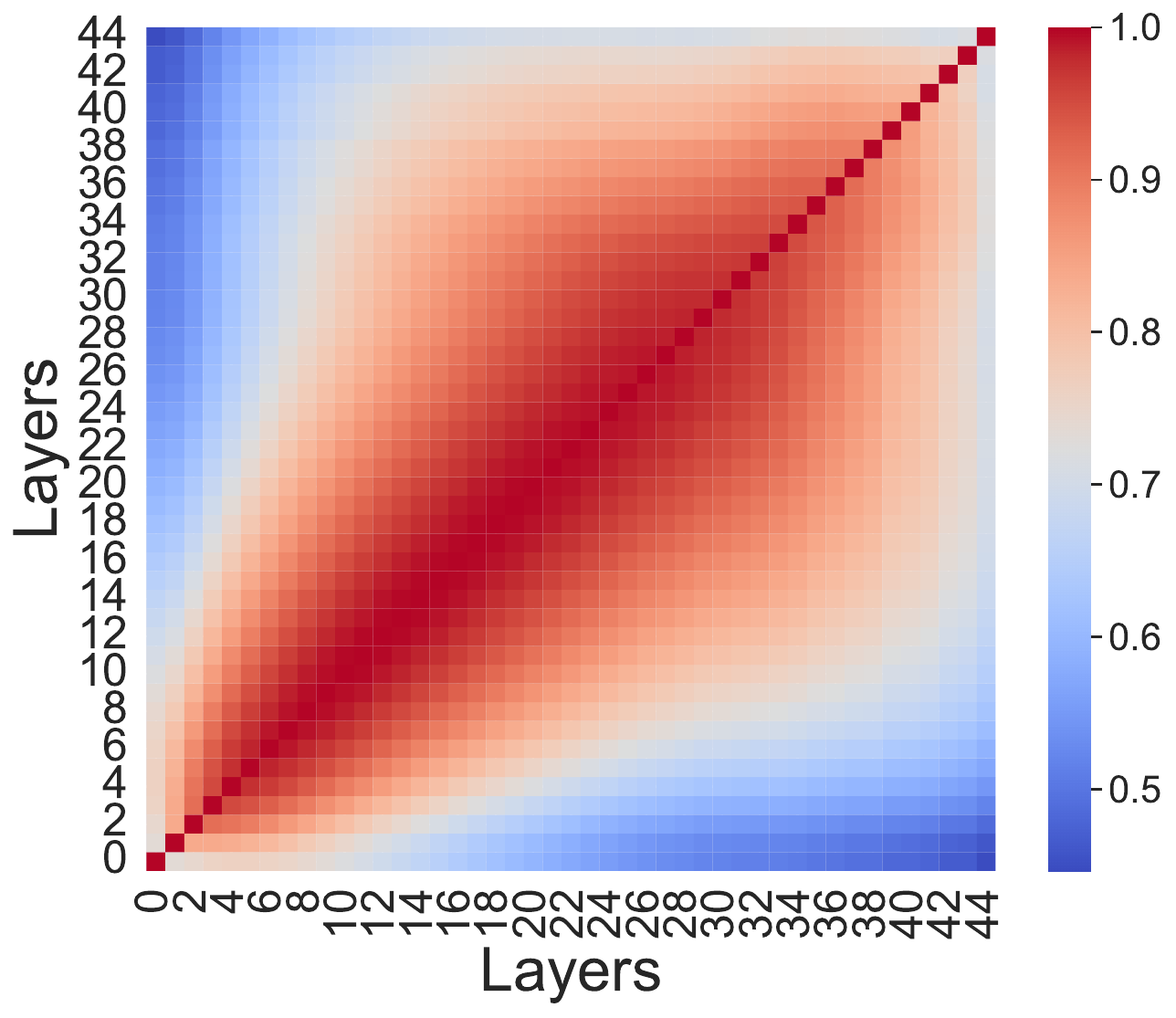}} 
\subfloat[CaloScore]
{\includegraphics[width=0.25\textwidth]
{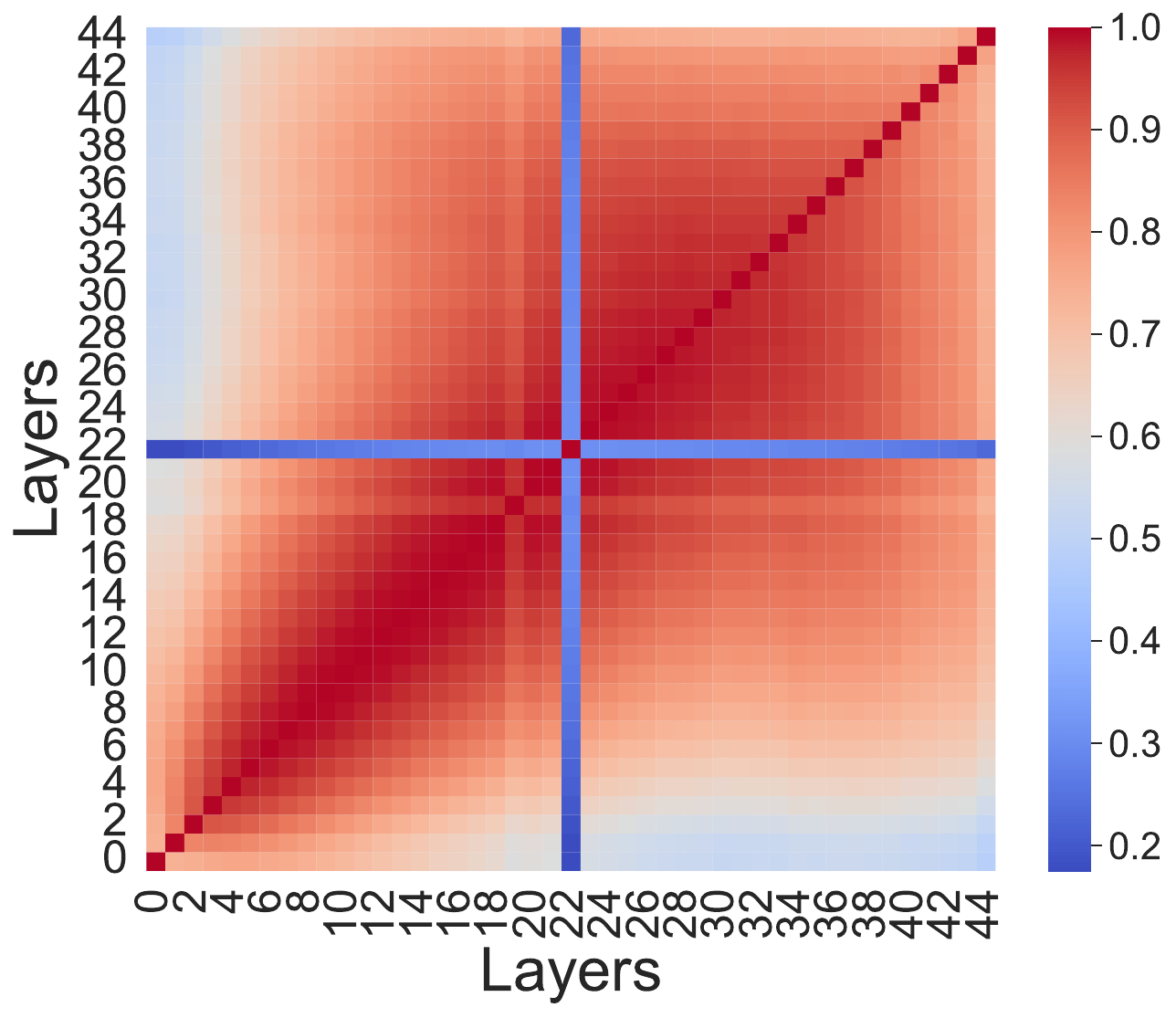}}
\subfloat[CaloDiffusion]
{\includegraphics[width=0.25\textwidth]
{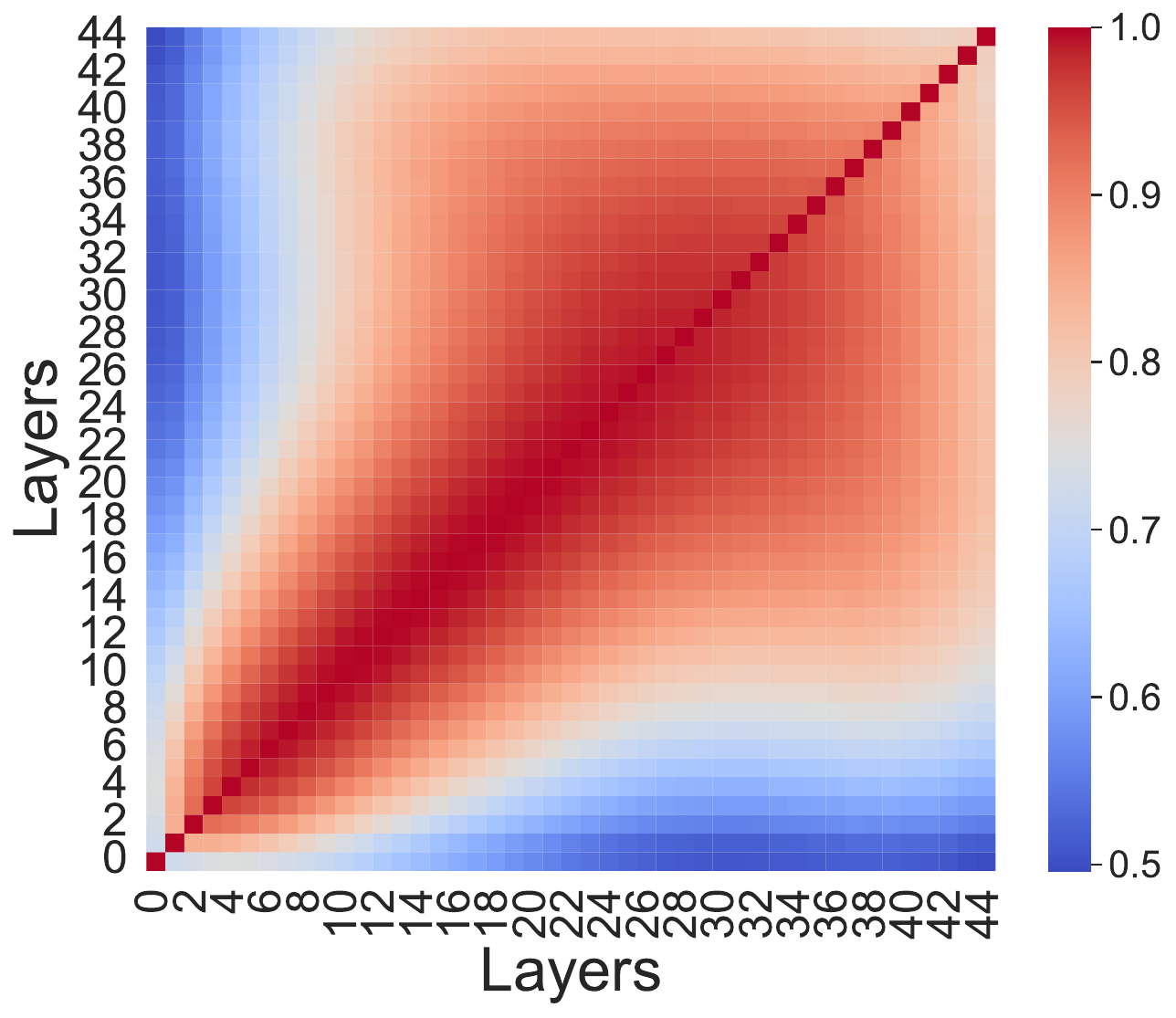}}
\subfloat[CaloINN]
{\includegraphics[width=0.25\textwidth]
{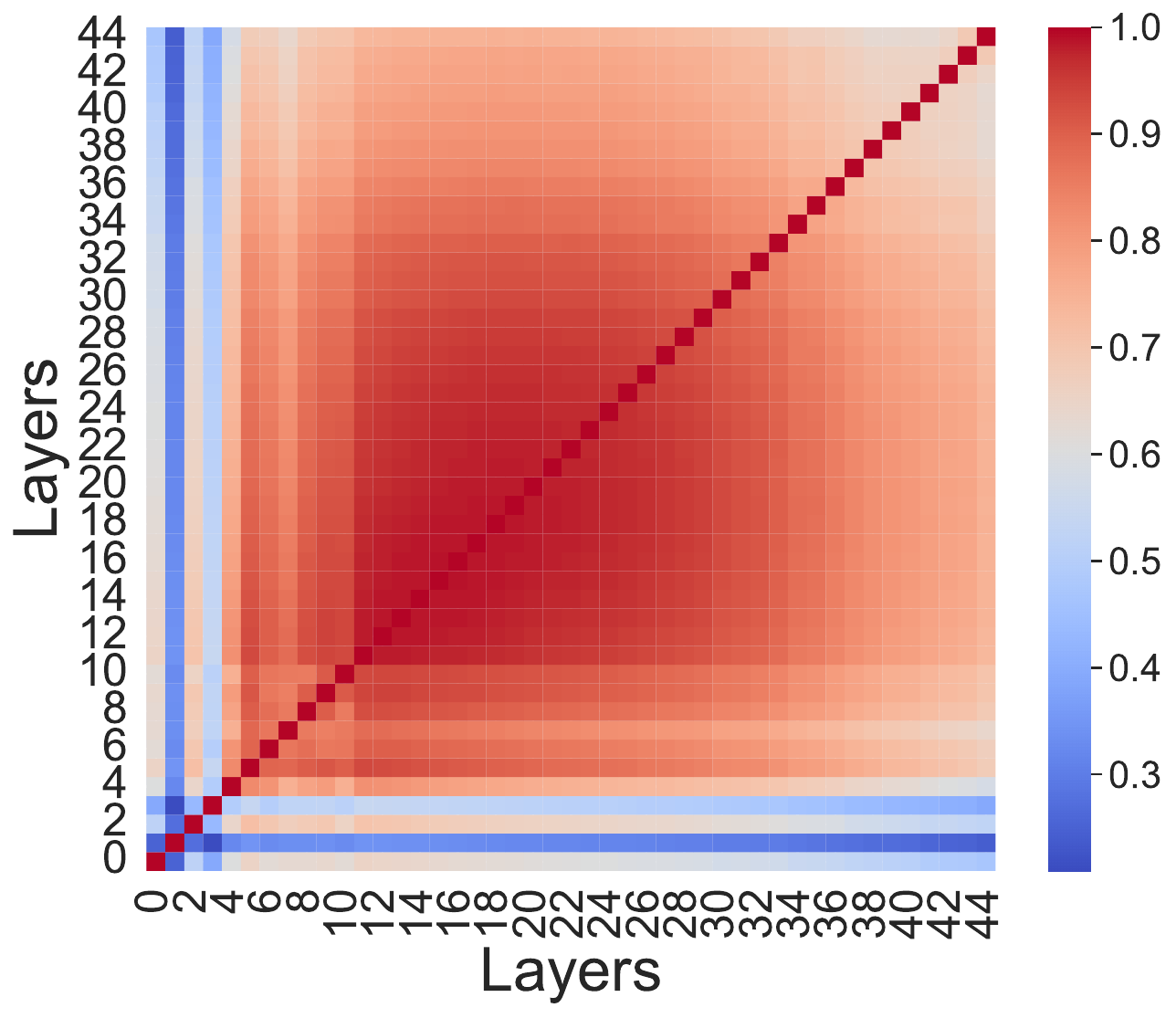}}
\caption{Layer wise correlation - dataset 2} \label{fig:corr_ds2}
\end{figure}

\begin{figure}  
\subfloat[Geant4]
{\includegraphics[width=0.25\textwidth]
{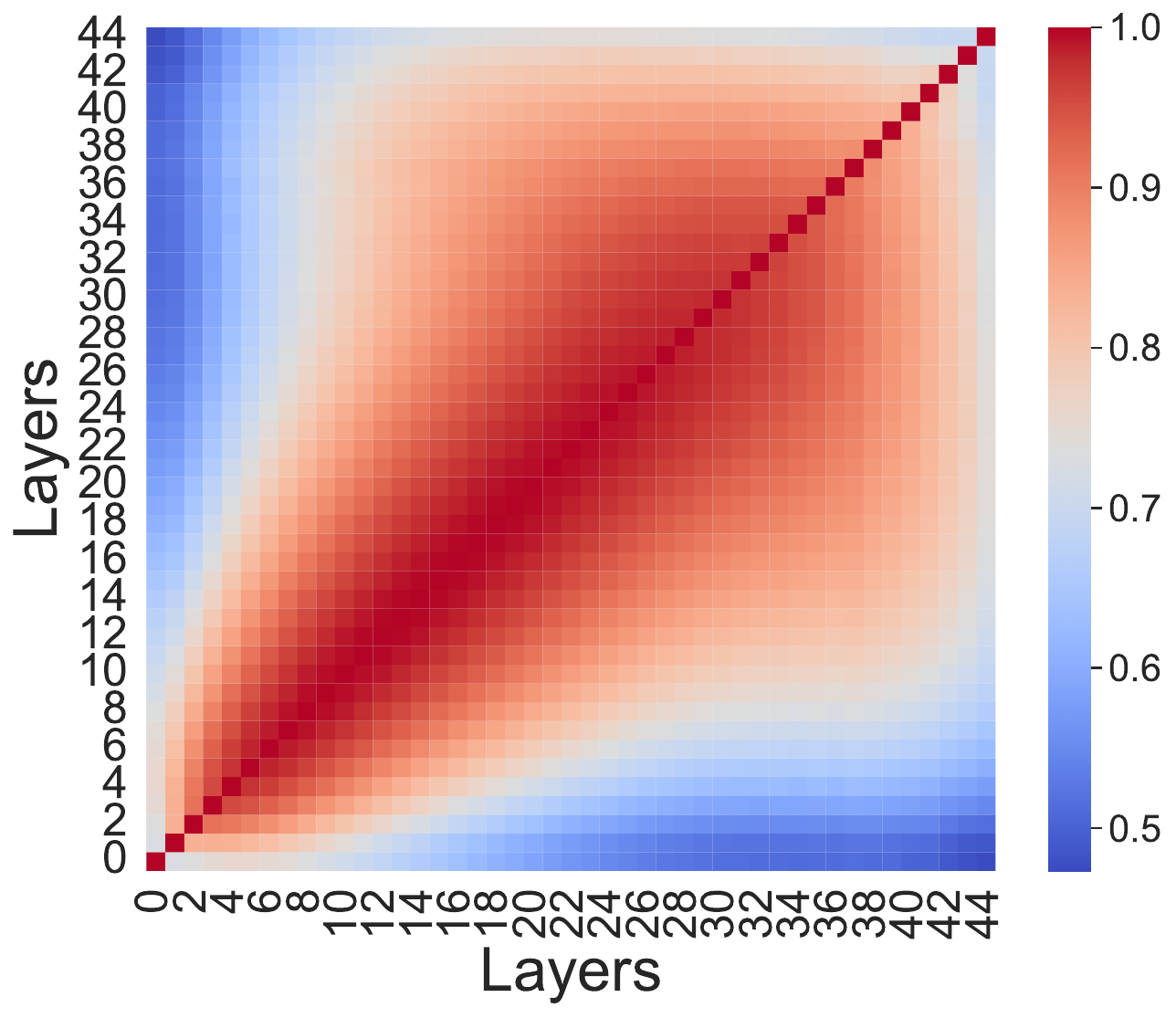}} 
\subfloat[CaloScore]
{\includegraphics[width=0.25\textwidth]
{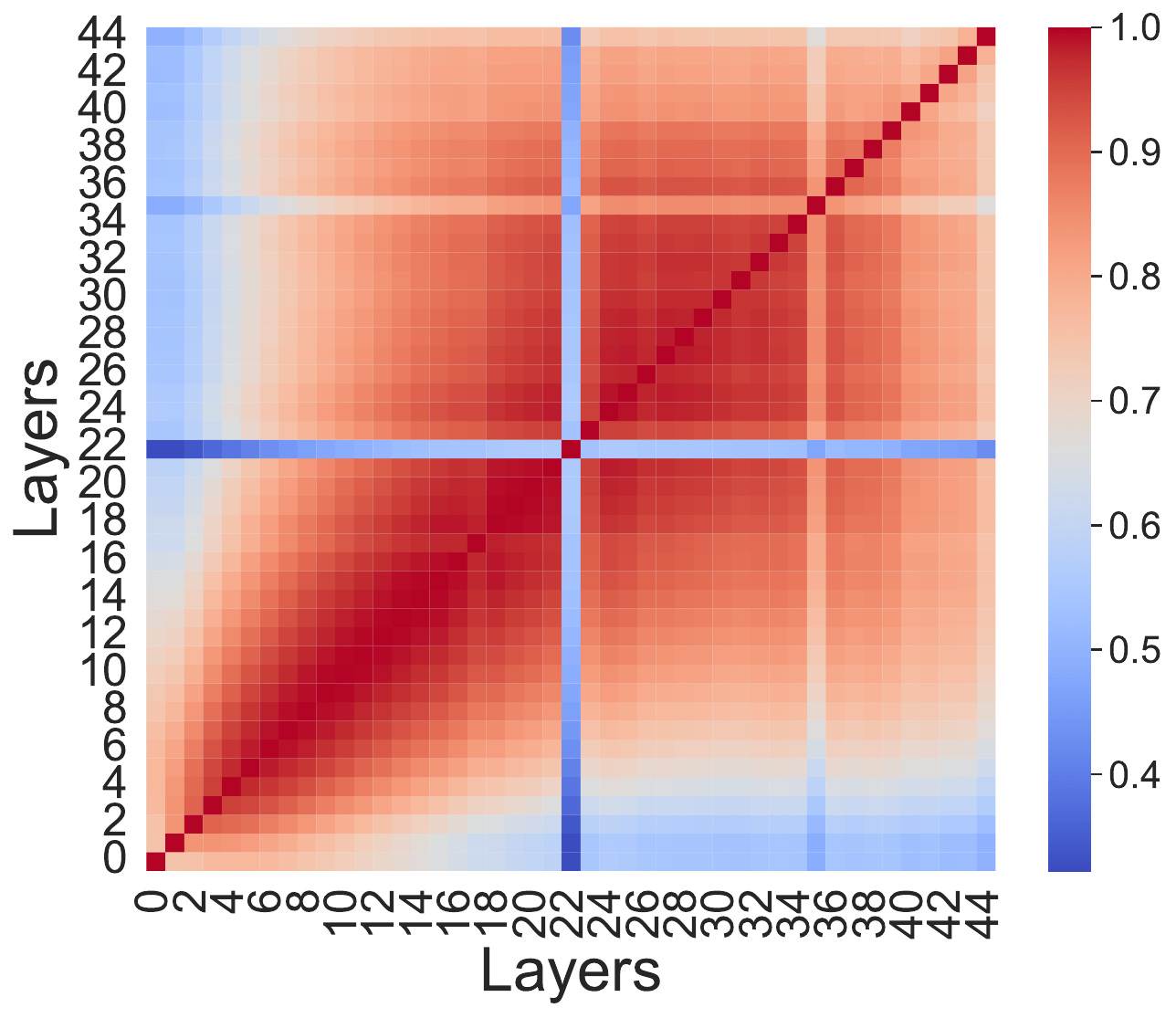}}
\subfloat[CaloDiffusion]
{\includegraphics[width=0.25\textwidth]
{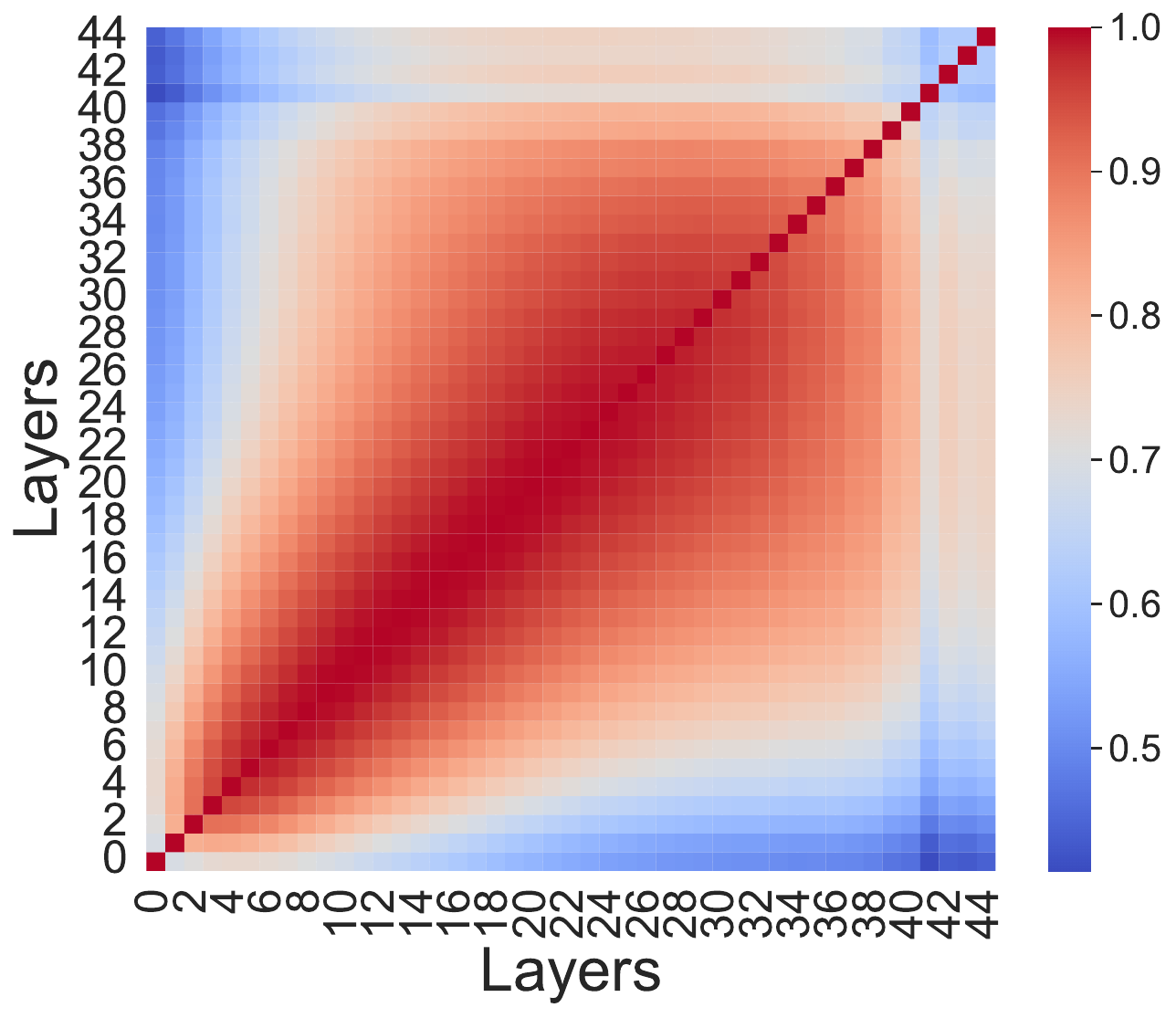}}
\subfloat[CaloINN]
{\includegraphics[width=0.25\textwidth]
{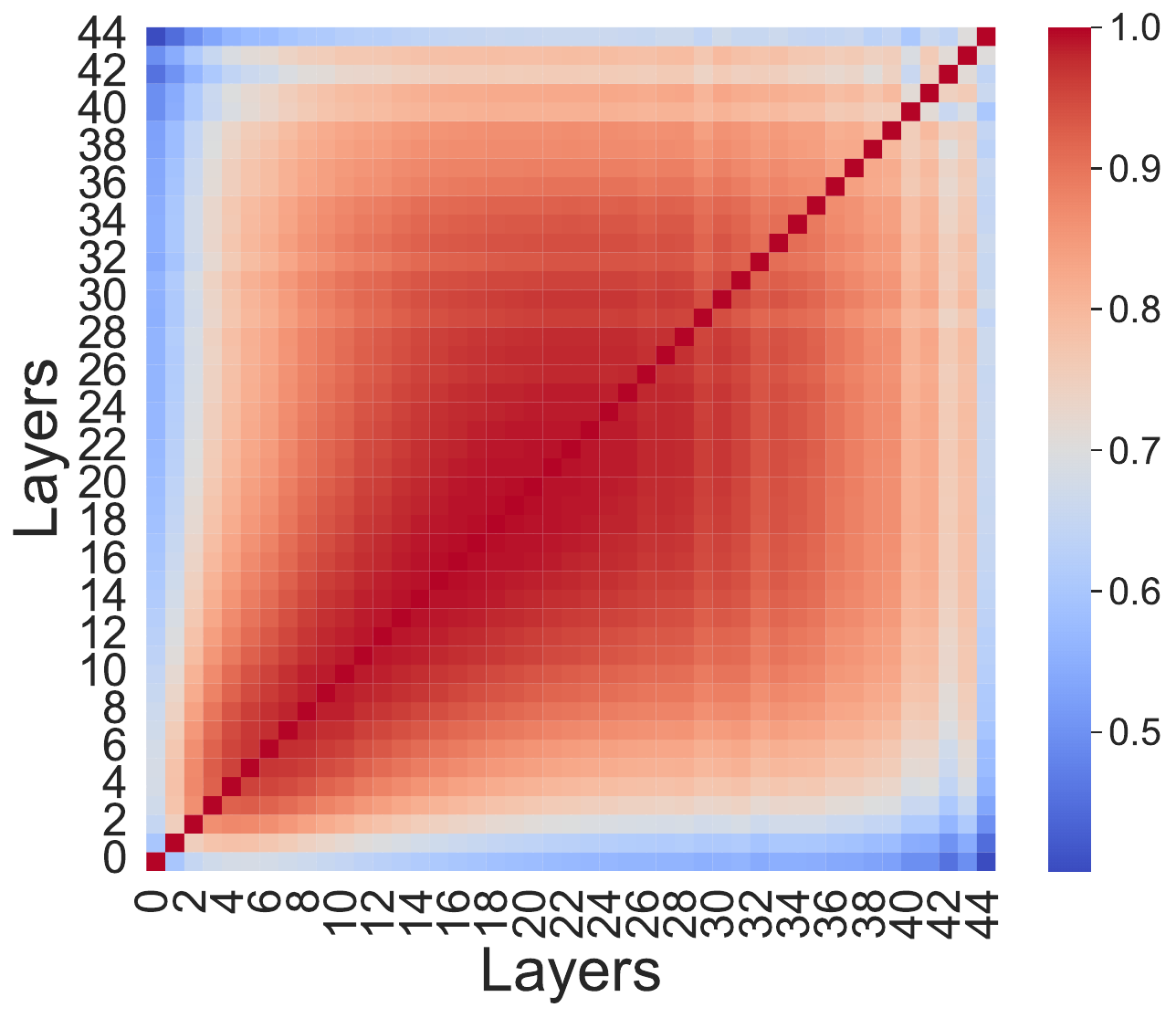}}
\caption{Layer wise correlation - dataset 3} \label{fig:corr_ds3}
\end{figure}

Figure~\ref{fig:corr_ds2} and~\ref{fig:corr_ds3} reports the PCC across layers for dataset 2 and 3, respectively. The distinctive layerwise PCC trend of Geant4 is well-modeled by CaloDiffusion. We do not notice any negative PCC in Geant4. However, when it comes to CaloScore and CaloINN, the pattern is overspread, failing to accurately capture it for specific layers (e.g., layer 22 for dataset 2 and 3 in CaloScore). Across multiple runs, this inconsistency persists for CaloScore for both dataset 2 and 3, indicating a potential modeling issue, which likely requires further investigation.

\subsection{Classifier test}
Table~\ref{tab:auc_jsd} enumerates the classifier test scores, measured as AUC and JSD, between Geant4 and showers generated by the models along with scores for mixed-precision inference for CaloDiffusion. The classifier learns a score proportional to the likelihood ratio between the synthetic and reference samples. The closer the likelihoods, the harder for the classifier to distinguish between the two samples. 

\begin{table}[tbh]
    \caption{AUC and JSD for CaloDiffusion, CaloScore and CaloINN (*CaloScore authors did not generate pion showers).}
    \label{tab:auc_jsd}
    \centering
    \resizebox{\textwidth}{!}{
    \begin{tabular}{llllllllll}
        \cline{1-10}
        \multicolumn{2}{c}{} & 
        \multicolumn{2}{c}{CaloDiffusion} &
        \multicolumn{2}{c}{CaloScore} &
        \multicolumn{2}{c}{CaloINN} &
        \multicolumn{2}{c}{CaloDiffusion (mix)} \\
        \cline{1-10}
         Dataset &Features &AUC & JSD & AUC & JSD & AUC & JSD & AUC & JSD\\ \hline
         \multirow{3}{*}{1(photons)} &low & $0.6265$ &$0.0409$ &$0.7736$ &$0.1758$ & $0.9258$ &$0.5203$ & $0.6865$ & $0.0782$ \\ \cline{2-10}
            &high & $0.6985$ &$0.1561$ &$0.5976$ &$0.0517$ & $0.8432$ &$0.3014$ & $0.7312$ & $0.1524$ \\ \cline{2-10}
            &normd & $0.6915$ &$0.1464$ &$0.7633$ &$0.1876$ & $0.9229$ &$0.5038$ & $0.7242$ & $0.1408$ \\ 
        \hline
         \multirow{3}{*}{1(pions)} &low & $0.6798$ &$0.0737$ &$*$ &$*$ & $0.9355$ &$0.5365$ & $0.6303$ & $0.0446$ \\ \cline{2-10}
            &high & $0.7263$ &$0.1530$ &$*$ &$*$ & $0.8886$ &$0.3921$ & $0.7020$ & $0.1598$ \\ \cline{2-10}
            &normd & $0.7264$ &$0.1416$ &$*$ &$*$ & $0.9424$ &$0.5606$ & $0.6972$ & $0.1592$ \\ 
        \hline
         \multirow{3}{*}{2(electrons)} &low & $0.5785$ &$0.0120$ &$0.6134$ &$0.0324$ & $0.9841$ & $0.7971$ & $0.5800$ & $0.0145$ \\ \cline{2-10}
            &high & $0.5629$ &$0.0110$ &$0.6326$ &$0.0433$ & $0.9861$ & $0.7908$ & $0.5540$ & $0.0092$ \\ \cline{2-10}
            &normd & $0.5448$ &$0.0079$ &$0.5787$ &$0.0152$ & $0.9662$ & $0.6610$ & $0.5544$ & $0.0092$ \\ 
        \hline
         \multirow{3}{*}{3(electrons)} &low & $0.5686$ &$0.0100$ &$0.8068$ &$0.2599$ & $0.3853$ &$0.4621$ & $0.5688$ & $0.0112$ \\ \cline{2-10}
            &high & $0.5920$ &$0.0246$ &$0.7797$ &$0.2014$ & $0.9974$ &$0.9167$ & $0.5805$ & $0.0177$ \\ \cline{2-10}
            &normd & $0.5463$ &$0.0091$ &$0.8132$ &$0.2527$ & $0.9265$ &$0.5139$ & $0.5558$ & $0.0117$ \\ 
        \hline
    \end{tabular}
    }
\end{table}

From Table~\ref{tab:auc_jsd}, CaloDiffusion generates samples most similar to Geant4, with the AUC score being as closest to $0.5$ and JSD being the least among other models. CaloScore is the second best but for dataset 3, where the AUC values are closer to $1$. CaloINN generates the least similar showers. CaloDiffusion does equally well using mixed-precision inference with the AUC and JSD scores being similar or better than full precision inference. This novel experimental result proves that mixed or lower precision inference does not affect the overall accuracy of the generated samples.

\subsection{EMD, FPD, KPD scores}\label{sec:emd-fpd-kpd}

 \begin{figure}  
\subfloat[EMD - Dataset 2 \label{fig:emd_sparsity_ds2}]
{\includegraphics[width=0.25\textwidth]
{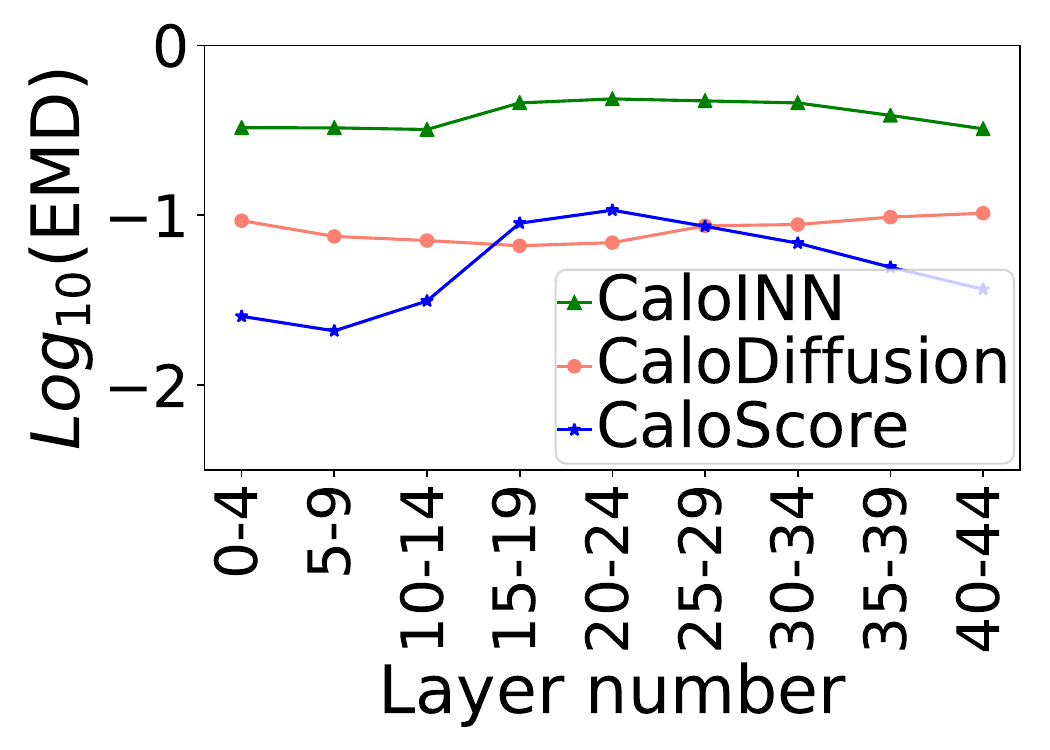}} 
\subfloat[EMD - Dataset 3 \label{fig:emd_sparsity_ds3}]
{\includegraphics[width=0.25\textwidth]
{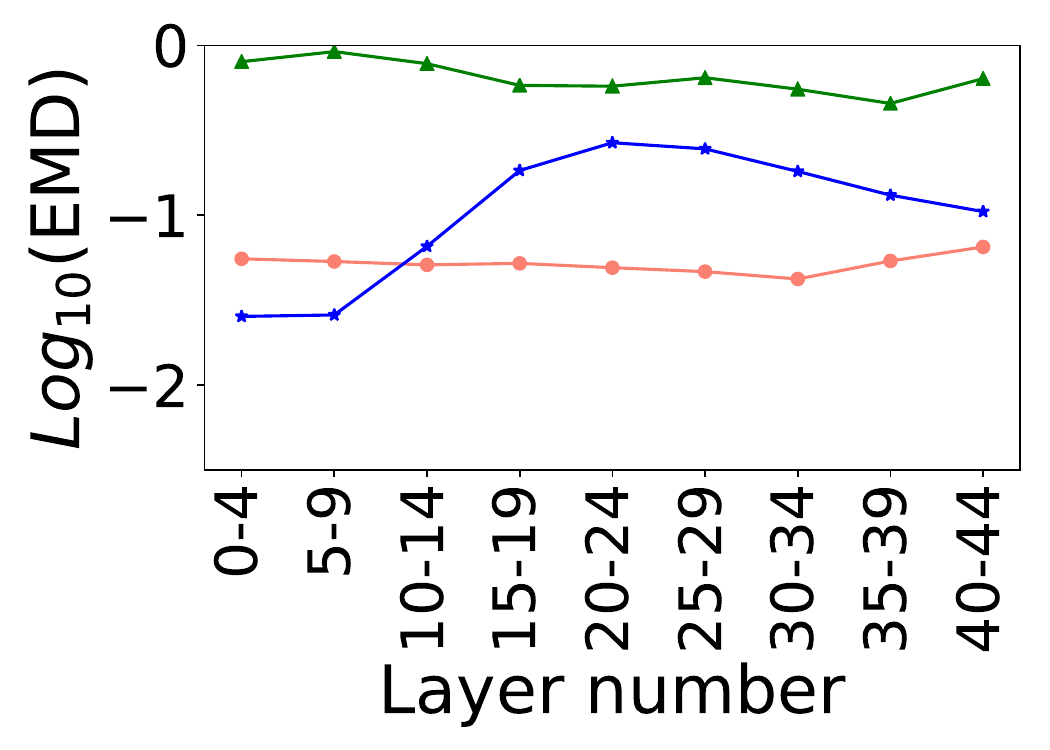}}
\subfloat[S.P. - Dataset 2 \label{fig:sp_sparsity_ds2}]
{\includegraphics[width=0.25\textwidth]
{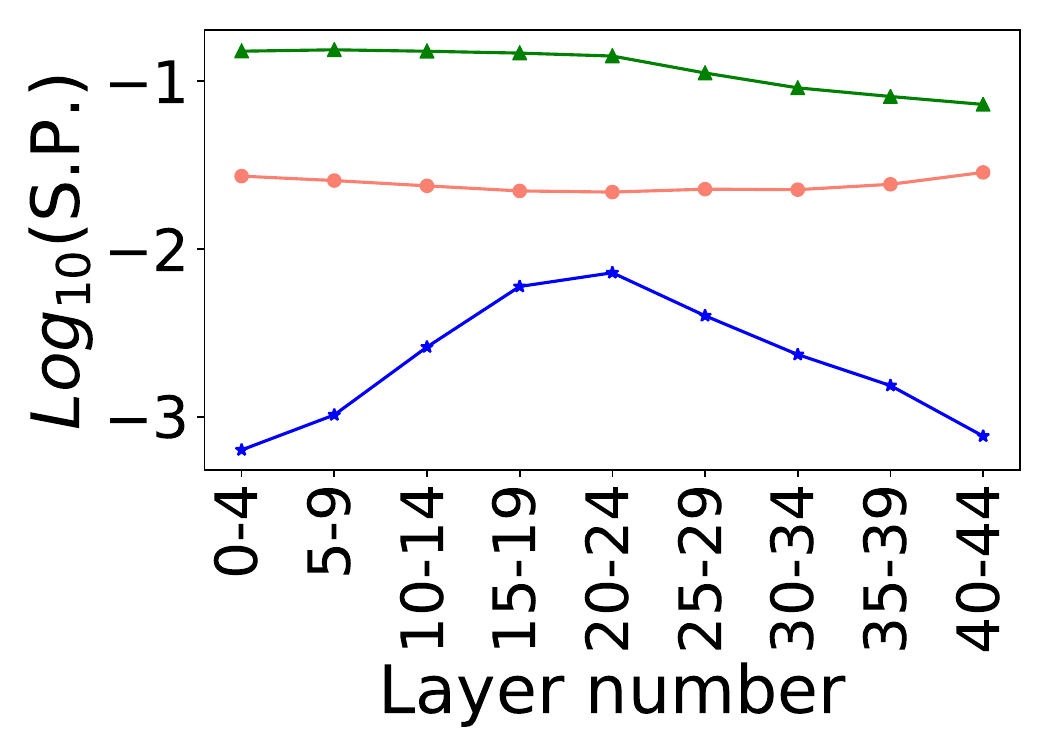}} 
\subfloat[S.P. - Dataset 3 \label{fig:sp_sparsity_ds3}]
{\includegraphics[width=0.25\textwidth]
{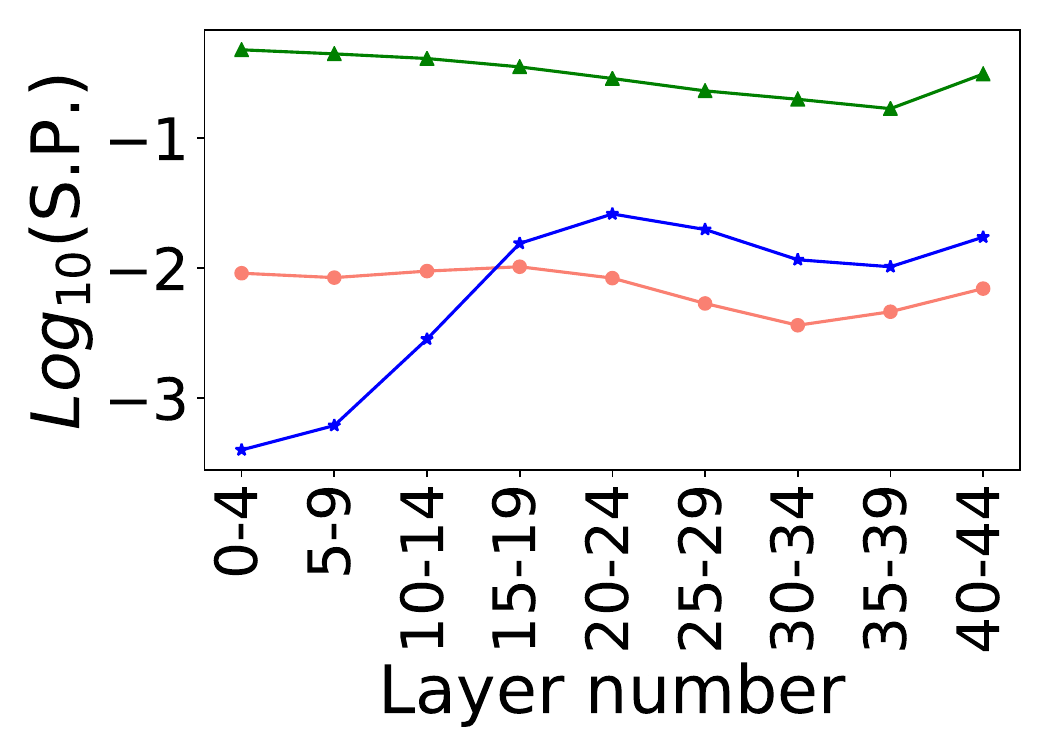}}
\caption{EMD scores and separation power for sparsity with datasets 2 and 3.} \label{fig:emd_main}
\end{figure}

%The smaller the EMD score, the closer the two distributions are to each other. Here we compare each physical observable relative to Geant4. 
Our observations indicate that the EMD score is more sensitive compared to histograms of various physical features. For instance, CaloDiffusion accurately models the shower shape and layer-wise energy distribution (Figures~\ref{fig:layer_engergy_all},~\ref{fig:ec_ds2},\ref{fig:ec_ds3},\ref{fig:sw_ds2_photon},\ref{fig:sw_ds3}). And it is consistent in EMD score plots. However, regarding the distribution of sparsity, CaloDiffusion underperforms than CaloScore in the initial layers for dataset 3 (Figure~\ref{fig:sparsity_ds3}) and later layers for dataset 2 (Figure~\ref{fig:sparsity}). In these layers, the EMD score is highly sensitive to differences compared to Geant4 (Figures~\ref{fig:emd_sparsity_ds2},~\ref{fig:emd_sparsity_ds3}). Refer to Appendix Figure~\ref{fig:ds2_emd} and~\ref{fig:ds3_emd} for EMD scores of the models relative to Geant4 across other physical features of the shower.
% The FPD and KPD measure the similarity of physics observables based on multidimensional correlated features, both in terms of quality and diversity between two distributions. The smaller the difference, the more similar two distributons are.

\begin{table}[tbh]
    \caption{FPD and KDP for CaloDiffusion, CaloScore and CaloINN (*CaloScore authors did not generate pion showers).}
    \label{tab:fpd_kpd}
    \centering
    \resizebox{\textwidth}{!}{
    \begin{tabular}{lllllllll}
        \toprule
        %\cline{1-9}
        \multicolumn{1}{c}{} & 
        \multicolumn{2}{c}{CaloDiffusion} &
        \multicolumn{2}{c}{CaloScore} &
        \multicolumn{2}{c}{CaloINN} &
        \multicolumn{2}{c}{CaloDiffusion (mix)} \\
        \cline{1-9}
         Dataset &FPD ($\times 10^{-3}$) & KPD ($\times 10^{-3}$) & FPD ($\times 10^{-3}$) & KPD ($\times 10^{-3}$) & FPD ($\times 10^{-3}$) & KPD ($\times 10^{-3}$) & FPD ($\times 10^{-3}$) & KPD ($\times 10^{-3}$) \\ \hline
         \multirow{1}{*}{1(photons)} & $25.2016\scriptstyle{\pm 0.4}$ & $5.2317\scriptstyle{\pm 1.5}$ &$4.7172\scriptstyle{\pm 0.1}$ &$1.1278\scriptstyle{\pm 0.6}$ & $85.5213\scriptstyle{\pm 1.1}$ &$26.729 \scriptstyle{\pm 3.5}$ & $27.6084 \scriptstyle{\pm 0.4}$ & $5.7032 \scriptstyle{\pm 1.4}$\\  
         \hline
         \multirow{1}{*}{1(pions)} & $34.5148 \scriptstyle{\pm 0.4}$ & $3.4485 \scriptstyle{\pm 0.8}$ &$*$ &$*$ &$60.0015\scriptstyle{\pm 1.5}$ & $4.4875\scriptstyle{\pm 0.9}$ & $33.1041 \scriptstyle{\pm 0.5}$ & $3.3565 \scriptstyle{\pm 0.7}$\\ 
         \hline
         \multirow{1}{*}{2(electrons)} & $52.7041 \scriptstyle{\pm 1.4}$ & $0.0992\scriptstyle{\pm 0.1}$ & $36.6782\scriptstyle{\pm 0.6}$ & $0.0710\scriptstyle{\pm 0.0}$ & $1044.9918\scriptstyle{\pm 9.3}$ & $5.314\scriptstyle{\pm 0.5}$ & $51.5933 \scriptstyle{\pm 1.1186}$ & $0.11\scriptstyle{\pm 0.2}$\\ 
        \hline
         \multirow{1}{*}{3(electrons)} & $46.2705\scriptstyle{\pm 1.5}$ & $0.215\scriptstyle{\pm 0.2}$ & $120.4597 \scriptstyle{\pm 1.5}$ & $0.3599\scriptstyle{\pm 0.1}$ &$1389.2166 \scriptstyle{\pm 2.5}$ & $4.8813\scriptstyle{\pm 0.3}$ & $45.5731\scriptstyle{\pm 1.4}$ & $0.19 \scriptstyle{\pm 0.1}$\\ 
        \hline
    \end{tabular}
    }
\end{table}

The FPD and KPD scores of the models are listed in  Table~\ref{tab:fpd_kpd}. We note that CaloScore generates accurate showers for datasets 1 and 2, compared to the other models. This indicates that CaloScore can capture the correlation among high level features well. However, when the correlation analysis was conducted at a more granular level, we observed it performing poorly (Section~\ref{sec:pcc}). For dataset 3, CaloDiffusion generates more accurate showers whereas CaloINN does poorly both for FPD and KPD measures on all datasets. Additionally, CaloDiffusion with mixed-precision inference shows improved FPD and KPD scores, evidence that lower precision does not necessarily compromise accuracy.

\subsection{Separation Power}  Figure~\ref{fig:sp_sparsity_ds2} and~\ref{fig:sp_sparsity_ds3} show that separation power is also sensitive compared to histogram of sparsity feature, similar to the observation in Section~\ref{sec:emd-fpd-kpd}. Please refer to Appendix-\ref{app_c}, Figure~\ref{fig:ds2_sep} and~\ref{fig:ds3_sep} for separation power analysis of other physical features. The figures indicate that the separation power between Geant4 and CaloINN is higher than other models, indicating CaloINN is less accurate in generating showers. On the other hand, CaloDiffusion and CaloScore perform well for datasets 2 and 3, CaloScore being slightly better.%CaloINN performs significantly worse than the other two models. 

\subsection{Timing}
Table~\ref{tab:inference_time_models} lists the inference time per event on CPU and GPU for all models. Missing entries indicate either hardware or modeling issues (e.g., lack of proper documentation). 
% Missing entries in the table indicate that we did not record the time for that configuration due to hardware limitations (e.g., models failing to load on the available GPU memory and incomplete documentation for CaloINN+VAE in recording time).
The last column reports sample generation time of CaloDiffusion in mixed-precision mode, where we observe significant improvement for datasets 2 and 3, which are high in dimensionality, especially for larger batch sizes. For low dimension dataset 1, mixed precision has a longer generation time, even with higher batch sizes. To investigate this further, we used \textit{wandb.ai}~\cite{wandb} to monitor real-time GPU utilization with different batch sizes (Refer to Appendix~\ref{app_c}, Figure~\ref{fig:gpu_uti}) and find that GPU is not properly utilized for dataset 1(pion). Mixed precision can speed up the sample generation time if the inference process can utilize the GPU fully~\cite{PyTorchAMPRecipe}. To accomplish this, it is recommended to use large batch size or high dimensional data. Since Dataset 1 is low in dimensionality compared to the other datasets, it performs poorly in mixed precision mode. 
%That is why it is not possible to speed up dataset 1 inference with mixed precision.

\begin{table}[tbh]
    \centering
    \caption{Shower generation time per event for CaloDiffusion on CPU and GPU for various batch sizes (BS) (*CaloScore authors did not generate pion showers).}
    \label{tab:inference_time_models}
    %\begin{tabular}{llllllll}
    \resizebox{\textwidth}{!}{
    \begin{tabular}{p{0.10\linewidth}p{0.02\linewidth}p{0.10\linewidth}
    p{0.11\linewidth}p{0.11\linewidth}p{0.10\linewidth}
    p{0.10\linewidth}p{0.10\linewidth}p{0.05\linewidth}}
        \cline{1-9}
        \multicolumn{2}{c}{} & 
        \multicolumn{2}{c}{CaloDiffusion (sec)} &
        \multicolumn{2}{c}{CaloScore (sec)} &
        \multicolumn{2}{c}{CaloINN (sec)} &
        \multicolumn{1}{c}{CaloDiffusion (sec)} \\
        \cline{1-9}
         Dataset &BS &CPU & GPU & CPU & GPU & CPU$\scriptstyle{(\times 10^{-3})}$ & GPU$\scriptstyle{(\times 10^{-3})}$ & GPU(mix)\\ \hline
         \multirow{3}{*}{1(photons)}& $1$ &$10.40 \scriptstyle{\pm 0.39}$ &$4.64 \scriptstyle{\pm 0.06}$ &$73.93 \scriptstyle{\pm 0.23}$ & $3.90 \scriptstyle{\pm 0.20}$ &$23.66 \scriptstyle{\pm 1.75}$ & $25.58 \scriptstyle{\pm 0.3}$ & $5.44 \scriptstyle{\pm 0.27}   $ \\ %\cline{2-8}
          &$10$&$2.85 \scriptstyle{ \pm 0.11}$&$0.48 \scriptstyle{\pm  0.01}$ & $73.65 \scriptstyle{\pm 0.26}$ & $4.04 \scriptstyle{\pm 0.18}$ & $ 5.78 \scriptstyle{\pm 0.27}$ & $2.57 \scriptstyle{\pm 0.00}$ & $0.57 \scriptstyle{\pm 0.02}  $ \\ %\cline{2-8}
          &$100$&$3.22 \scriptstyle{\pm 0.08}$ &$0.07 \scriptstyle{ \pm 0.0005}$ & $73.67 \scriptstyle{\pm 0.20}$ & $3.88 \scriptstyle{\pm 0.19}$ & $2.36 \scriptstyle{\pm 0.2}$ & $0.27 \scriptstyle{\pm 0.004}$ & $0.07 \scriptstyle{\pm 0.7\times10^{-3}}$ \\
         \hline
         \multirow{3}{*}{1(pions)}& $1$ &$11.13 \scriptstyle{\pm 0.21}$ &$4.65 \scriptstyle{\pm 0.03}$ & $*$ & $*$ & $30.63 \scriptstyle{\pm 0.59}$ & $ 26.06 \scriptstyle{\pm 0.14}$ & $5.67\scriptstyle{\pm 0.08}  { }$  \\ %\cline{2-8}
          &$10$&$2.96 \scriptstyle{\pm 0.17}$&$0.49 \scriptstyle{\pm 0.0035}$& $*$ & $*$ & $9.03 \scriptstyle{\pm 0.05}$ & $2.67 \scriptstyle{\pm 0.04}$ & $0.57 \scriptstyle{\pm 0.01}  $ \\ %\cline{2-8}
          &$100$&$3.49\scriptstyle{ \pm 0.27}$ & $0.08 \scriptstyle{\pm 0.0005}
$ & $*$ & $*$ & $3.24 \scriptstyle{\pm  0.03}$ & $0.30 \scriptstyle{\pm 0.004}$ & $0.07 \scriptstyle{\pm 0.9 \times 10^{-3}}$ \\
         \hline
         \multirow{3}{*}{2 (electrons)}& $1$ &$37.21 \scriptstyle{ \pm 3.83}$ &$4.55 \scriptstyle{\pm 0.16}$ & $251.38 \scriptstyle{\pm 0.19}$ & $39.79 \scriptstyle{\pm 0.21}$ & $282.16 \scriptstyle{\pm 2.14}$ & $  42.83 \scriptstyle{\pm 1.64}$ & $4.86 \scriptstyle{\pm 0.04}  $ \\ %\cline{2-8}
          &$10$&$10.92 \scriptstyle{\pm 0.17}$&$0.54 \scriptstyle{\pm 0.01}$ & $251.55 \scriptstyle{\pm 0.19}$ & $39.34 \scriptstyle{\pm 0.22}$ & $111.92 \scriptstyle{\pm 0.55}$ & $5.22  \scriptstyle{\pm 0.06}$ & $\mathbf{0.50 \scriptstyle{\pm 0.01}}$ \\ %\cline{2-8}
          &$100$& $22.71 \scriptstyle{\pm  7.03}$&$0.22 \scriptstyle{\pm 0.0001}$ & $251.11 \scriptstyle{\pm 0.17}$ & $39.55 \scriptstyle{\pm 0.21}$ & $53 \scriptstyle{\pm 0.28}$ & $1.49 \scriptstyle{\pm 0.10}$ & $\mathbf{0.12 \scriptstyle{\pm 0.1\times 10^{-3}}}$ \\
         \hline
         \multirow{3}{*}{3 (electrons)}& $1$ &$167.11 \scriptstyle{\pm 4.04}$ &$5.80\scriptstyle{\pm 0.04}$ & $1242.45 \scriptstyle{\pm 0.19}$ & $198.58 \scriptstyle{\pm 0.17}$ & $-$ & $-$ & $\mathbf{5.61\scriptstyle{\pm 0.06}}$ \\ %\cline{2-8}
          &$10$&$166.54 \scriptstyle{ \pm 5.50}$&$2.47 \scriptstyle{ \pm 0.0006}$ & $1241.74 \scriptstyle{\pm 0.19}$ & $199.98 \scriptstyle{\pm 0.20}$ & $-$ & $-$ & $\mathbf{1.46\scriptstyle{\pm 04\times 10^{-3}}}$ \\ %\cline{2-8}
          &$100$&$172.25 \scriptstyle{ \pm  6.31}$ &$1.92 \scriptstyle{\pm 0.02}$ & $1241.89 \scriptstyle{\pm 0.23}$ & $199.10 \scriptstyle{\pm 0.18}$ & $-$ & $-$ & $\mathbf{1.09 \scriptstyle{\pm 1.9 \times 10^{-3}}  }$ \\
         \hline
    \end{tabular}
    }
\end{table}

\subsection{Mixed precision inference evaluation}

\begin{figure}  
\subfloat[Layer energy distribution (GeV)  \label{fig:layer_engergy_all_mix}]
{\includegraphics[width=0.5\textwidth]
{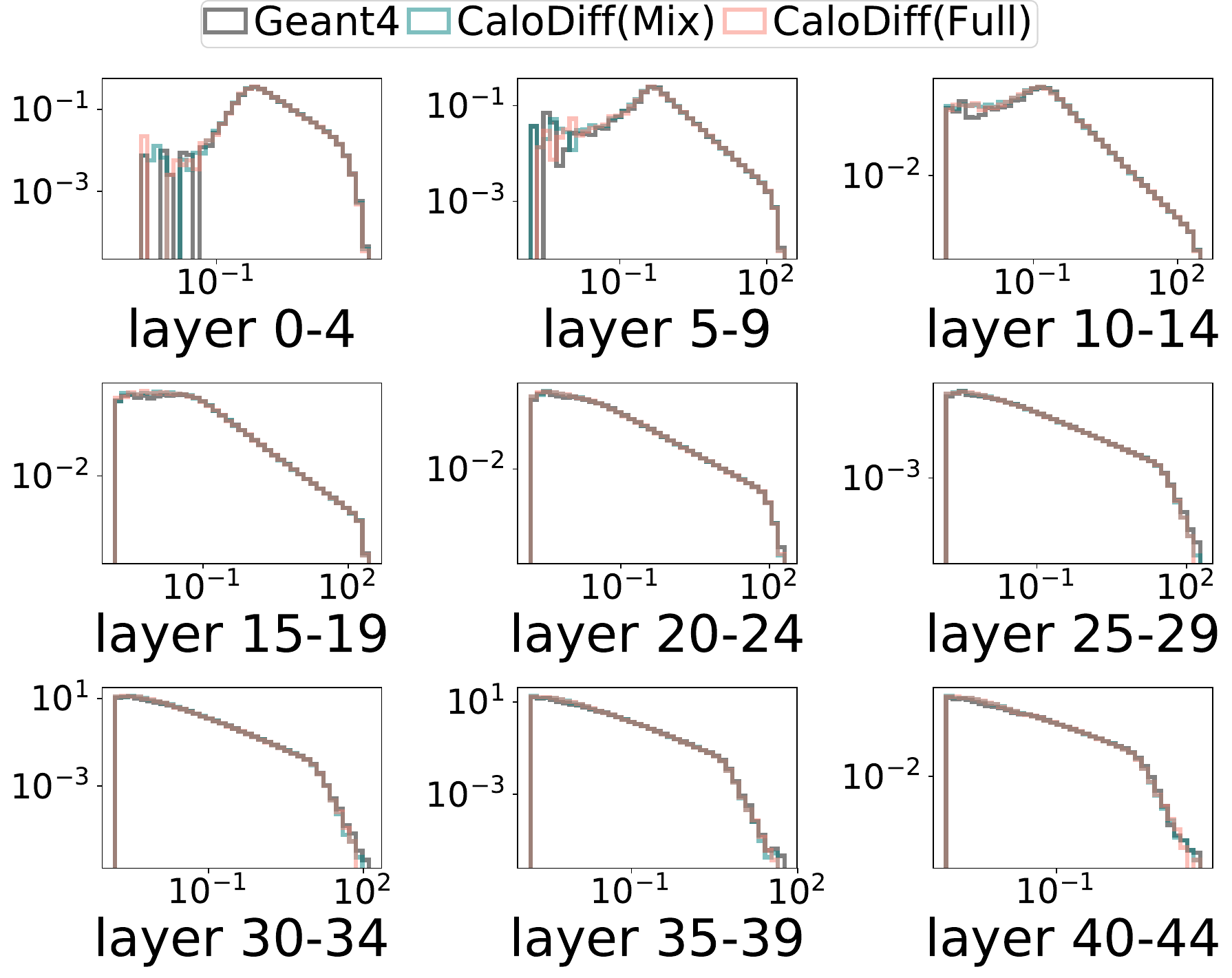}} 
\subfloat[Distribution of sparsity \label{fig:sparsity_mix}]
{\includegraphics[width=0.5\textwidth]
{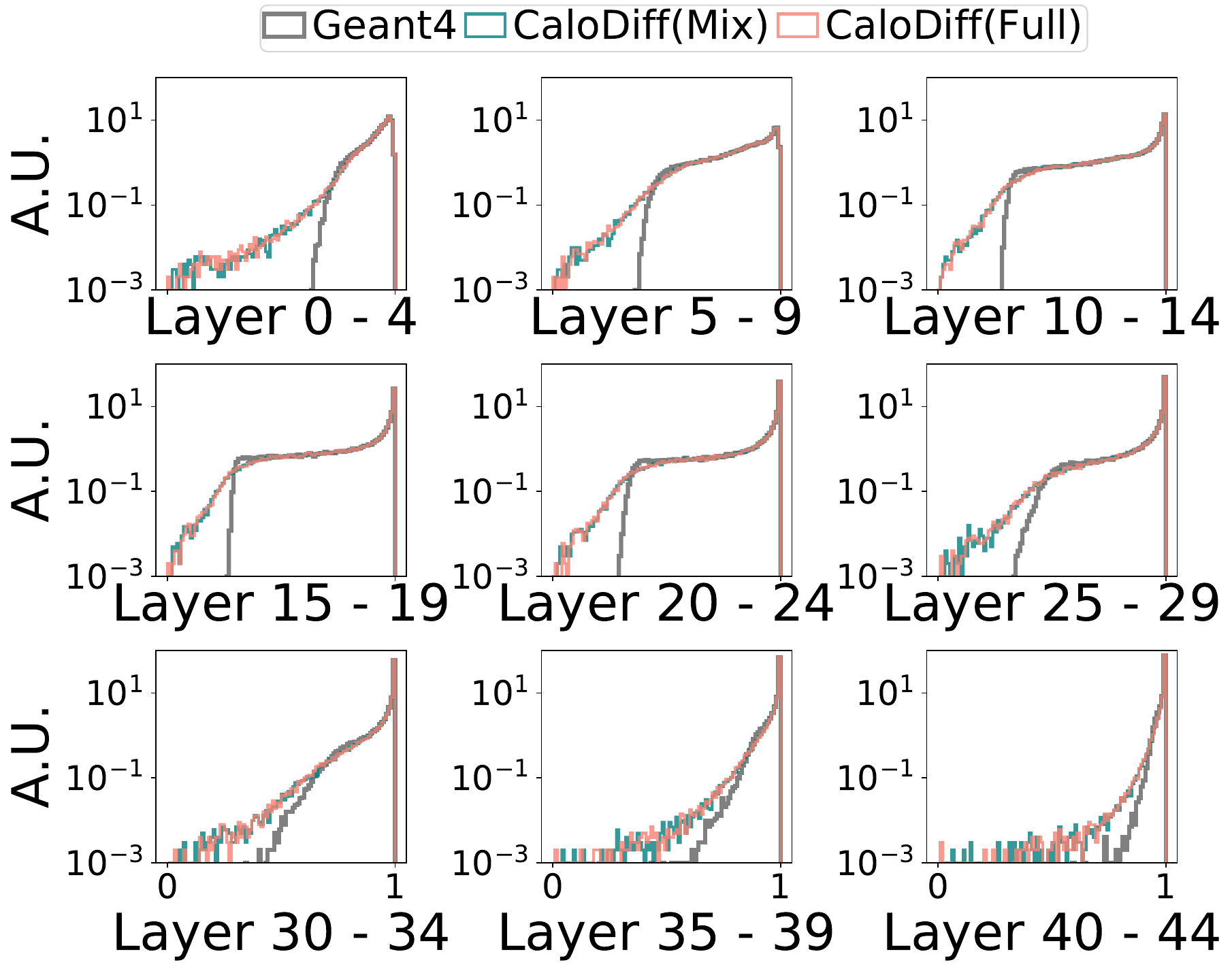}} 
\caption{Comparison of histogram of two physics observables with dataset 3 between full precision and mixed precision mode using CaloDiffusion.} \label{fig:histogram_observable_mix}
\end{figure}

 \begin{figure}  
\subfloat[EMD - Layer energy distribution \label{fig:emd_mix_main}]
{\includegraphics[width=0.25\textwidth]
{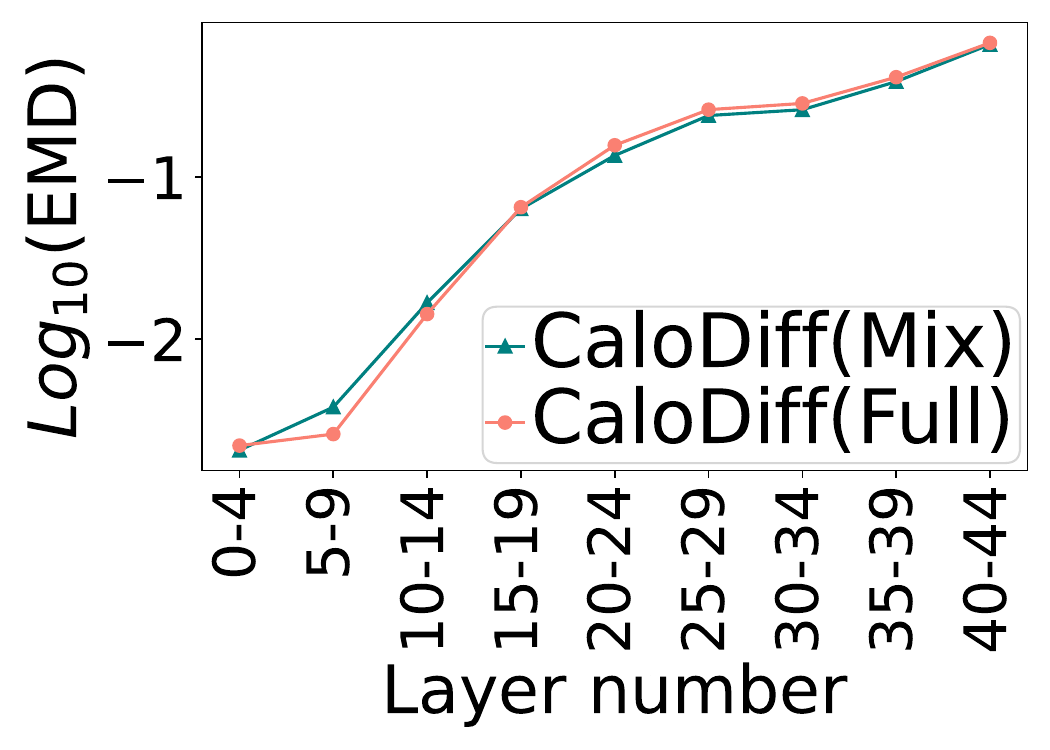}} 
\subfloat[EMD - Sparsity \label{fig:emd_sparsity_mix_ds3}]
{\includegraphics[width=0.25\textwidth]
{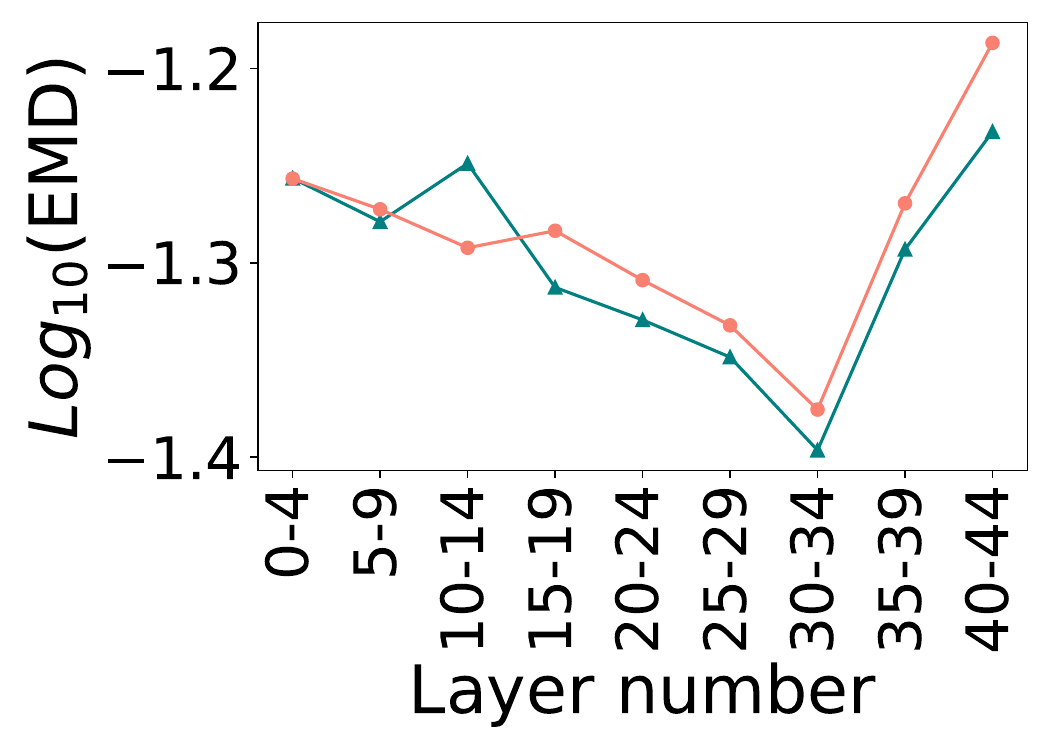}}
\subfloat[S.P. - Layer energy distribution\label{fig:sep_mix_main}]
{\includegraphics[width=0.25\textwidth]
{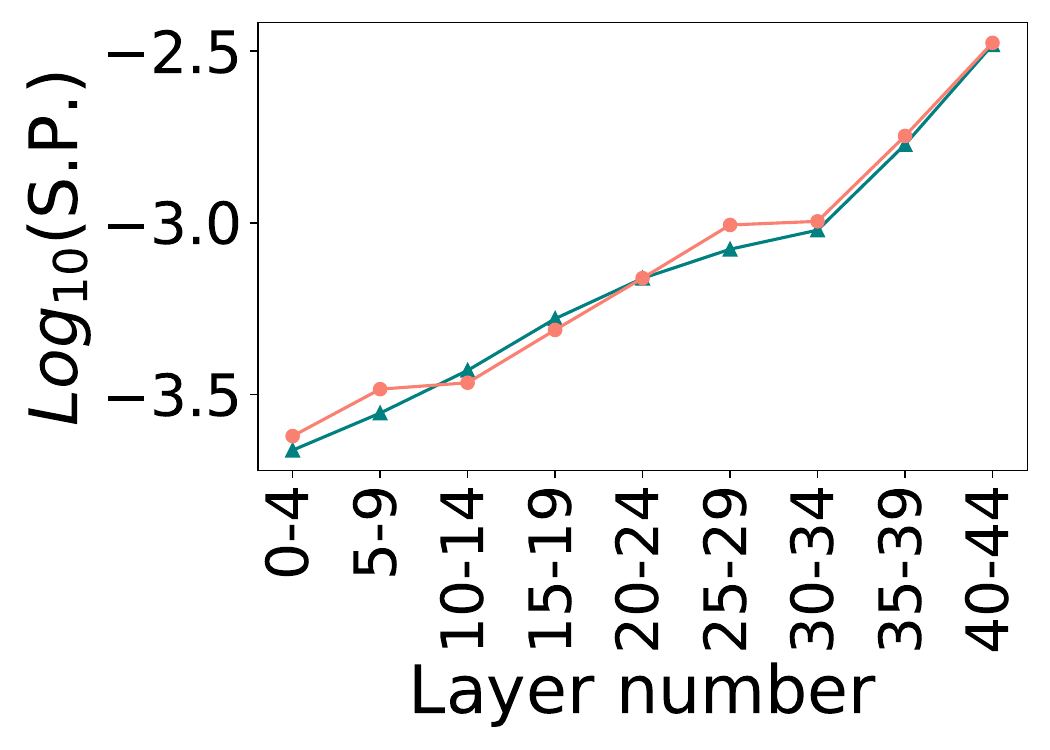}}
\subfloat[S.P. - Sparsity \label{fig:sep_sparsity_mix_ds3}]
{\includegraphics[width=0.25\textwidth]
{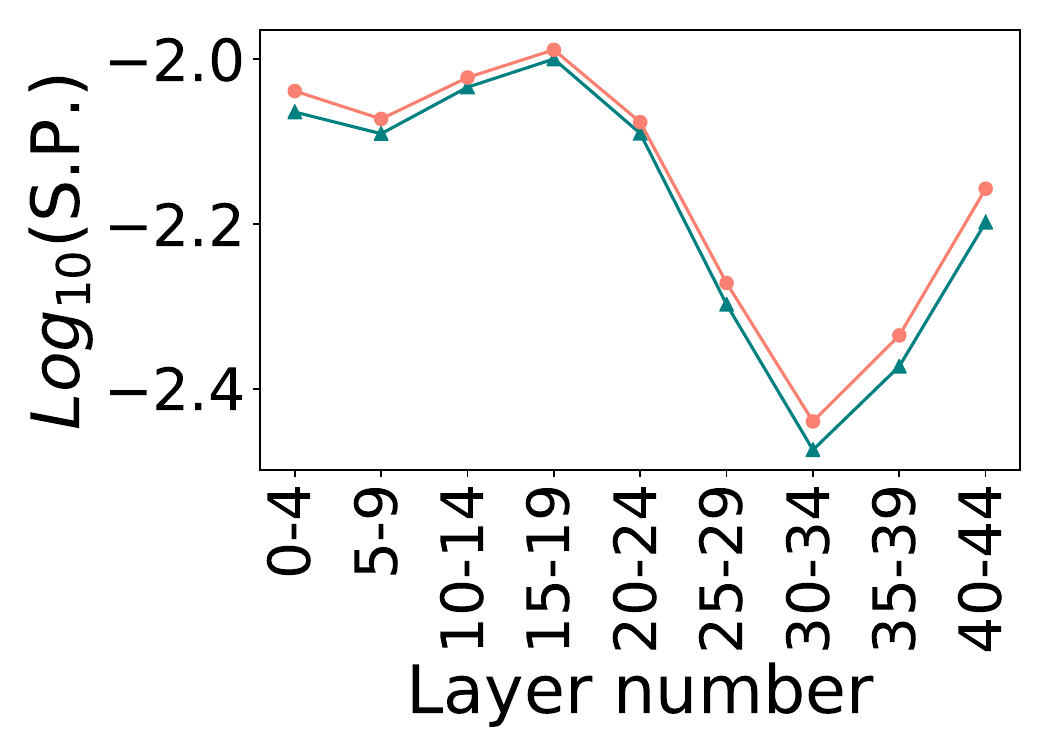}}
\caption{Comparison of EMD scores and separation power for  layer energy distribution and sparsity with dataset 3 between full precision and mixed precision mode using CaloDiffusion.} \label{fig:mix_main}
\end{figure}

Observing a speed-up in sample generation for datasets 2 and 3 with batch sizes of 10 and 100, we further examine whether the generated samples align with Geant4. To quantify these differences, we generate histograms, EMD scores, and separation power from CaloDiffusion using  mixed precision inference on GPU, and compare these with samples generated using full precision inference. Figure~\ref{fig:histogram_observable_mix} shows the histograms for the distribution of layer energy and sparsity for dataset 3 with batch size 100. We observe no significant difference in the performance of the model between full precision and mixed precision modes.  Consequently, Figure~\ref{fig:mix_main} shows the EMD score and separation powers for layer energy distribution and sparsity. These plots also reveal that there is minimal degradation in the similarities between Geant4 data and mixed precision inference samples when compared to full precision inference. Refer to Appendix Figures~\ref{fig:ds2_emd_mix},~\ref{fig:ds2_sep_mix},~\ref{fig:ds3_emd_mix} and~\ref{fig:ds3_sep_mix} for analysis on dataset 2 and dataset 3 with all physical features.

%display the quantitative performance of mixed precision inference on the GPU for datasets 2 and 3. 

%These datasets are highlighted as they show significant speed improvements. 

\section{Related work}\label{related_work}

Despite the development of numerous generative techniques for calorimeter shower simulation, research on comprehensive evaluation metrics remains limited. The generative models in~\cite{ho2020denoising, song2020generative, papamakarios2017masked,lipman2022flow,goodfellow2020generative,vaswani2017attention,kingma2017improving,ho2019flow,kingma2019introduction} are primarily designed for image and text data. When applied to calorimeter shower simulation, where data are represented as voxel images or point clouds, applying the same evaluation metrics used for image or text data proves insufficient. 

Although~\cite{hashemi2023deep,buhmannadvancing} discuss the characteristics of different surrogate models in calorimeter simulation, they lack in comprehensive analyses of these models. In~\cite{krause2021caloflow, das2024understand}, the authors introduce classifier test and explain the meaning of the weights of these classifiers in calorimeter shower simulation. However, this metric is imperfect because the classifier itself a black box making it difficult to identify dominating features during training. FPD and KPD are recommended by ~\cite{kansal2023evaluating} to capture correlations among high level features and EMD to see the dissimilarities in individual feature. CaloDiffusion is the only model to adopt FPD, KPD and CaloScore consider EMD. We believe these metrics should be evaluated by all models. Although histograms are common in this field, the use of separation power to quantify the differences between the histograms of generated and reference data~\cite{diefenbacher2020dctrgan} is rare. Finally, not all generative models adopted all of the metrics, discussed in~\ref{eval_metrics}), to access the performance. Our work provides a systematic analysis of three state-of-the-art generative models to assess their performance using various metrics. 

\section{Conclusion}\label{conclusion}

 "Fast Simulation" using deep-generative models is pivotal in overcoming computational bottlenecks of the calorimeter shower generation. In this study, we conducted a rigorous evaluation of three generative models using standard datasets and a diverse set of metrics derived from physics, computer vision, and statistics. Additionally, we explored the impact of using full vs. mixed precision modes during inference on GPUs. Our evaluation revealed that the CaloDiffusion and CaloScore generative models demonstrate the most accurate simulation of particle showers, yet there remains substantial room for improvement. We identified areas where the evaluated models fell short in accurately replicating Geant4 data. The benchmarks used in this work will be integrated with the FAIR Surrogate Benchmark Initiative~\cite{SBIFAIR} and the MLCommons Science working group~\cite{MLCommonsScience}. We believe, identifying standard metrics will facilitate in development and comprehensive evaluation of generative AI methods extending to new fields~\cite{DiffDA}. We identified some \textit{limitations} of our work. \emph{First}, we did not explore the memory usage of various models which is an important consideration when incorporating "Fast simulation" into the calorimeter pipeline. \emph{Second}, Leave-one-out-cross validation is used to measure the performance of model when one makes predictions on data not used to train the model. We did not investigate the leave-one incident energy validation strategy to measure the efficacy of the models on unseen data. \emph{Third}, other generative methods like VAE, GAN, and point cloud are not examined in this work. Future studies will explore these avenues.

\clearpage
\section*{Acknowledgements}
The authors gratefully acknowledge the partial support of DE-SC0023452: FAIR Surrogate Benchmarks Supporting AI and Simulation Research and the Biocomplexity Institute at the University of Virginia.

\bibliographystyle{plain}
\bibliography{ref}

\clearpage

\clearpage
\begin{appendices}
\renewcommand\thefigure{\thesection\arabic{figure}}
\section{Background}\label{app_related}
Various generative methods, including Generative Adversarial Networks (GANs)~\cite{goodfellow2020generative}, Normalizing Flows (NF)~\cite{kingma2017improving,ho2019flow}, Variational Autoencoders (VAEs)~\cite{kingma2019introduction}, Diffusion and Score Based models~\cite{ho2020denoising,sohldickstein2015deep,song2020generative} have been proposed in recent years as alternative approaches for surrogate modeling in full calorimeter simulation. 

Although FastCaloGAN~\cite{atlas2020fast} was the first full generative model for calorimeter shower simulation, GAN has been in use for calorimeter shower simulation~\cite{de2017learning,paganini2018accelerating, paganini2018calogan, erdmann2018generating, erdmann2019precise, carminati2018three} since the beginning of the emergence of generative methods in High Energy Physics (HEP). GANs have two competing networks; generator network and discriminator network. The generator network tries to generate artificial samples from Gaussian noise and the discriminator tries to differentiate the fake samples from the real ones. The two networks are trained against each other simultaneously until the Nash Equilibrium is reached. GAN-based models can generate samples orders of magnitude faster than Geant4 with reasonable accuracy compared to the previous parameterized models. However, GAN models struggle to converge due to the competing nature of the two networks getting stuck in a saddle point instead of the global minimum. Also, in GANs, 'mode collapse' happens when the generator produces samples from only a small part of the distribution to fool the discriminator.

Variational Autoencoder (VAE)~\cite{cresswell2022caloman,buhmann2021decoding,buhmann2021getting,diefenbacher2023new,salamani2023metahep,madulacalolatent,ernst2023normalizing} are also instrumental in generating calorimeter showers. VAE consists of two networks, an encoder and a decoder. The encoder tries to map the real data distribution to a latent space that follows a multivariate Gaussian distribution. To generate new data, VAE will sample a point from this multivariate Gaussian distribution and decode it to a point in real data distribution using the decoder network. Although VAE can generate calorimeter shower samples faster than GAN and Geant4, the generated samples lack in quality for complex detector geometry. Moreover, unlike GAN, VAE does not have the expressive power. Even so, VAE combined with GAN ~\cite{buhmann2021decoding,buhmann2021getting,diefenbacher2023new} , Normalizing Flow (NF)~\cite{ernst2023normalizing,cresswell2022caloman} and Diffusion~\cite{madulacalolatent} and GPT~\cite{liu2024calo} models is effective in generating quality showers to a certain degree.

To address the problems in GANs and VAEs - including training instability, mode collapse, inexpressiveness, and the generation of low-quality samples - Normalizing Flow (NF) emerges as a promising alternative for both density estimation and sampling tasks. NF operates by learning a bijective transformation between the observed data space and a latent space. Since the true distribution of the data space is often unknown, NF seeks to estimate this density. NF was first introduced in CaloFlow~\cite{krause2022caloflow, krause2021caloflow, krause2021caloflow_2} and later for complex detectors~\cite{diefenbacher2023l2lflows,buckley2023inductive,ernst2023normalizing,favaro2024calodream}. Although NF exhibits impressive performance in generating precise shower images faster than Geant4, it needs to preserve the dimensionality in each stage of the flow because of bijective mapping. Therefore, it is difficult to scale NF with higher dimensional complex geometry of detectors. 

Diffusion and Score based models~\cite{amram2023denoising,mikuni2022score,mikuni2024caloscore,madulacalolatent,kobylianskii2024calograph} are some of the well-developed generative models. Diffusion and Score-based models usually follow the same approaches in estimating data distribution and sampling. In general, an image is perturbed by a noising process, until it becomes pure noise. In the inverse process, a new sample is generated by sampling from the pure noise space, and pass it through the denoising process to map it to the original space. These models have shown remarkable performance in generating high-quality samples. However, slower in sampling compared to all other generative models due to a large number of denoising steps. They can generate high-quality samples for complex detector geometry but with higher memory and time complexity.

Point-cloud in generative models is used for high granular calorimeter shower simulation because this method naturally handles sparse shower simulation, and preserves all the information about the calorimeter shower. Since it is geometrically independent, generative models implemented on the point cloud make it transferable to different detector datasets. Although NF models and Diffusion models exhibit a good agreement with the real world data, these models are not applicable for high granular complex detector since NF models do not have scalability and diffusion models are time-consuming compared to other methods for 3D image-based dataset. NF~\cite{schnake2022generating,schnake2024calopointflow} and Diffusion models~\cite{buhmann2023caloclouds,buhmann2024caloclouds,acosta2023comparison} using the point cloud representation of calorimeter shower are proposed. However, such methods require thorough exploration as the number of points in the sampling process is not predetermined, and multiple generated points can end up in the same calorimeter~\cite{schnake2022generating} cell. ~\cite{schnake2024calopointflow} have resolved `multiple hit' problem.

\section{Evaluated models}\label{app_a}

\subsection{CaloINN}
% a few line about calogan
To address the challenges encountered by GAN-based models ~\cite{paganini2018calogan,giannelli2023caloshowergan} including instability in training and mode collapse, Normalizing Flow presents a promising alternative for both density estimation and sampling. Normalizing Flow learns the bijective transformation between the data space $x$ and the base or latent space $z$. The data space $x$ is unknown and we want to determine the density of data space $x$ defined as $p(x)$. 
CaloINN is an application of Normalizing Flow for high granular calorimeter detectors. Unlike CaloFlow~\cite{krause2021caloflow}, coupling layer-based flows are used instead of autoregressive flows. These coupling layer-based flows are referred to as invertible neural networks. For both datasets 1 and 2, this model indicates an acceleration in generation time compared to other proposed models~\cite{amram2023denoising, mikuni2024caloscore}.However, it leads to a noticeable decline in the quality of samples. Moreover, the INN model is not directly applicable for higher dimensional data as dataset 3. Due to its incompetency for dataset 3, the authors adopted a VAE along with CaloINN. Since it reduces the dimensionality for the CaloINN mapping resulting in a faster sampling process for dataset 3.  In VAE, the preprocessing block is used to normalize, which is later compressed to latent space by the encoder of the VAE. Then with the help of INN, it learns the energy distribution and latent variables. For sample generation, INN samples are decoded back to the shower phase space from the space of the VAE. This is also important to mention that this adopt two step generation process where in the first step it estimates the energy distribution for each layer conditioned on the incident energy of the primary particle and in the second step, it estimates the normalized voxel distribution conditioned on the incident energy and the layer energies from the previous step.

\subsection{CaloDiffusion}

We choose CaloDiffusion to evaluate due to its high fidelity and well established performances in generating high quality samples of calorimeter showers. CaloDiffusion follows denoising diffusion process. To train a denoising diffusion model, initially a batch of images is corrupted by iteratively adding Gaussian noise, a random time step is chosen for this. Then a neural network is trained to learn how to denoise these corrupted images. Later for sampling, a noisy sample $x_T$ is denoised by the trained model until it reaches to be $x_0$ by repeatedly evaluating $p(x_{t-1}|x_t)$. CaloDiffusion applies denoising diffusion model on the voxel-based image data of calorimeter. Initially a preprocessing step is done to normalize voxels of each shower. Here for conditioning parameter, incident energy is considered.For diffusion process, $400$ noising steps are used along with cosine noise scheduler. Primary inputs to the network are noisy representation of the calorimeter shower as well as the encoded conditional information of incident energy of primary particles and the noise level of the current diffusion step. This encoding is conducted by a two layer fully connected network.A U-net model consisting convolution layers and ResNet blocks is used for denoising model. Cylindrical convolutions are used to maintain the information of the cylindrical geometry of the detectors for dataset 2 and 3. Apart from that, to cope with the challenges introduced by the irregular binning of dataset 1, authors propose the Geometry Latent Mapping(GLaM) which can learn a mapping from the data geometry to a perfectly regular geometric structure where it is possible to apply cylindrical convolution. Unlike CaloINN and CaloScore, it does not adopt two step generation process. We selected CaloDiffusion for assessment because of its model interpretability, superior sample quality, and accuracy as demonstrated by various evaluation metrics. However, this model lags in terms of generation speed.

\subsection{CaloScore}

CaloScore~\cite{mikuni2022score, mikuni2024caloscore} is the first model that is applied to all three dataset of CaloChallenge. This model is based on Score based generative model where instead of estimating the data density, the gradient of the data density(score) is learned. This makes score based more flexible compared to Normalizing Flow based models. A neural network is trained to approximate the data score for high dimensional data. A stochastic process corrupt the data distribution over time based on a specific drift and diffusion coefficient. The goal here is to reverse this process to generate new observation starting from a noisy distribution. For sample generation they use Euler-Maruyama algorithm. This model is also implemented using U-net architecture incorporating 3d Convolution operation for dataset 2 and 3 where for dataset 1, 1D convolution operations are used. Similar to CaloDiffusion, here they encoded the conditional information such as time component of SDE and incident energy of primary particles. CaloScore is faster than Geant4 however, it is slower than GAN based generative model in case of sample generation in calorimeter shower. In our benchmarking models, we are considering CaloScore v2 which is a modified version of CaloScore~\cite{mikuni2022score}. To make the sampling faster it uses progressive distillation which reduces the number of timesteps required in diffusion process. It also adopts two step genertaive model where in the first step they estimate each layers energy distribution conditioned on the incident energy of the primary particle and in the second step they employ both incident energy and deposited energy in each layer as conditional parameters to estimate the normalzied voxel distribution. Similar to previous version of CaloScore a U-net architecture with additional attention layers is implemented in CaloScore V2. Overall, it has improved in performance compared to the previous version. It has shown a speed up in sample generation by a factor of $500-2000$ compared to caloscore v1.

\subsection{Training details}\label{system}
For CaloDiffusion, we used the trained models, published by authors, for all three datasets. For CaloINN, we follow the instructions on their GitHub page~\cite{caloinn}, without any changes. For dataset 3, CaloINN gives `Out Of Memory' error which is consistent with the authors' claim in their paper. The authors suggest using their VAE+INN model for dataset 3. We did not do any grid search on hyperparameters as our goal is not to optimize the current models. 

The training and validation datasets are prepared following the instructions from CaloChallenge-2022 in this regard. Dataset 2 consists on two files, one of the two files is used for training and the other one is used to evaluate the generative models. Dataset-3 has 4 files, each containing 50k Geant4 simulated electron showers with energies sampled from a log-uniform distribution ranging from 1 GeV to 1 TeV. The first two files are for training and the remaining two were combined for evaluation purposes. 
All of the GPU inferences were executed on machines with NVIDIA V100 GPUs with 32 GB VRAM. The CPU inferences were executed on Intel(R) Xeon(R) Gold 6248R CPU @ 3.00GHz machines. For mixed-precision experiments, we run the mixed-precision inference on NVIDIA V100 GPUs. We study the effects of mixed-precision inference using the CaloDiffusion model. CaloDiffusion is implemented using the Pytorch framework. We utilize \textit{torch.amp}, an automatic mixed precision package from Pytorch for inference from the trained models supplied by the authors of the paper. By full precision, we mean $32$ bit floating point and mixed precision transitions between float32 and float16 based on the operations.

\begin{table}[ht]
    \centering
    \begin{tabular}{lll}
    \hline
    Model Name & Dataset    & Num of Params  \\ \hline
    \multirow{3}{*}{CaloDiffusion} & DS1 & $\sim 520$K\\
    & DS2 & $\sim 520$K \\ 
    & DS3 & $\sim 1.2$M \\ 
    \hline
    \multirow{3}{*}{CaloScore} & DS1 & $\sim 2.25$M \\ 
    & DS2 & $\sim 3.76$M\\ 
    & DS3 &$ \sim 3.76$M\\ 
    \hline
    \multirow{4}{*}{CaloINN} & DS1(photon) & $\sim 18.8$M \\ 
    & DS1(pion) & $\sim 26.5$M \\
    & DS2 & $\sim 193.5$M \\ 
    & DS3 & $\sim 199$M \\
    \hline
    \end{tabular}
    \caption{Number of parameters in CaloDiffusion, CaloScore, CaloINN models.}
    \label{tab:params}
\end{table}
\section{Evaluation metrics}\label{app_b}
\subsection{Histograms of physics observables}
The histograms of the generated showers can provide insights into how well they replicate the distribution of real calorimeter shower samples. We consider them as qualitative analysis. We investigate five such physics observables, described below.

\textbf{Layer wise energy distribution:}
% Define Layer-wise energy distribution
Calculating the energy distribution per layer is simple for dataset 1, where histograms are generated based on the deposited energy of each layer. Whereas, dataset 2 and 3, consists of a total of 45 layers, summarizing the layer-wise energy distribution involves accumulating the deposited energy of five consecutive layers and generating histograms accordingly. %Consequently, nine plots are generated for each dataset to summarize the understanding of layer-wise energy distribution.

For datasets 2 and 3, we examine the distribution of energy (1 Gev to 1000Gev) as a function of the layer of the calorimeter in Fig 2 through Fig 9. For most cases, the energies below 1 MeV are not significant due to the presence of noise. Therefore, the energy distributions plotted in the figures have energies above 1 MeV. 

The $E_{tot}/E_{inc}$ histogram represents the total deposited energy in an event, which has been normalized by the incident energy.

\textbf{Center of energy in $\eta$ and $\phi$ direction}
The center of energy along a coordinate is defined as the sum of energy deposited in each voxel times the voxel's coordinate distance from the origin, normalized by the total energy deposited. This is computed in both radial and angular directions. The equation to compute it is shown below:
\begin{equation}
    \Bar{x}=\frac{\langle x_i E_i \rangle}{\sum E_i}
\end{equation}
where $x_i$ is the cell location and $E_i$ is the energy at that location. For dataset 1, we generate the center of energy distribution for each layer. For datasets 2 and 3 we generate it by taking the average of five consecutive layers to summarize the histograms. Since we get nine plots for each dataset.

\textbf{Shower width in $\eta$ and $\phi$ direction}
Another spatial property of a shower can be defined by shower width. Equation \ref{eq:width} defines the shower width. Similar approaches as Center of Enegry are followed here too.
\begin{equation}
    \sqrt{\frac{\langle x_i^2 E_i \rangle}{\sum E_i}-\Bar{x}^2}
    \label{eq:width}
\end{equation}

\textbf{Sparsity}
The sparsity is defined as the ratio of the number of voxels with non-zero deposition to the total number of voxels in each layer. Similar approaches as Center of Energy are used for dataset 2 and 3. 

\subsection{Classifier test}
One widely used metric is based on training a classifier to distinguish between the synthetic (generated) and reference samples. An optimal classifier will learn a score proportional to the likelihood ratio between the synthetic and reference samples. The closer the two samples, the closer the likelihoods, and the classifier will fail to distinguish between the two samples. Performance can be measured based on the area under the curve (AUC) of the receiver operating characteristic (ROC) curve of the classifier evaluated on a statistically independent dataset. An AUC of $1$ indicates that there is a significant difference in the generated sample, that is, the classifier can always distinguish the generated sample from a reference sample. An AUC of 0.5 would indicate the classifier cannot distinguish between the two samples.

To measure the similarity of the two distributions, we use Jensen-Shannon divergence (JSD).

\begin{equation}\label{eq:JSD}
    D_{JS} (g,p) = \frac{1}{2} D_{KL}\left(g\bigg|\frac{g+p}{2}\right) + \frac{1}{2}D_{KL}\left(p\bigg|\frac{g+p}{2}\right)
\end{equation}

The JS can be understood as a symmetric version of the Kullback–Leibler (KL)-divergence
\begin{equation}\label{KL_div}
D_{KL}(g|p) = \int g(x) log \frac{g(x)}{p(x)} dx
\end{equation}

\subsection{EMD, FPD, KPD score}
We compare the total deposited energy between the models using the 1-Wasserstein distance referred to as the Earth Mover’s Distance (EMD), between generated samples and the Geant simulation.  EMD measures the dissimilarity between two distributions. Informally it can be interpreted as the minimum energy cost of moving and transforming a pile of dirt in the shape of one probability distribution to the shape of the other distribution.Formally, the EMD between probability distributions P and Q can be defined as an infimum over joint probabilities:
\begin{equation}\label{eq:emd}
EMD(P, Q) = \inf_{\gamma \in {\Pi(P,Q)}} E_{(x,y)\sim\gamma}[d(x,y)]
\end{equation}
where $\Pi(P,Q)$ is the set of all joint distributions whose marginals are 
P and Q.

We use this similarity measures for all of the physics observables. This metric is sensitive to both the quality and diversity of the generated samples. When EMD can be used for quantifying the differences between individual feature distribution, we need an evaluation metric to quantify the distributions similarity based on multidimensional correlated features. ~\cite{kansal2023evaluating} suggests to adopt Fr\'echet and Kernel Physics Distances(FPD, KPD) as two evaluation metrics to compare generative models` performances in High Energy Physics in this regard. 

The Fr\'echet inception distance (FID) is one of the metrics used to measure the quality of images generated by a generative model, e.g., GANs. Unlike inception score (IS), which evaluates the distribution of generated images, the FID compares the distribution of generated images with the distribution of a set of true images. Formally, for any two probability distributions $\mu$, $\upsilon$ over $\mathcal{R}^n$ having finite mean and variances, the Fr\'echet distance is

\begin{equation}\label{eq:fpd}
d_F(\mu, \upsilon) = \bigg(\inf_{\gamma \in \Gamma(\mu, \upsilon)} \int_{\mathcal{R}^n \times \mathcal{R}^n} {\parallel x - y \parallel}^2 \, d\gamma(x, y) \bigg)^\frac{1}{2}
\end{equation}

where $\Gamma(\mu, \upsilon)$ is the set of all measures on $\mathcal{R}^n \times \mathcal{R}^n$ with marginals $\mu$ and $\upsilon$ on the first and second factors respectively. Fr\'echet distance between Gaussian distributions fitted to the features of interest is called Fr\'echet Gaussian Distance(FGD). A form of it, known as Fr\'echet Inception Distance(FID) is usually used as well-established evaluation metric in determining the similarity between real and generated images. Here it uses the activations of the I{\small NCEPTION} v3 convolution neural network models. A lower FID score means higher quality in synthetic images. It is sensitive to both quality and mode collapse. However, it assumes that the features are from Gaussian distribution. Although FGD is a biased estimator, FGD$_{\infty}$ is an unbiased estimator. Kernel Inception Distance(KID) is an improvement over FID as it does not assume that Gaussian feature distribution. Using a polynomial kernel, KID calculates the squared Maximum Mean Discrepancy (MMD) between the Inception representations of the generated and real samples. Since this test is non-parametric, it does not make the rigorous Gaussian assumption; instead, it assumes that the kernel represents a reliable similarity metric. Since the quadratic covariance matrix does not need to be fitted, fewer samples are also needed. According to~\cite{kansal2023evaluating}, FGD$_\infty$ is sensitive to various distortions applied to their test distributions. When this is applied to some physical features, it is called Fr\'echet Physics Distance(FPD). Similarly for KID, if it is applied to the physical features it is called Kernel Physics Distance. Among EMD, FPD and KPD, FPD is the most efficient as it is an interpretable and highly sensitive metric for evaluating generative models in HEP. Moreover, this FPD can be applied to different data structures such as voxel-based images and point clouds. Besides that, here we have full control over the features we are using to evaluate compared to the features of the classifier test. FPD and KPD are implemented in J{\small ET}N{\small ET} library~\cite{kansal2023jetnet} For physical features we are using the same high-level features such as Center of Energy in both $\eta$ and $\phi$ direction, Shower Widths in both $\eta$ and $\phi$ direction, Layer wise energy, incident energy. FPD and KPD are efficient in capturing the correlation between different features of this multidimensional feature space.

\subsection{Separation power}
To quantify the similarities between two different histograms we use separation power. In information theory, it is known as triangular discrimination. To compute separation power, we use the equation~\ref{separation_power}. In this equation, $h1$ and $h2$ denote two different histograms that we want to compare. Separation power $0$ means two histograms are the same and $1$ means they do not have any overlapping bins.
\begin{equation}\label{separation_power}
    \langle S^2 \rangle=\frac{1}{2}\sum_{i=1}^{n_{\text{bins}}} \frac{(h_{1,i} - h_{2,i})^2}{h_{1,i} + h_{2,i}}
\end{equation}
\subsection{Timing}
Timing is another significant metric in calorimeter shower simulation due to the efficacy of these proposed models depends on the acceleration of the sample generation time. That is why, we consider the time requirement to generate an event in seconds as an evaluation metric. We investigate the sampling process from the trained models of CaloDiffusion, CaloScore and CaloINN by considering different batch sizes ranging from 1 to 100. We conduct the generation process three times for each batch size to calculate the mean and standard deviation time taken per event generation in seconds.
\section{Additional results}\label{app_c}

\subsection{Histogram of physics observables}

\begin{figure}  
\subfloat[Photon (GeV)\label{fig:energy_dist_ds1_photon}]
{\includegraphics[width=0.9\textwidth]
{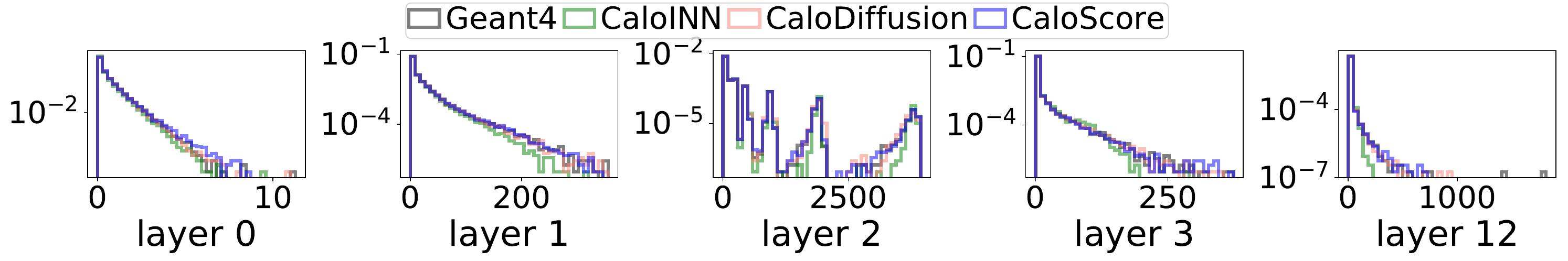}}
\hfill
\subfloat[Pion (GeV)\label{fig:energy_dist_ds1_pion}]
{\includegraphics[width=0.9\textwidth]
{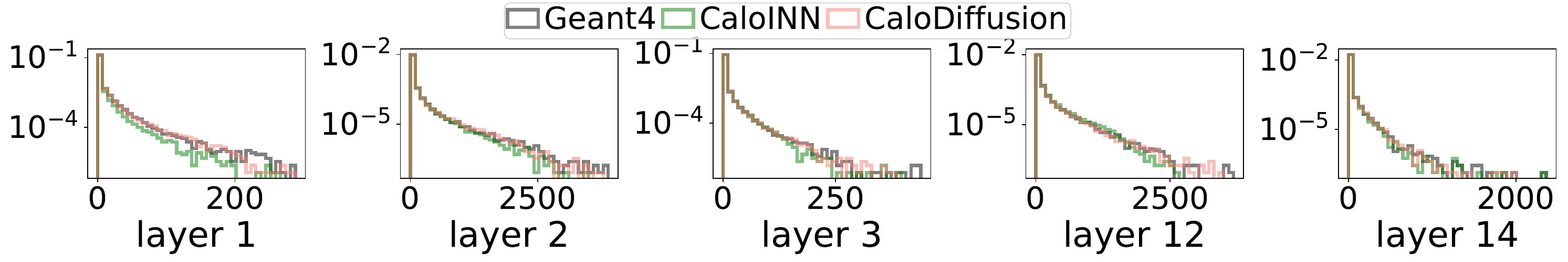}}
\caption{Layer wise energy distribution for Dataset 1 for Photon and Pion. CaloScore did not generate samples for pion.} \label{fig:energy_dist_ds1}
\end{figure}

\begin{figure}  
\subfloat[Dataset 2]
{\includegraphics[width=0.5\textwidth]
{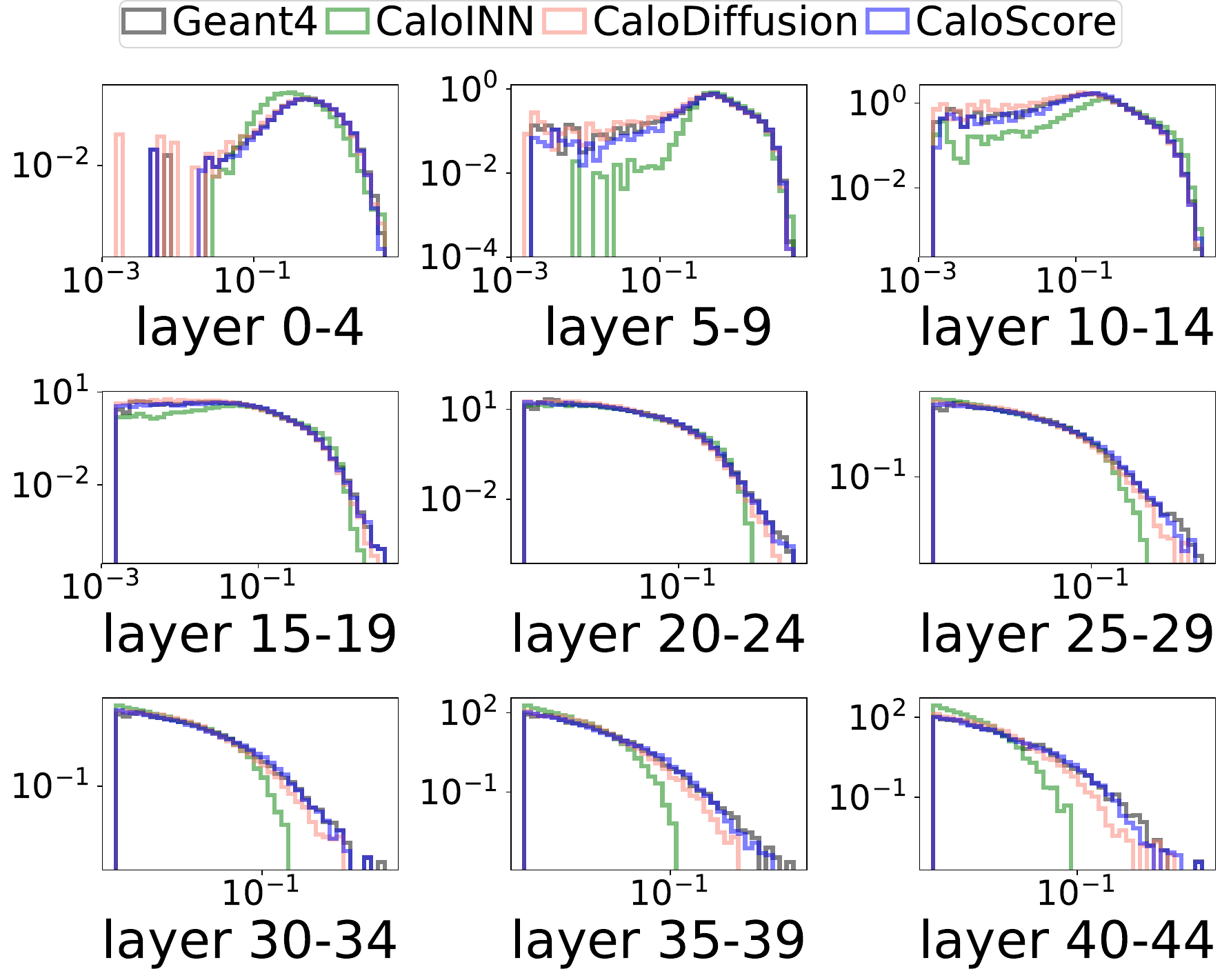}} 
\subfloat[Dataset 3]
{\includegraphics[width=0.5\textwidth]
{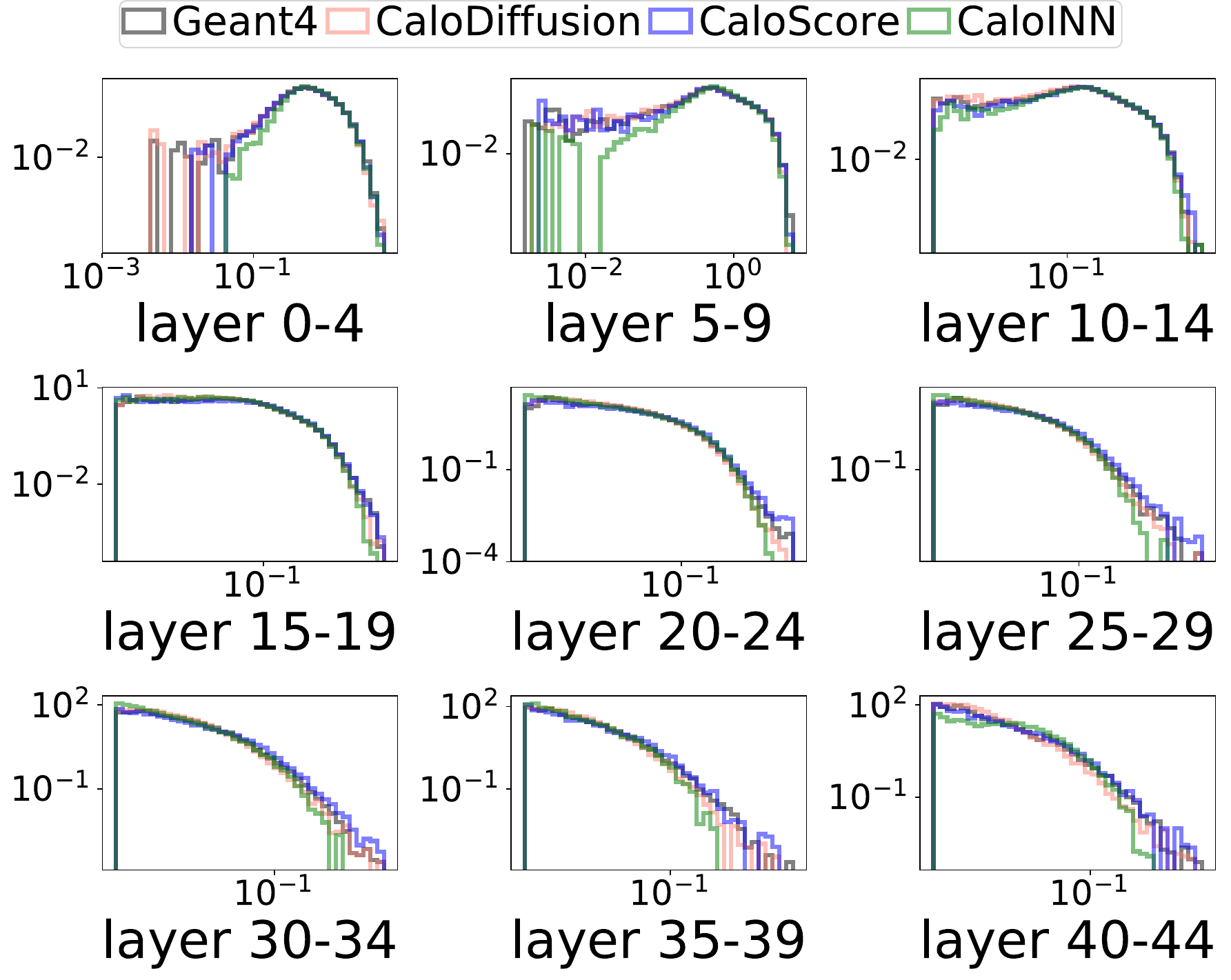}}
\caption{Layer wise energy distribution in a range of 1 GeV to 10 GeV } \label{fig:layer_engergy_1_10}
\end{figure}
\begin{figure}  
\subfloat[Dataset 2]
{\includegraphics[width=0.5\textwidth]
{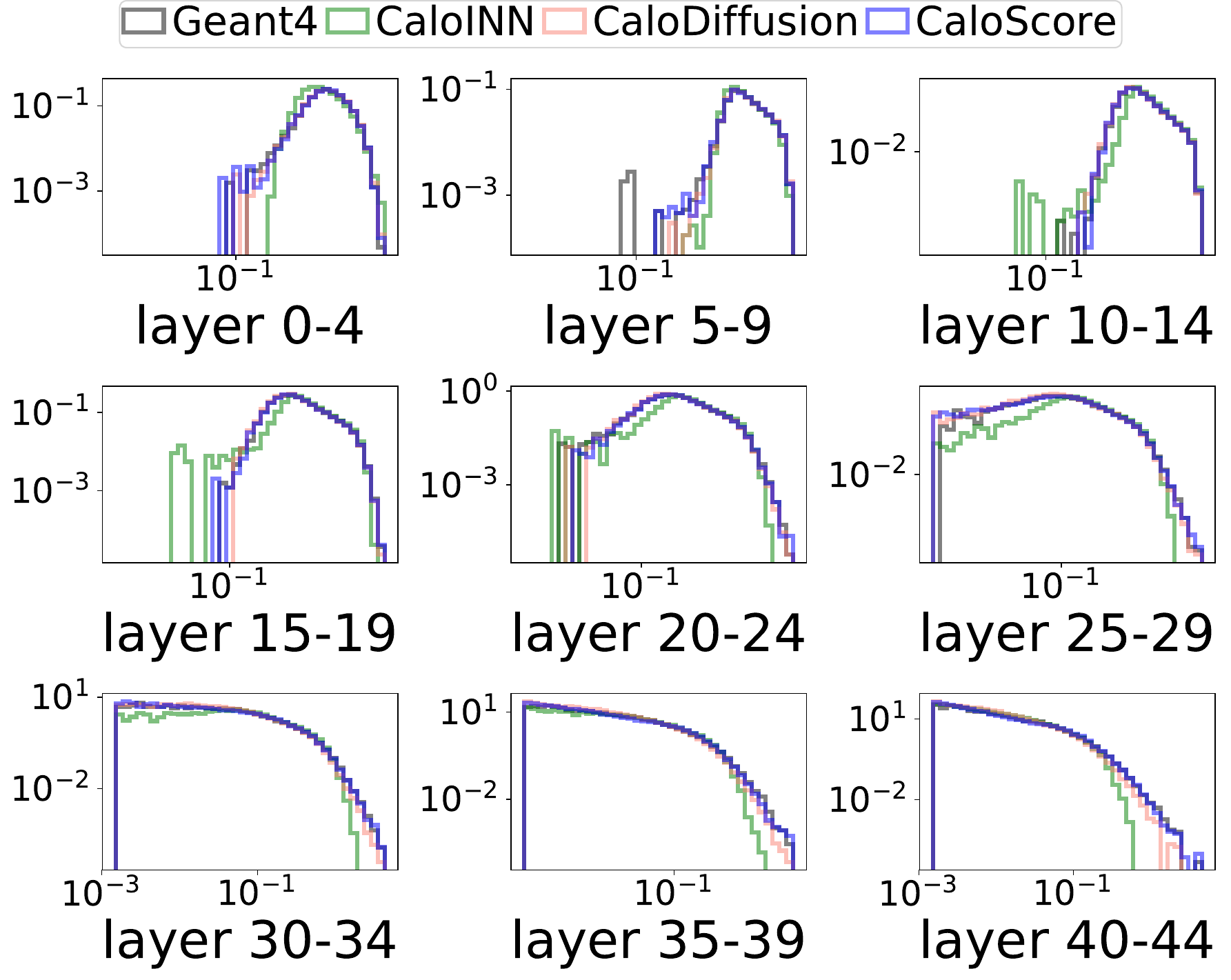}} 
\subfloat[Dataset 3]
{\includegraphics[width=0.5\textwidth]
{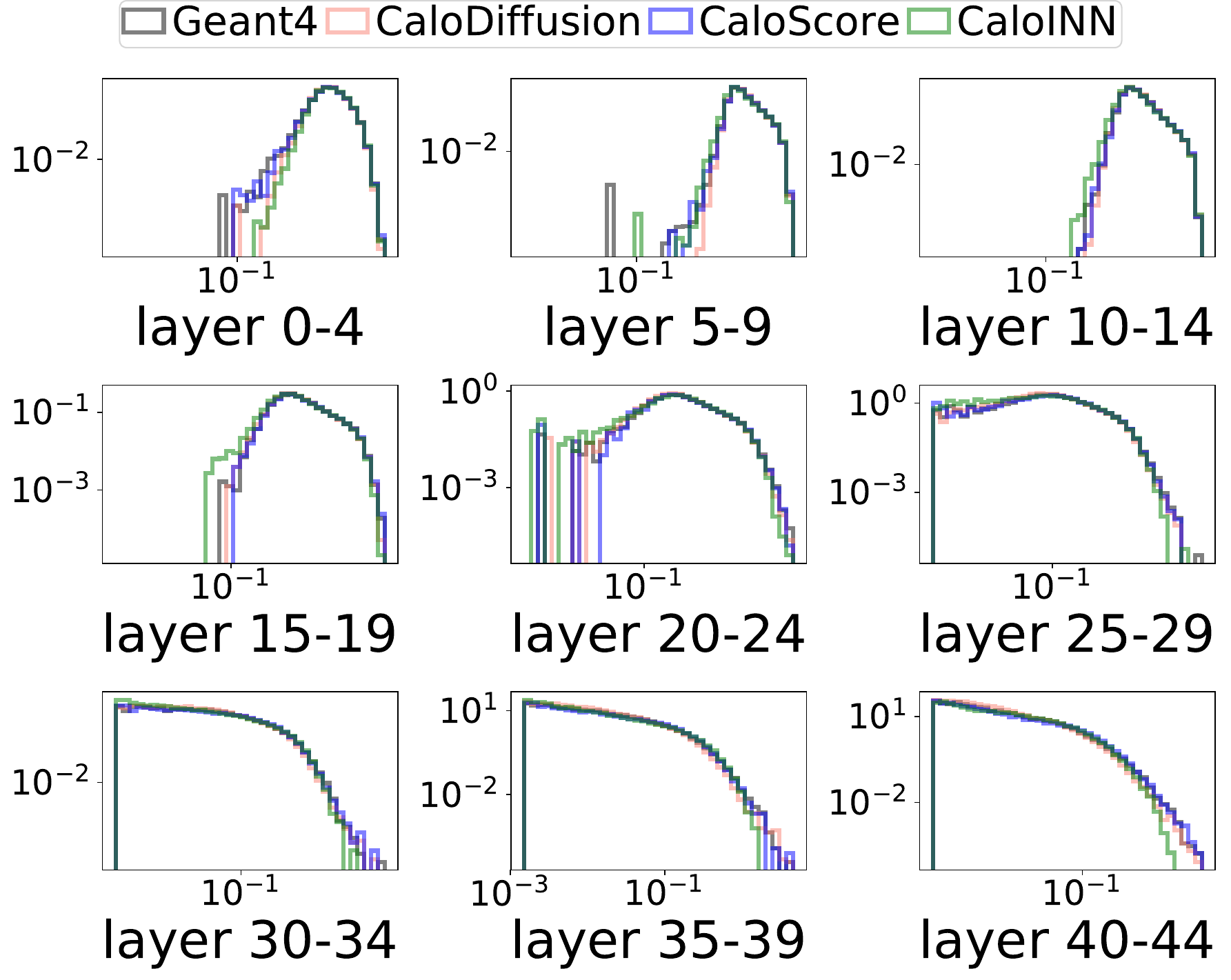}}
\caption{Layer wise energy distribution in a range of 10 GeV to 100 GeV } \label{fig:layer_engergy_10_100}
\end{figure}

\begin{figure}  
\subfloat[Dataset 2]
{\includegraphics[width=0.5\textwidth]
{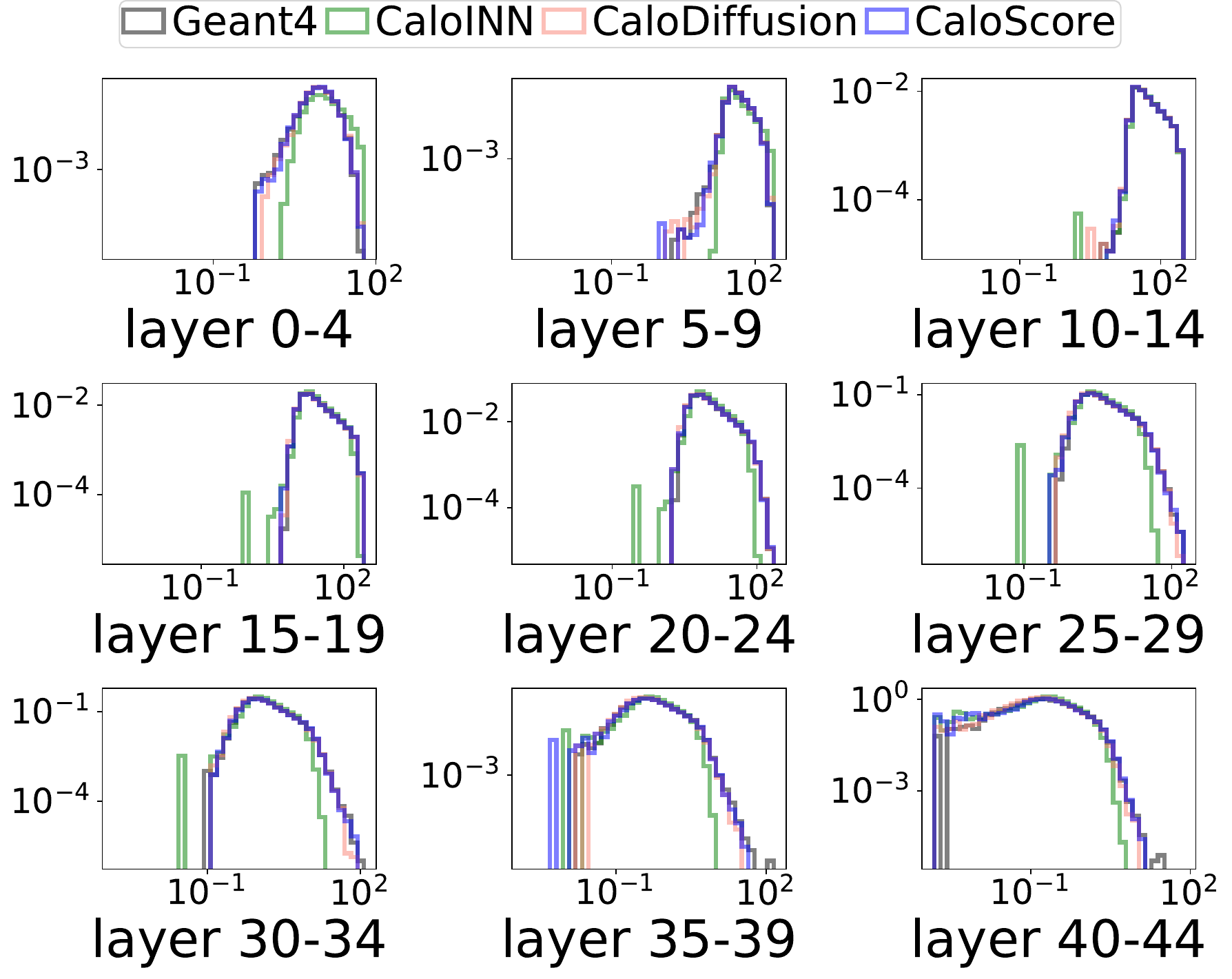}} 
\subfloat[Dataset 3]
{\includegraphics[width=0.5\textwidth]
{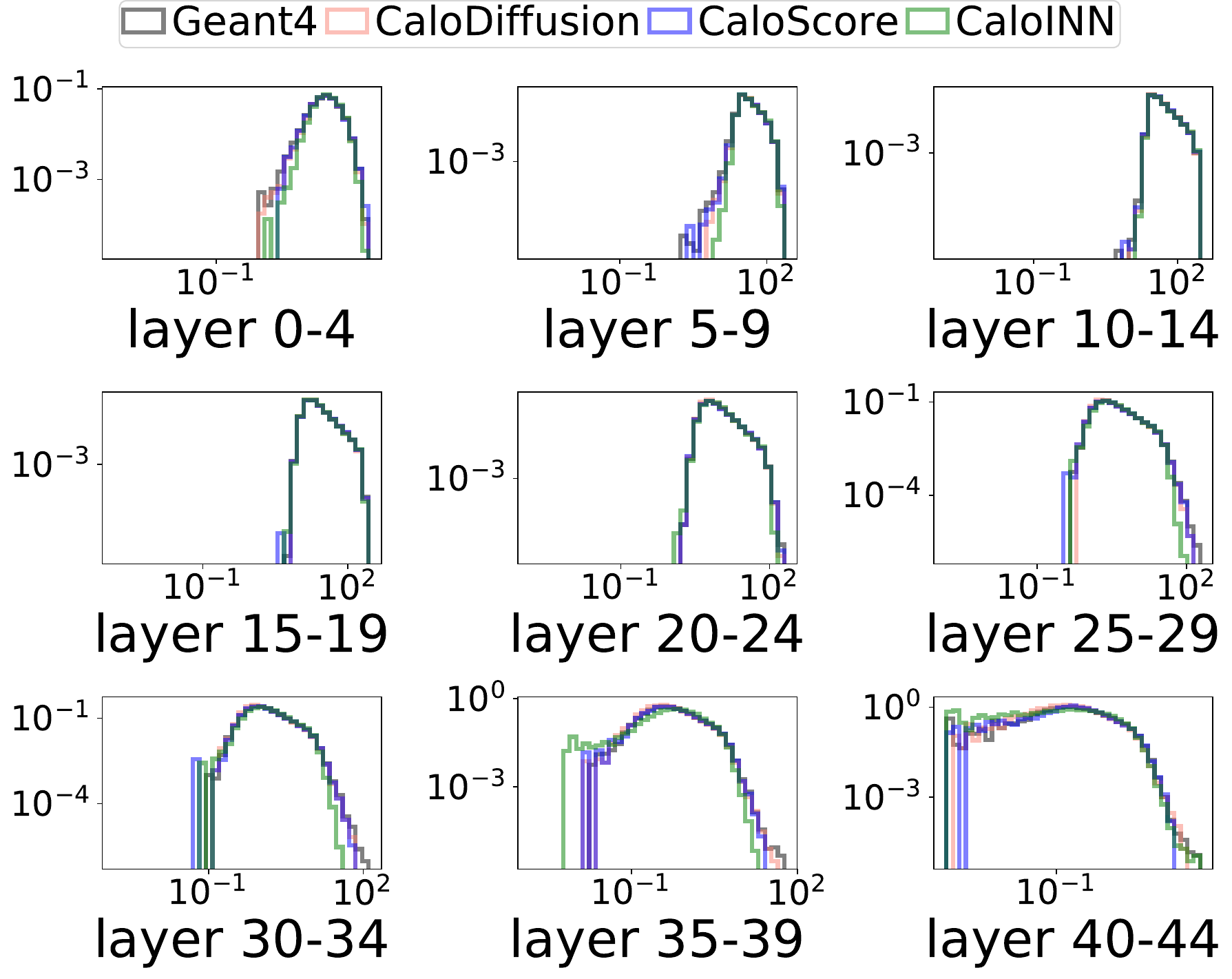}}
\caption{Layer wise energy distribution in a range of 100 GeV to 1000 GeV } \label{fig:layer_engergy_100_1000}
\end{figure}

\begin{figure}[h]
\centering
\subfloat[Dataset 1 - Photon]
{\includegraphics[width=0.24\textwidth]
{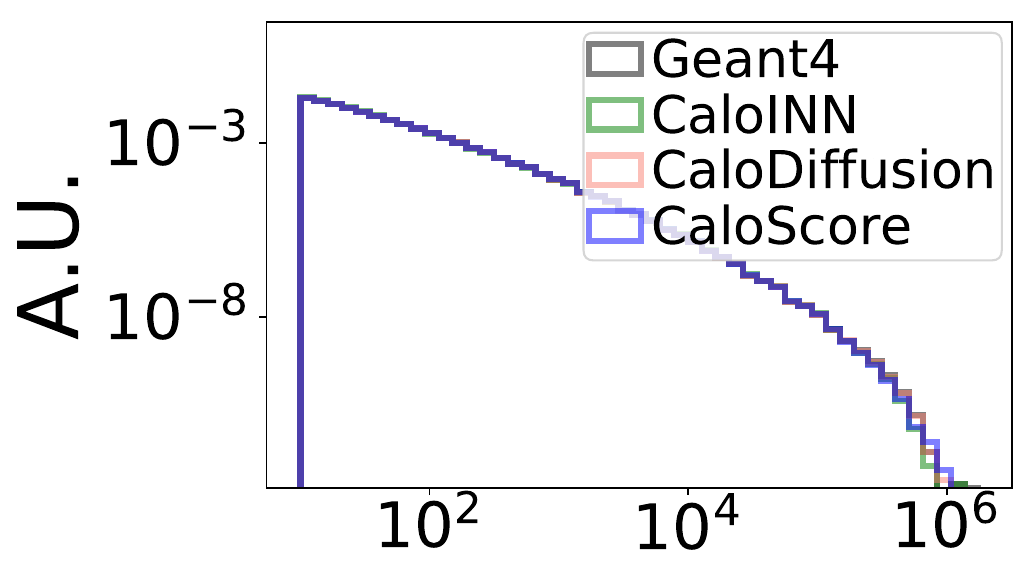}} 
\subfloat[Dataset 1 - Pion]
{\includegraphics[width=0.24\textwidth]
{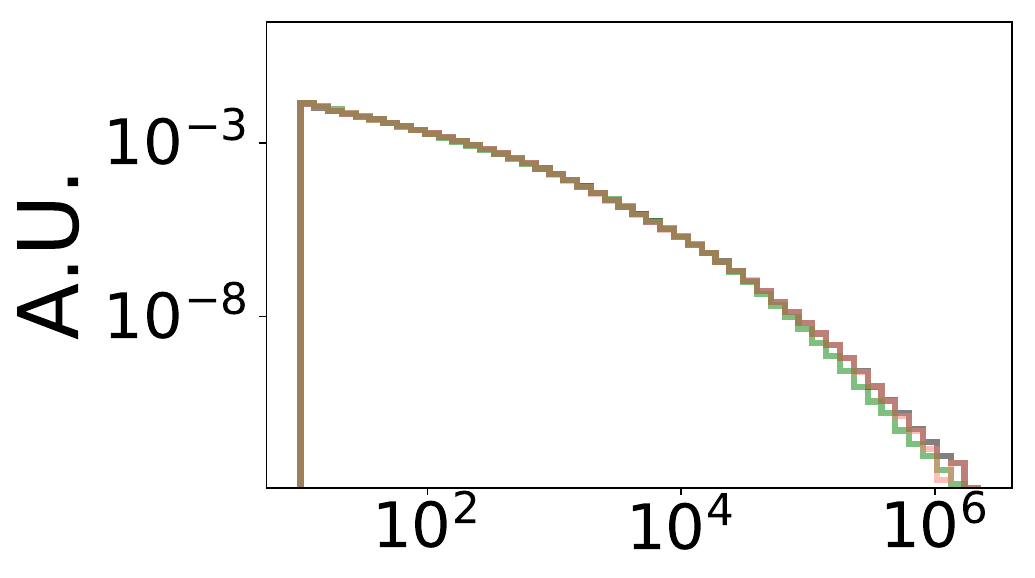}}
%\hspace{2mm}
\subfloat[Dataset 2]
{\includegraphics[width=0.24\textwidth]{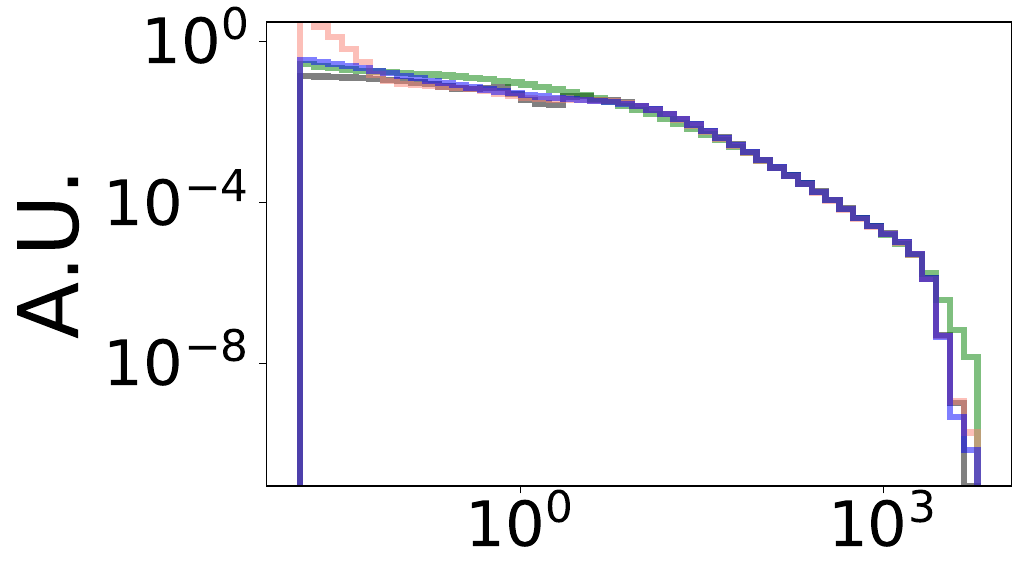}}
\subfloat[Dataset 3]
{\includegraphics[width=0.24\textwidth]{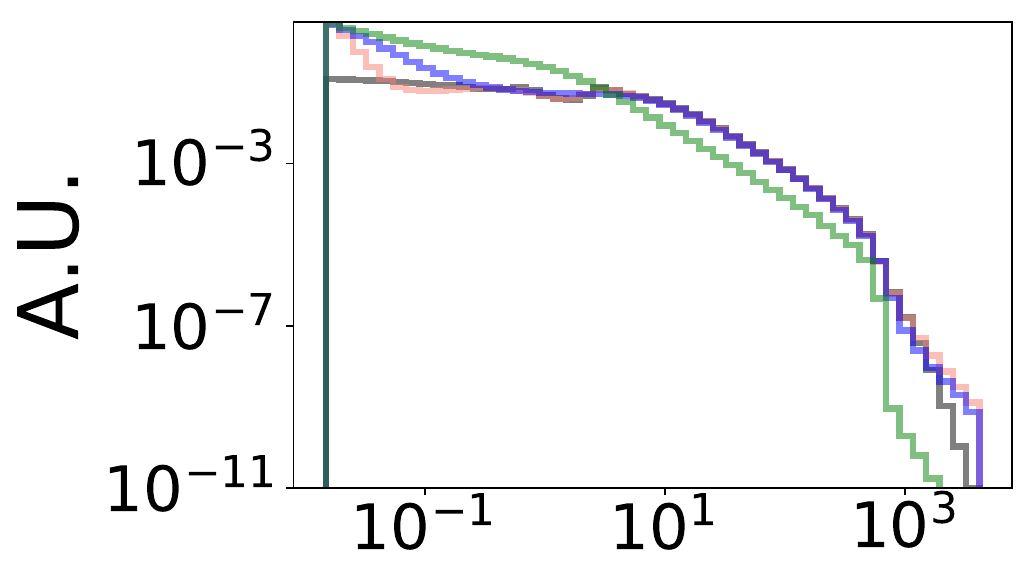}}
\vspace{3mm}
\caption{Voxel Energy [MeV]} \label{fig:voxel_energy}
\end{figure}

\textbf{Voxel energy distribution:}
Figure~\ref{fig:voxel_energy} shows the voxel energy distribution for all three datasets.

\textit{Analysis:} We observe that all three models effectively capture the energy deposited in each voxel.

\begin{figure}[h]
\centering
\subfloat[Dataset 1 - Photon]
{\includegraphics[width=0.24\textwidth]
{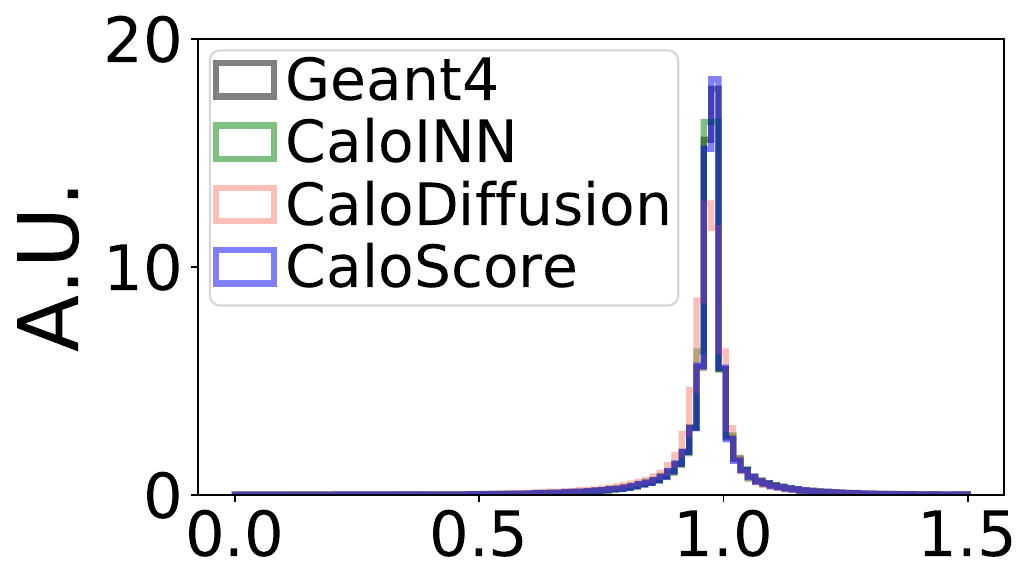}} 
\subfloat[Dataset 1 - Pion]
{\includegraphics[width=0.24\textwidth]
{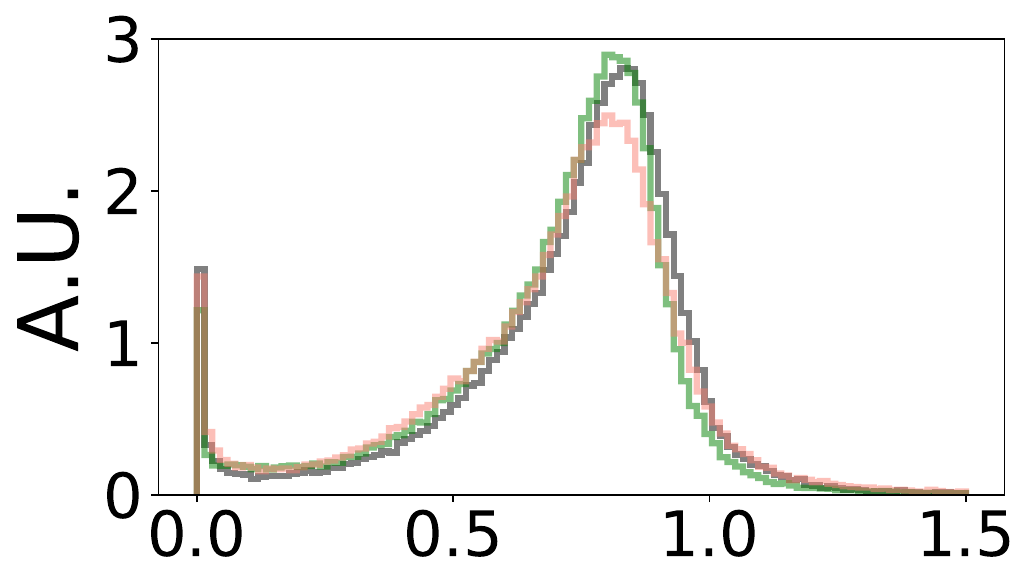}}
%\hspace{2mm}
\subfloat[Dataset 2]
{\includegraphics[width=0.24\textwidth]{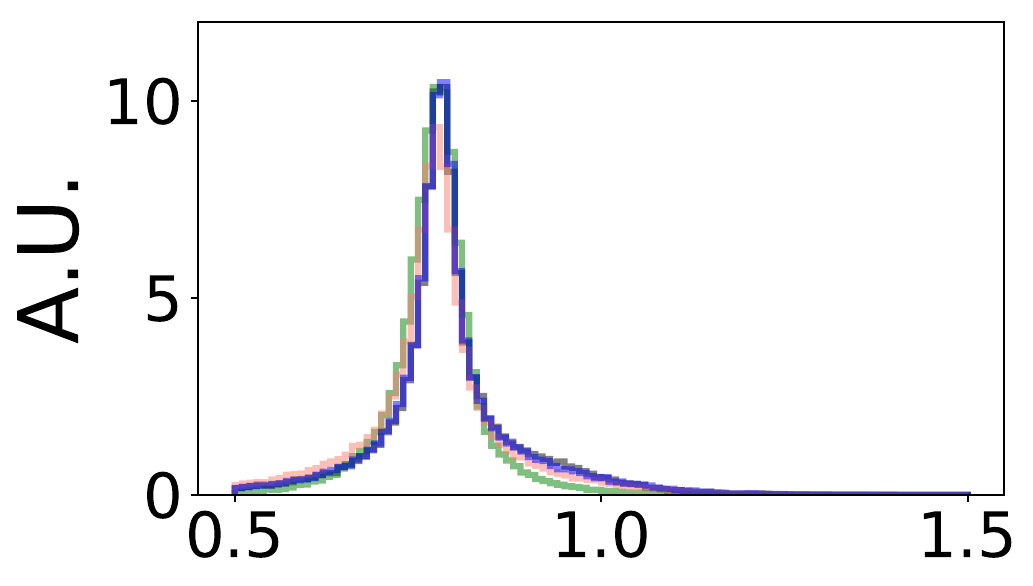}}
\subfloat[Dataset 3]
{\includegraphics[width=0.24\textwidth]{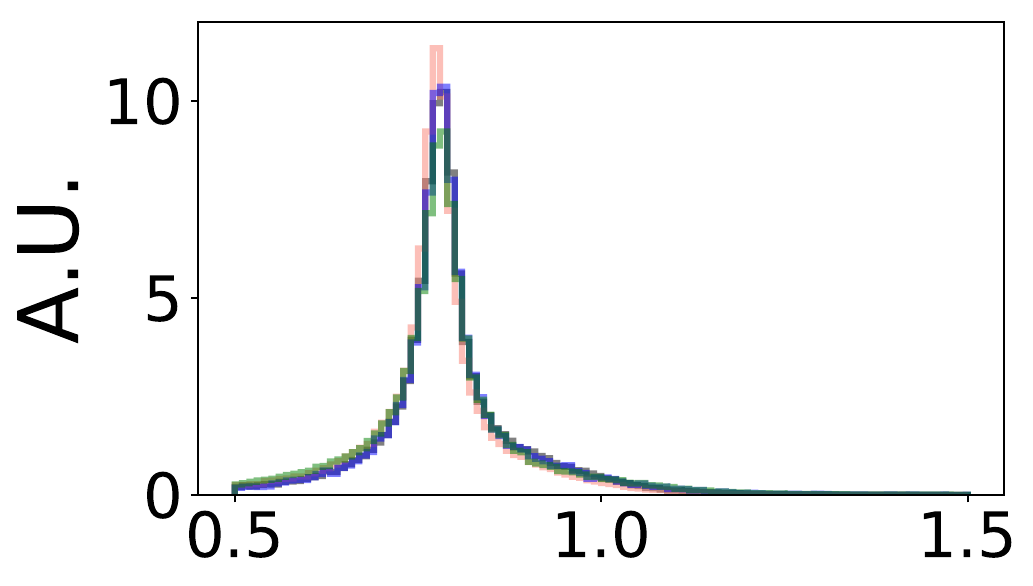}}
\vspace{3mm}
\caption{Energy ratio - $E_{tot} / E_{inc}$.} \label{fig:energy_ratio}
\end{figure}

$\mathbf{E_{ratio}}$:
Figure~\ref{fig:energy_ratio} represents the distribution of energy ratio.

\textit{Analysis:} All three models accurately capture the distribution of energy ratios.

\begin{figure}[h]
\centering
\subfloat[$\eta$ (mm) direction]
{\includegraphics[width=0.5\textwidth]
{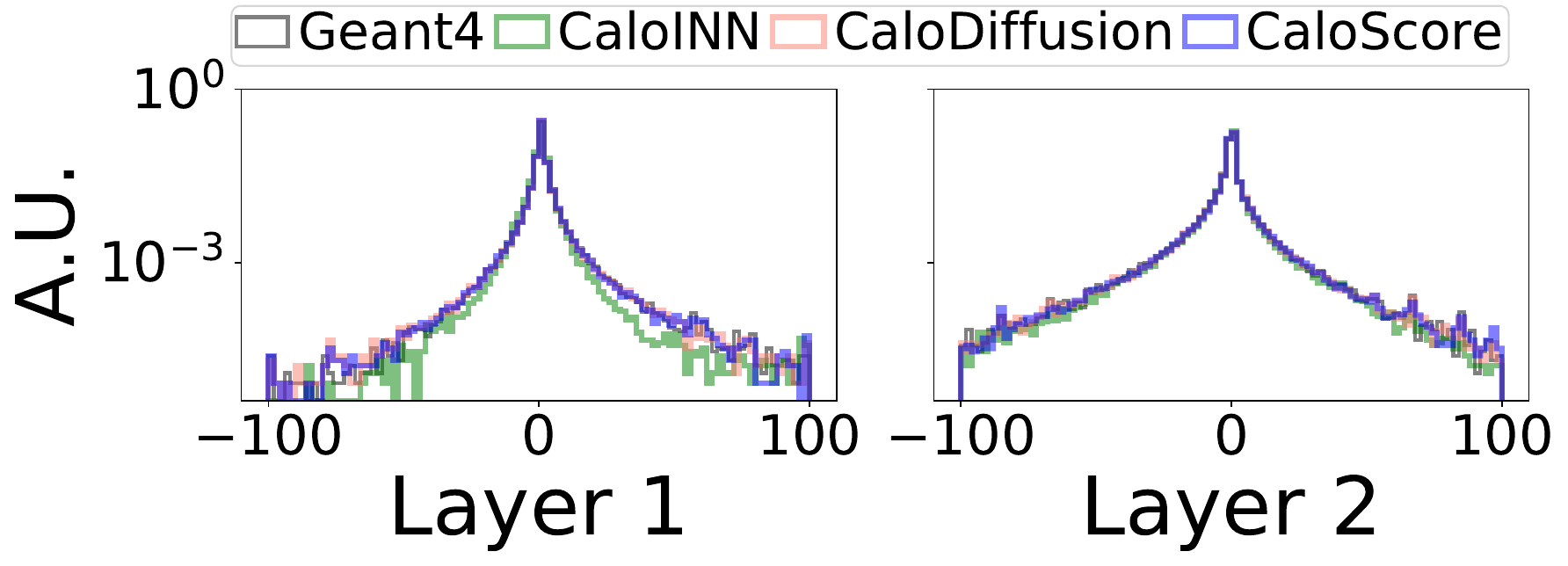}} 
\subfloat[$\phi$ (mm) direction]
{\includegraphics[width=0.5\textwidth]
{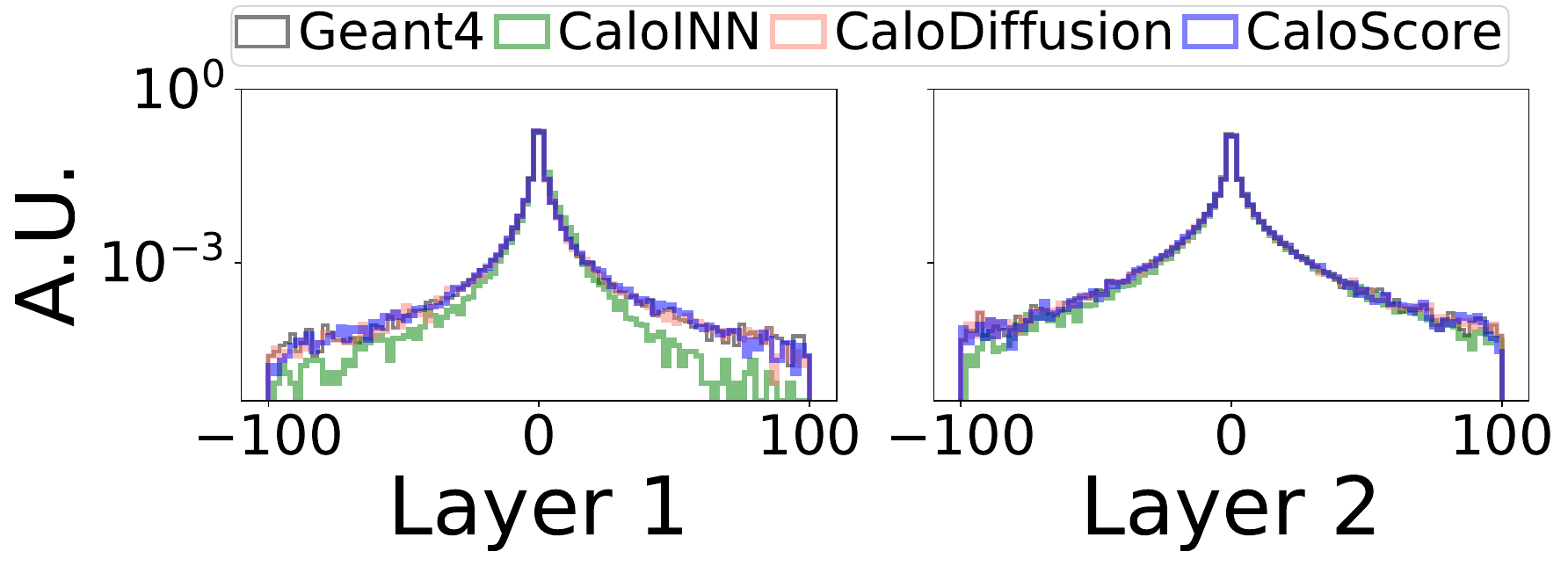}}
\hspace{2mm}
\caption{Center of Energy for dataset-1(photons).} \label{fig:ec_ds1_photon}
\end{figure}

\begin{figure}[h]
\centering
\subfloat[$\eta$ (mm) direction]
{\includegraphics[width=0.5\textwidth]
{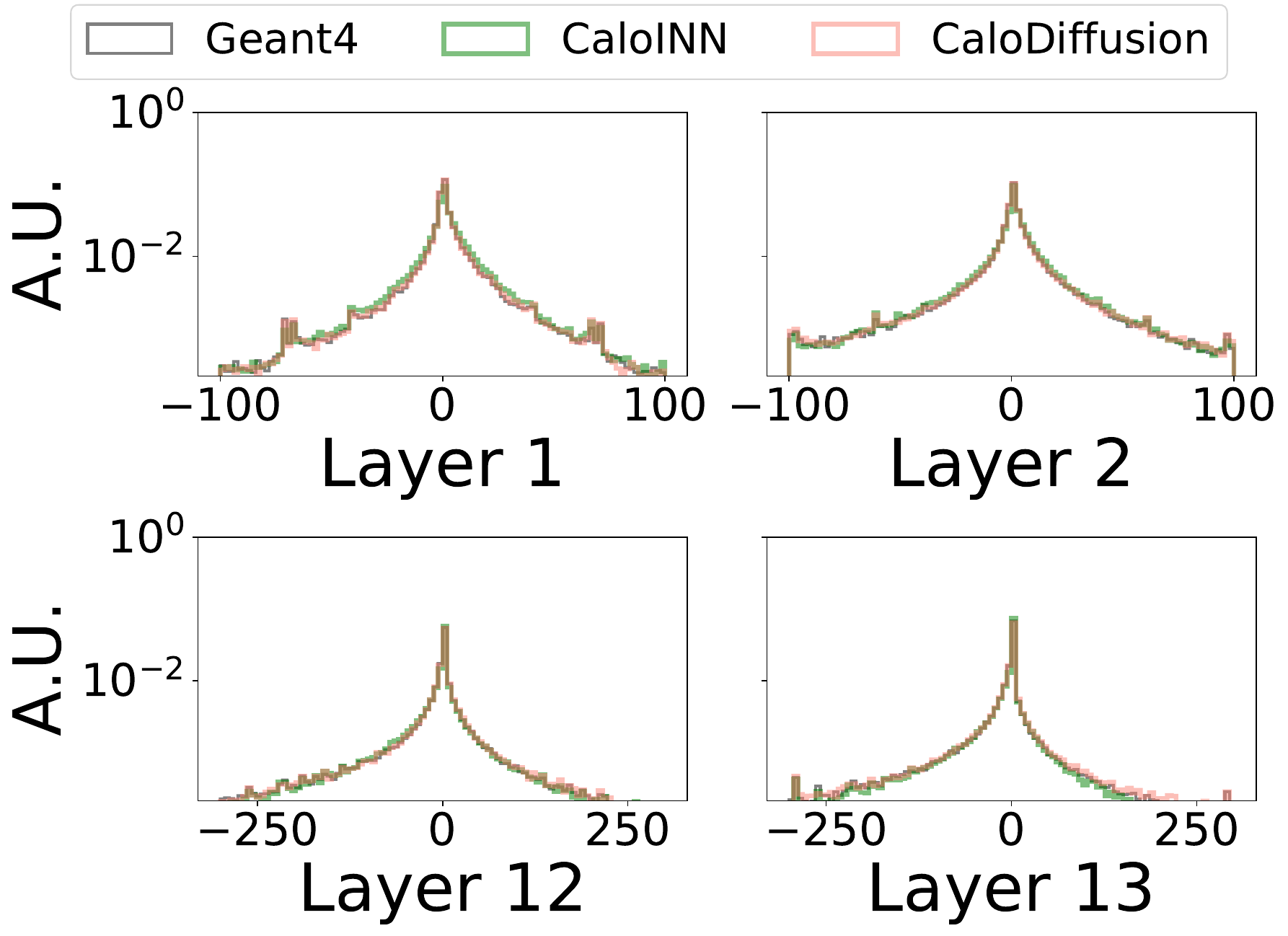}} 
\subfloat[$\phi$ (mm) direction]
{\includegraphics[width=0.5\textwidth]
{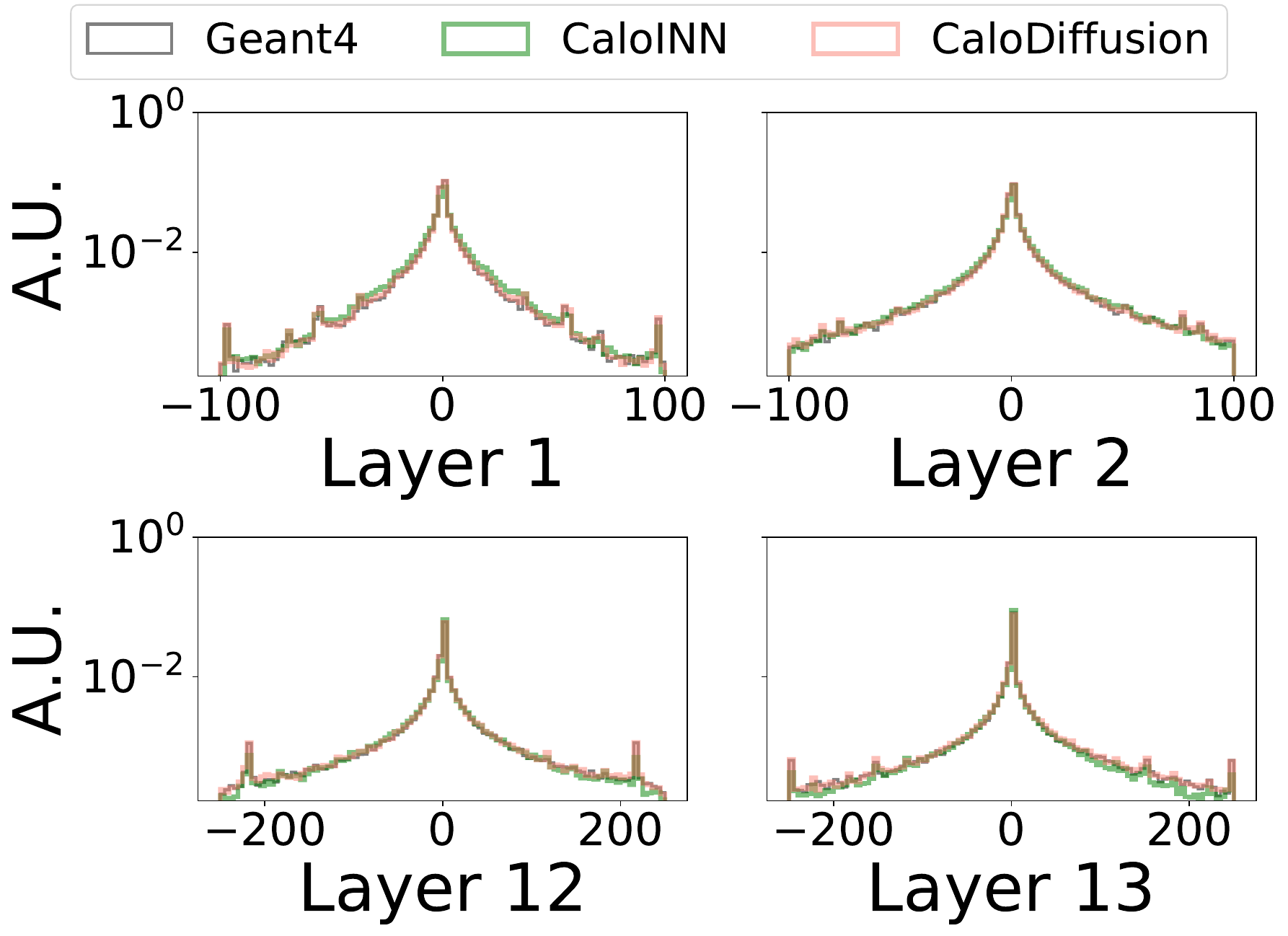}}
\hspace{2mm}
\caption{Center of Energy for dataset-1(pions).} \label{fig:ec_ds1_pion}
\end{figure}

\begin{figure}[h]
\centering
\subfloat[$\eta$ (mm) direction]
{\includegraphics[width=0.5\textwidth]
{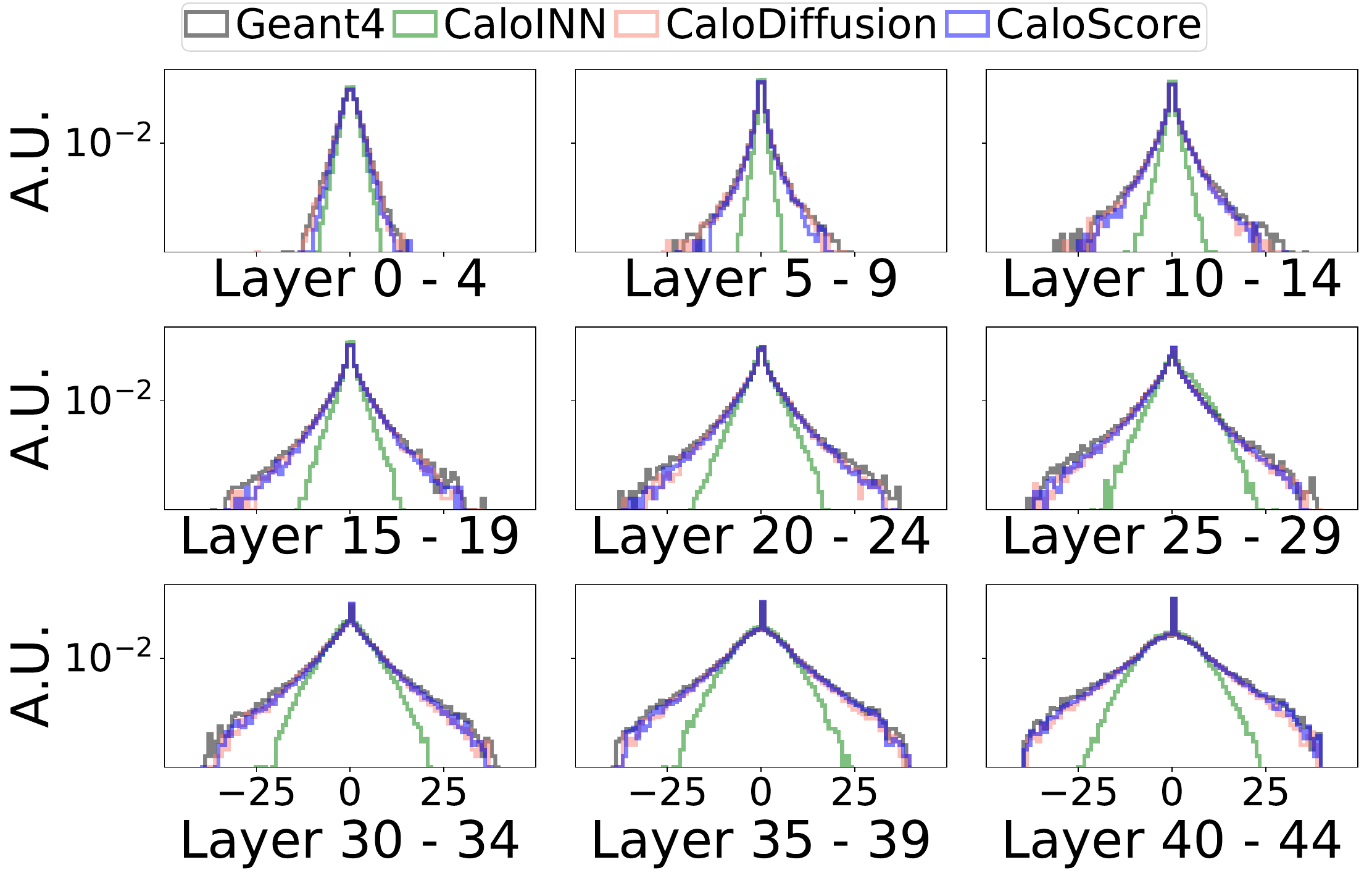}} 
\subfloat[$\phi$ (mm) direction]
{\includegraphics[width=0.5\textwidth]
{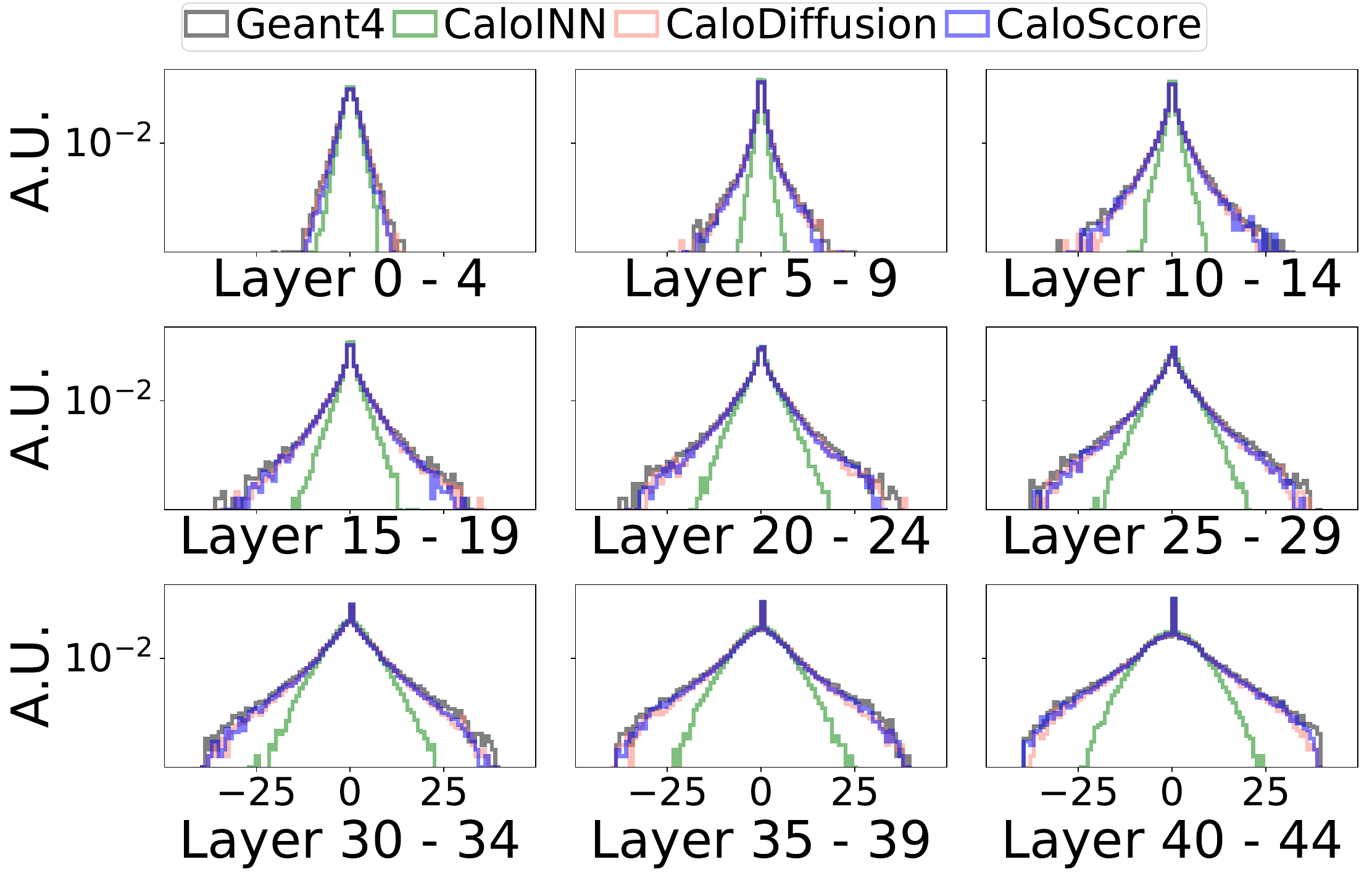}}
\hspace{2mm}
\caption{Center of Energy for dataset-2.} \label{fig:ec_ds2}
\end{figure}

\begin{figure}[h]
\centering
\subfloat[$\eta$ (mm) direction]
{\includegraphics[width=0.5\textwidth]
{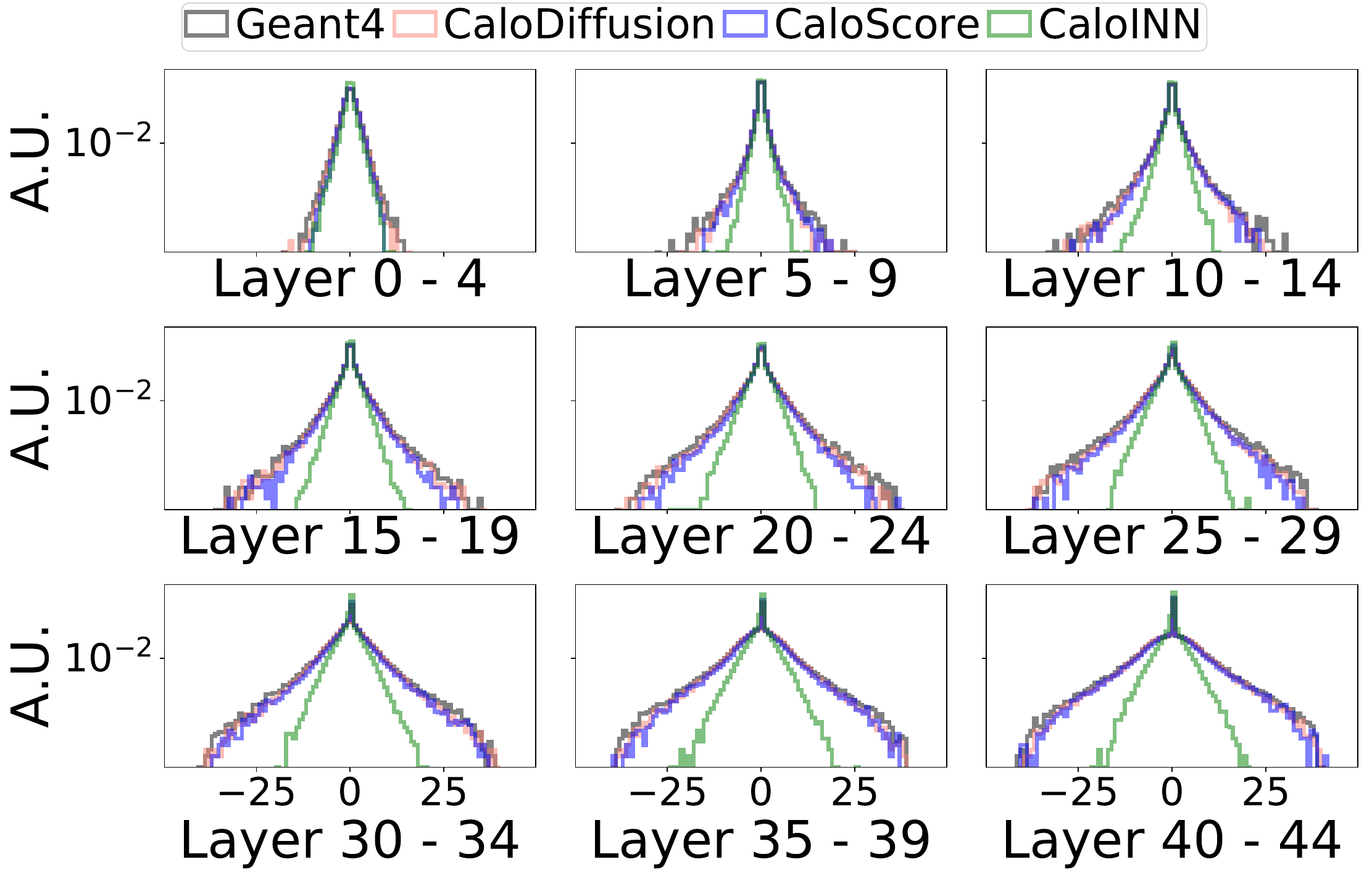}} 
\subfloat[$\phi$ (mm) direction]
{\includegraphics[width=0.5\textwidth]
{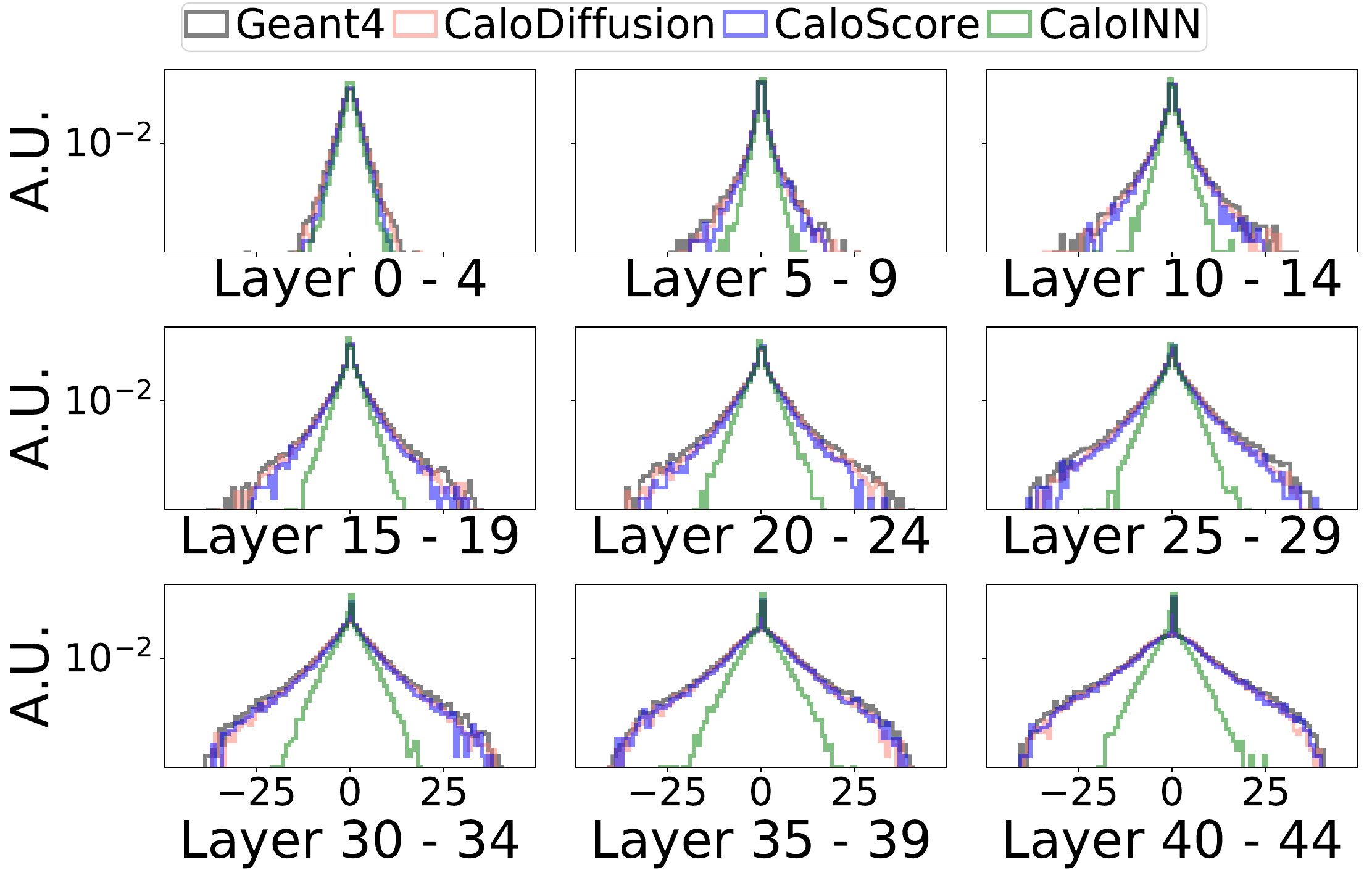}}
\hspace{2mm}
\caption{Center of Energy for dataset-3.} \label{fig:ec_ds3}
\end{figure}

\textbf{Center of energy:}
Figures~\ref{fig:ec_ds1_photon} through \ref{fig:ec_ds3} illustrates the center of energy deposition in each layer for all the datasets. Typically, the major part of the energy deposition is concetrated near $r=0$ as incident particles are generated at the center, orthogonal to the detector plane. As the shower progress, the interactions with the detector material leading in energy deposition away from the center. For instance, with layer 44 the majority of the energy deposited is spread over higher values of $r$. 

\textit{Analysis:} In all cases, CaloScore samples are able to reproduce shower trend most accurately and without any noticeable mismodeling. CaloDiffusion generates the second best reproduction. CaloINN does really well with dataset 1 pion shower generation but struggles with all the other datasets displaying significant discrepancies.

\begin{figure}[h]
\centering
\subfloat[$\eta$ (mm) direction]
{\includegraphics[width=0.5\textwidth]
{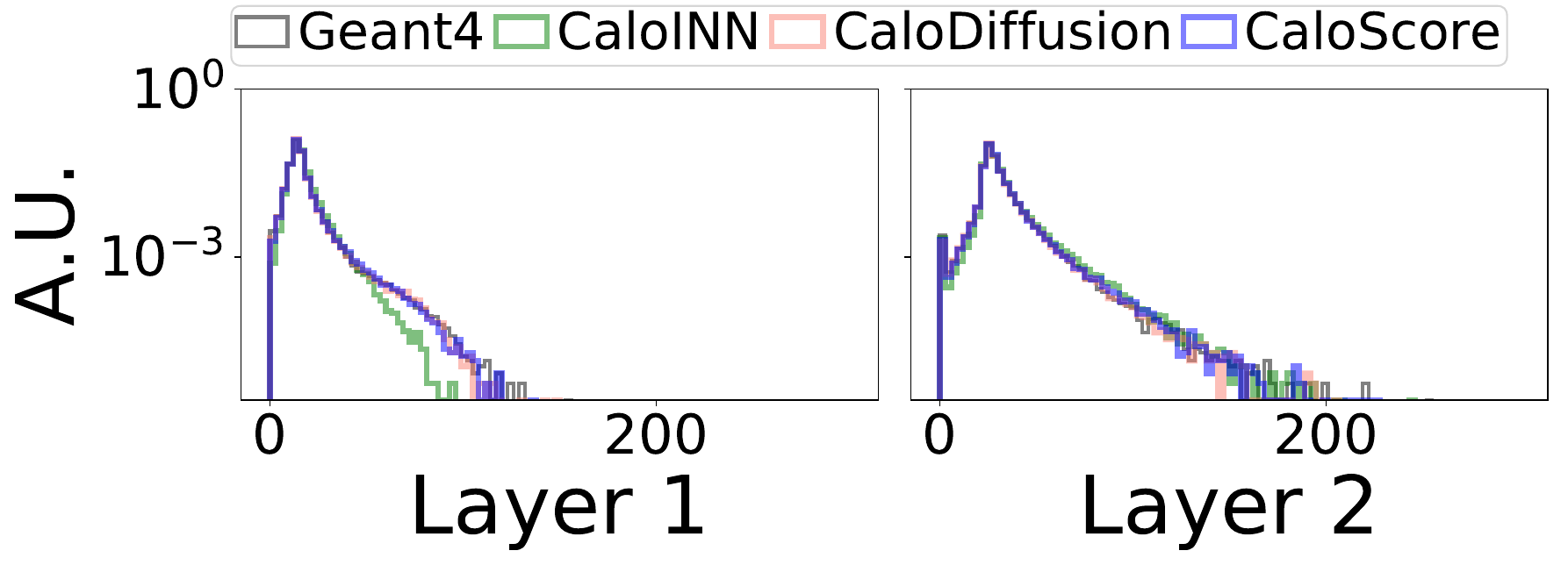}} 
\subfloat[$\phi$ (mm) direction]
{\includegraphics[width=0.5\textwidth]
{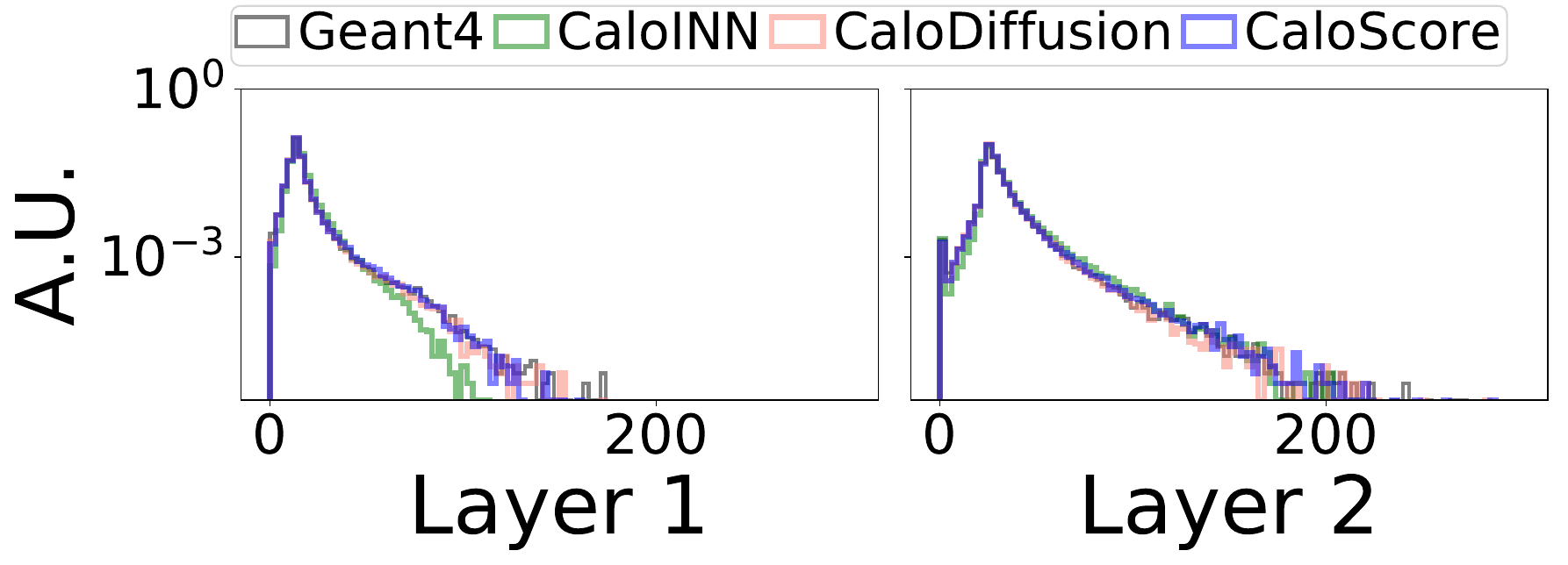}}
\hspace{2mm}
\caption{Shower width for dataset-1(photon).} \label{fig:sw_ds1_photon}
\end{figure}

\begin{figure}[h]
\centering
\subfloat[$\eta$ (mm) direction]
{\includegraphics[width=0.5\textwidth]
{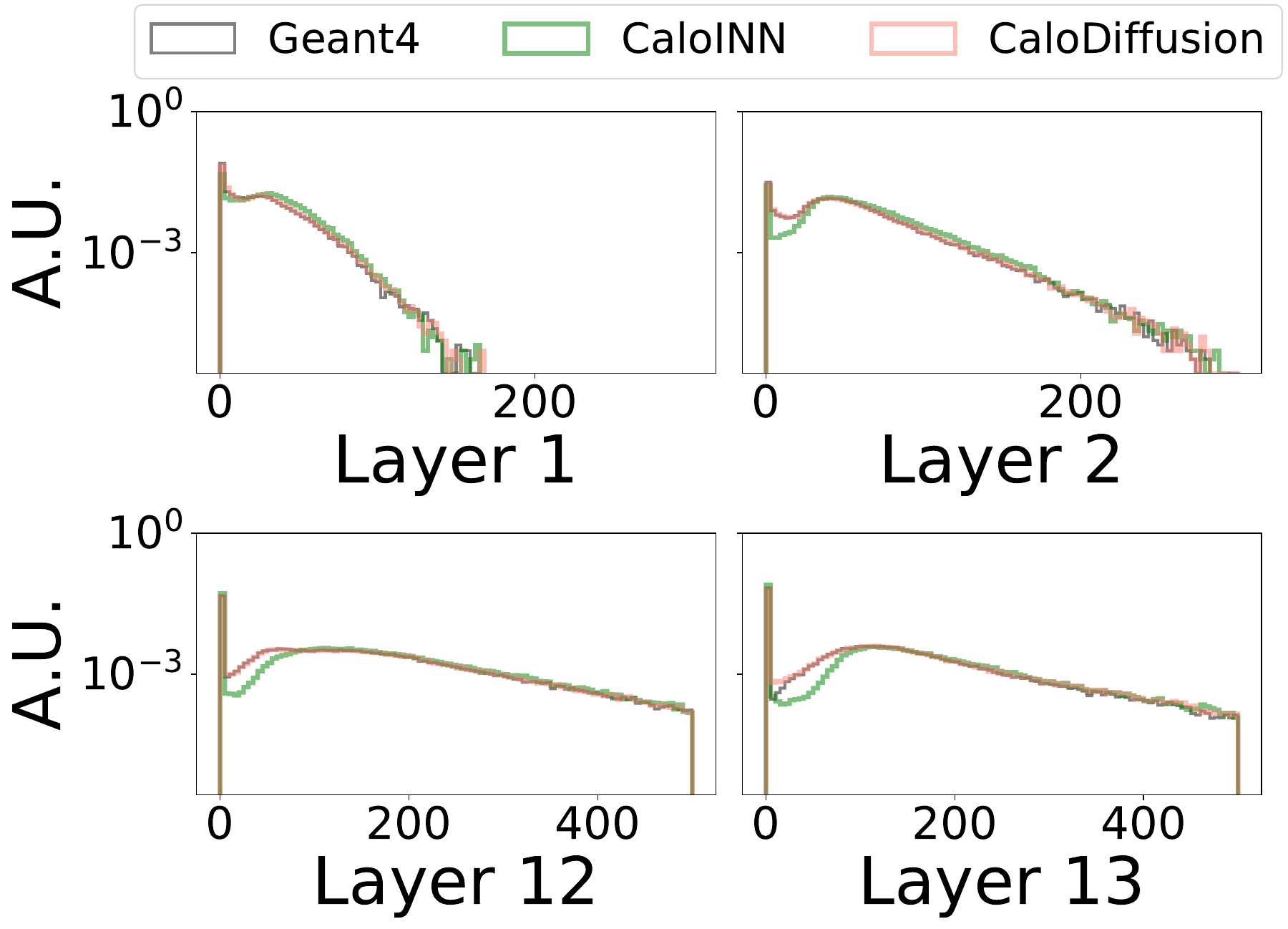}} 
\subfloat[$\phi$ (mm) direction]
{\includegraphics[width=0.5\textwidth]
{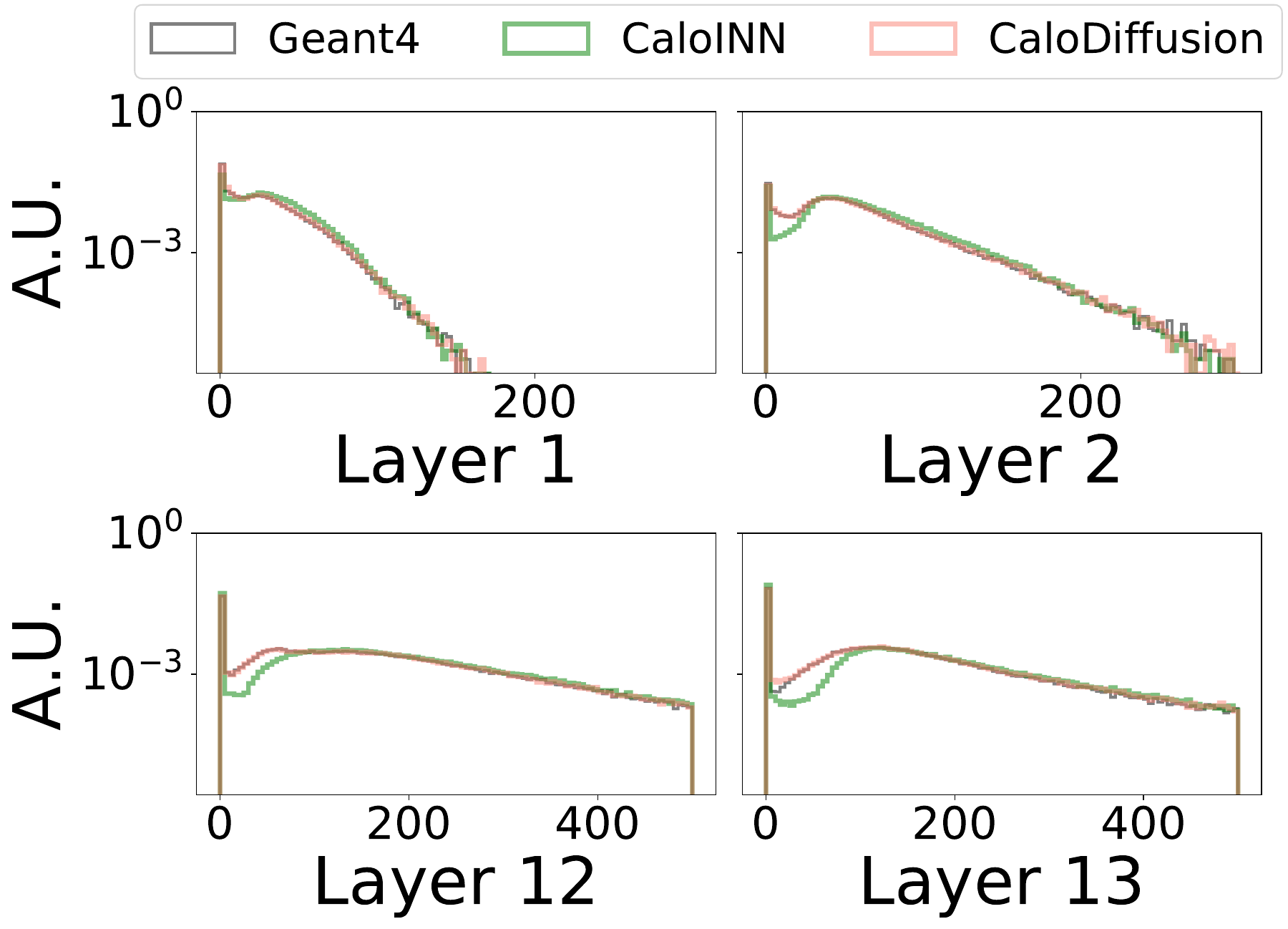}}
\hspace{2mm}
\caption{Shower width for dataset-1(pions).} \label{fig:sw_ds1_pion}
\end{figure}

\begin{figure}[h]
\centering
\subfloat[$\eta$ (mm) direction]
{\includegraphics[width=0.5\textwidth]
{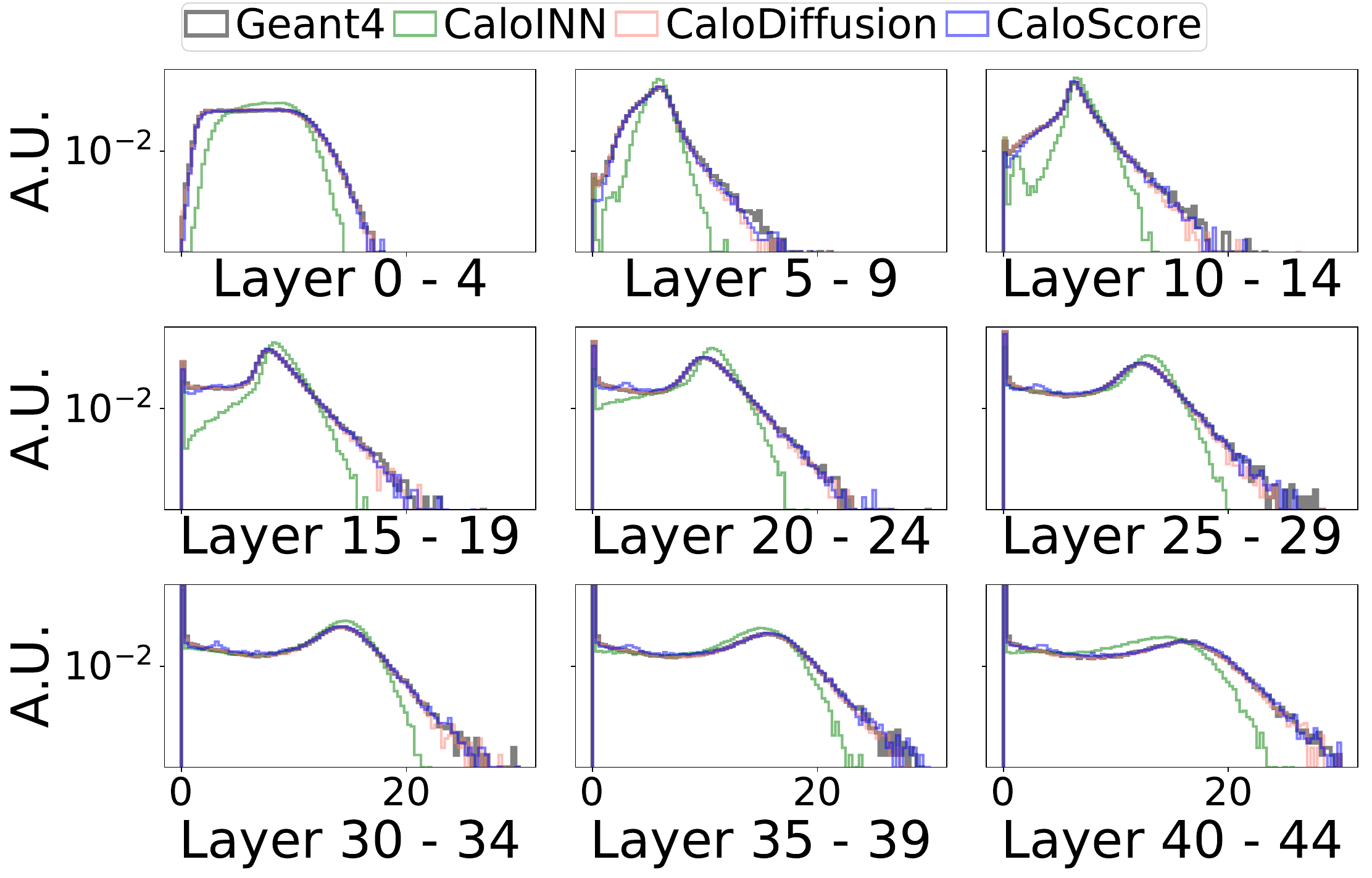}} 
\subfloat[$\phi$ (mm) direction]
{\includegraphics[width=0.5\textwidth]
{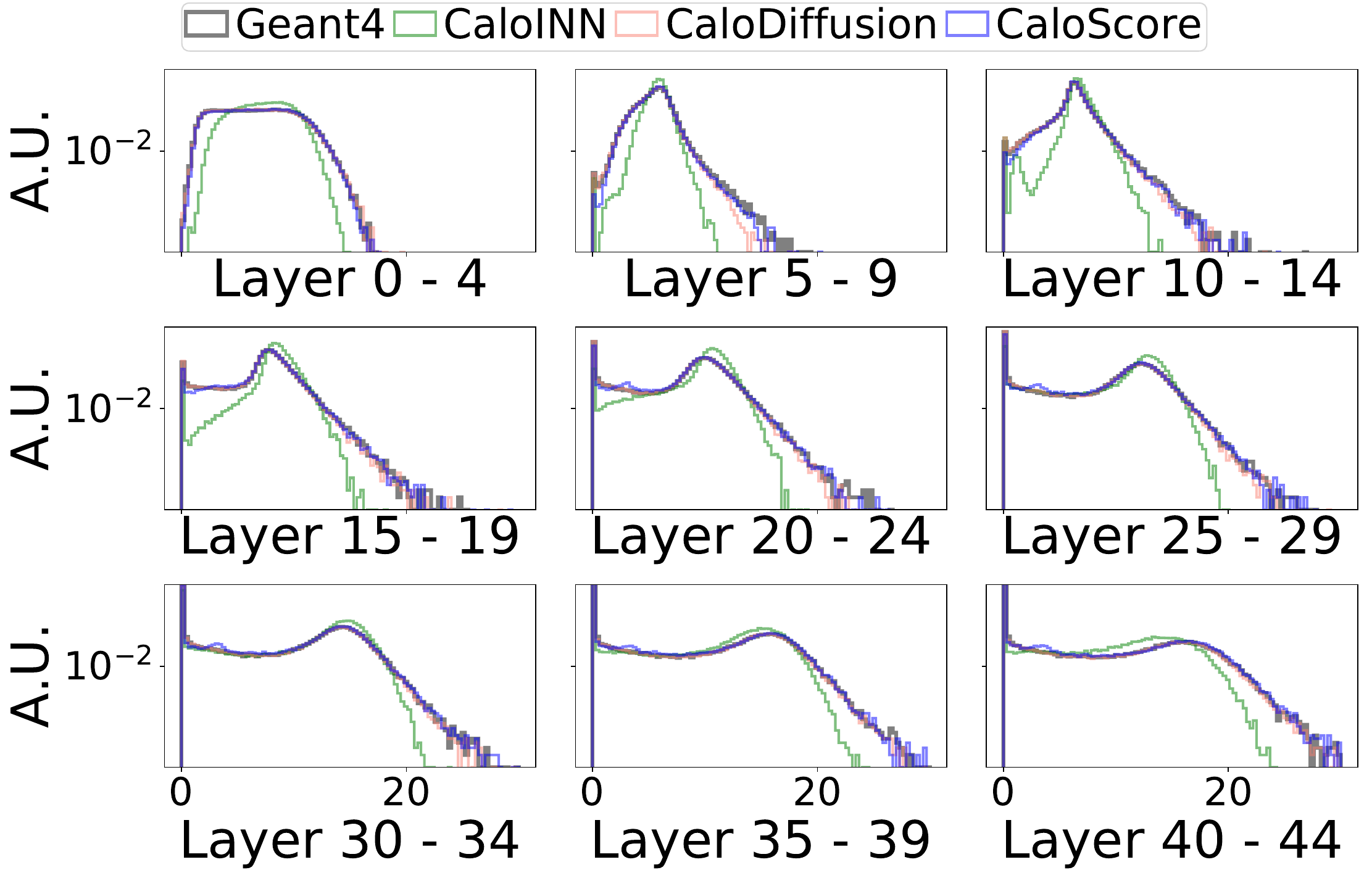}}
\hspace{2mm}
\caption{Shower width for dataset-2.} \label{fig:sw_ds2_photon}
\end{figure}

\begin{figure}[h]
\centering
\subfloat[$\eta$ (mm) direction]
{\includegraphics[width=0.5\textwidth]
{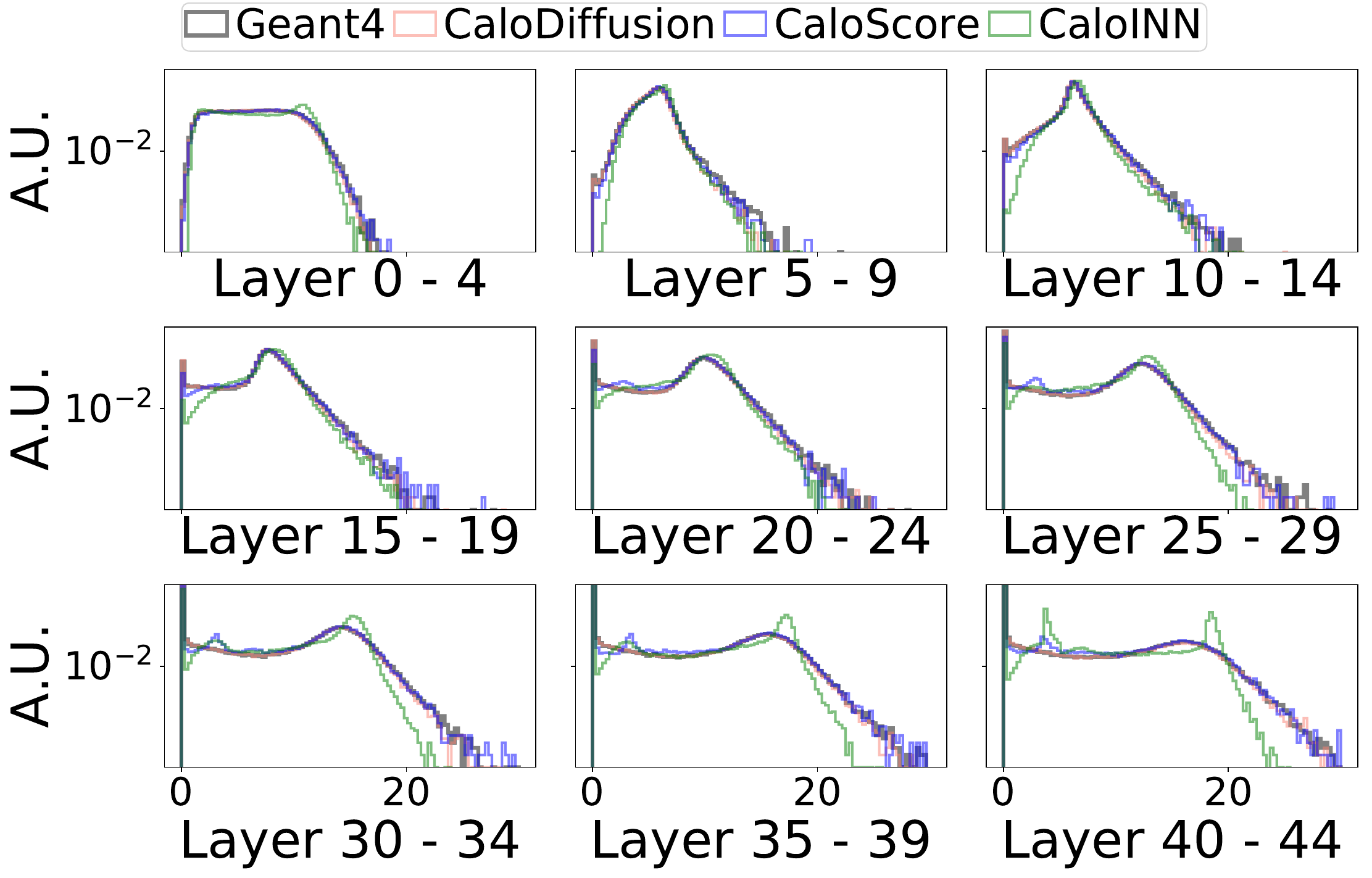}} 
\subfloat[$\phi$ (mm) direction]
{\includegraphics[width=0.5\textwidth]
{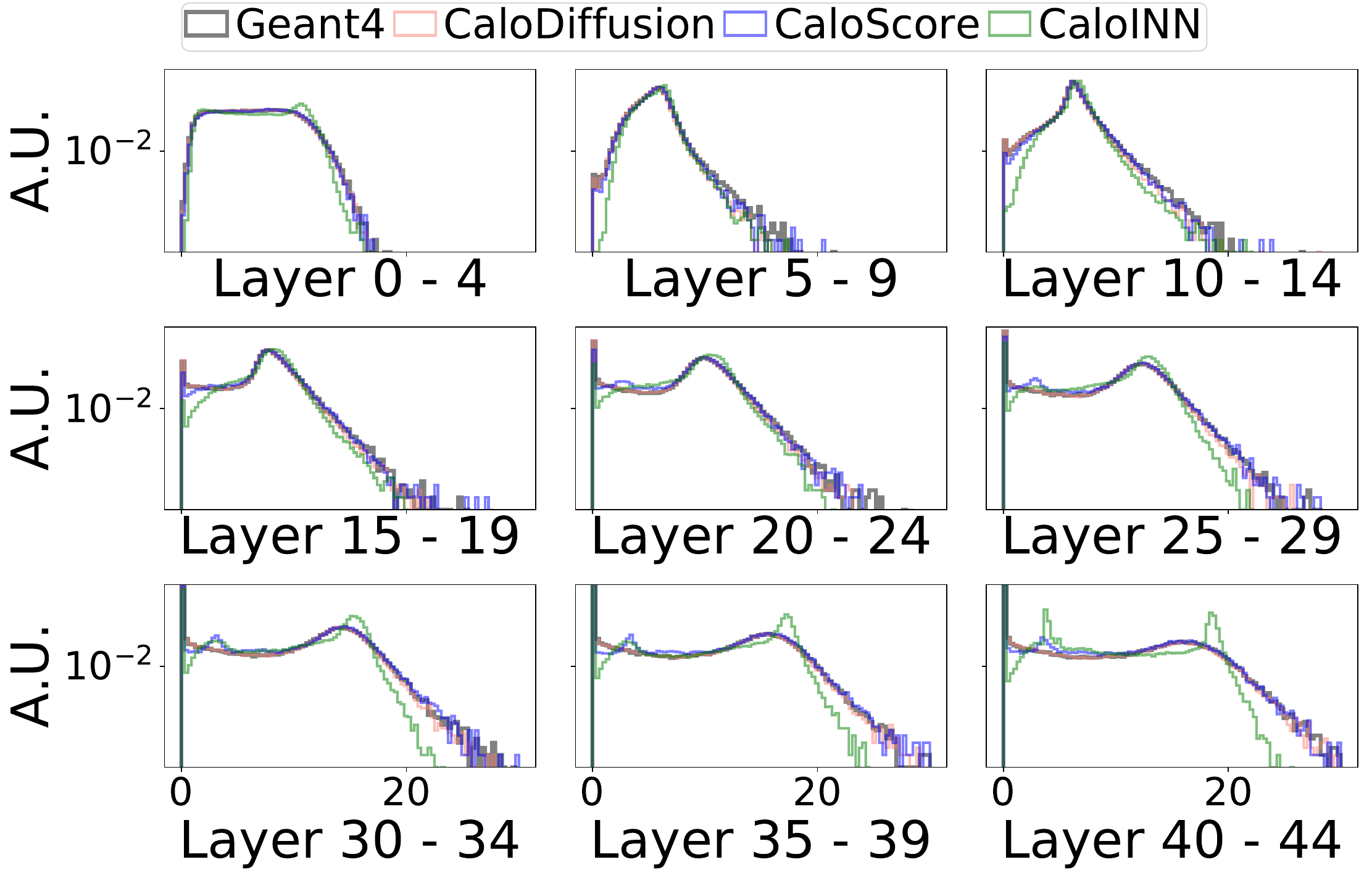}}
\hspace{2mm}
\caption{Shower width for dataset-3.} \label{fig:sw_ds3}
\end{figure}

\textbf{Shower width:}
We studied the angular distributions of the calorimeter showers in datasets 1, 2, and 3 in terms of the shower width, illustrated in Figures ~\ref{fig:sw_ds1_photon} through ~\ref{fig:sw_ds3}. Shower width measures how far secondary particles spread perpendicular to the direction of the initial particle from the perspective of $\eta$ and $\mu$, measured in mm.

%%% not good image. need to regenerate

%%% not good image. need to regenerate

\textit{Analysis:} For dataset 1 (photons), both CaloScore and CaloDiffusion perform similarly in terms of shower width while CaloINN shows significant mismodeling for all three datasets. We note that both CaloScore and CaloDiffusion show some degree of mismodeling (right end of the plots). For datsets 2 and 3, this mismatch is especially evident in layers 25-30 for both CaloDiffusion and CaloScore.

% \textbf{Sparsity:} Figure~\ref{fig:sparsity} illustrates the sparsity metric which is a measure of ratio of the number of voxels with nonzero deposition to the total number of voxels, for datasets 2 and 3. 
%%% need to be fixed same image for testing

\begin{figure}  
\subfloat[Layer energy distribution (GeV)]
{\includegraphics[width=0.5\textwidth]
{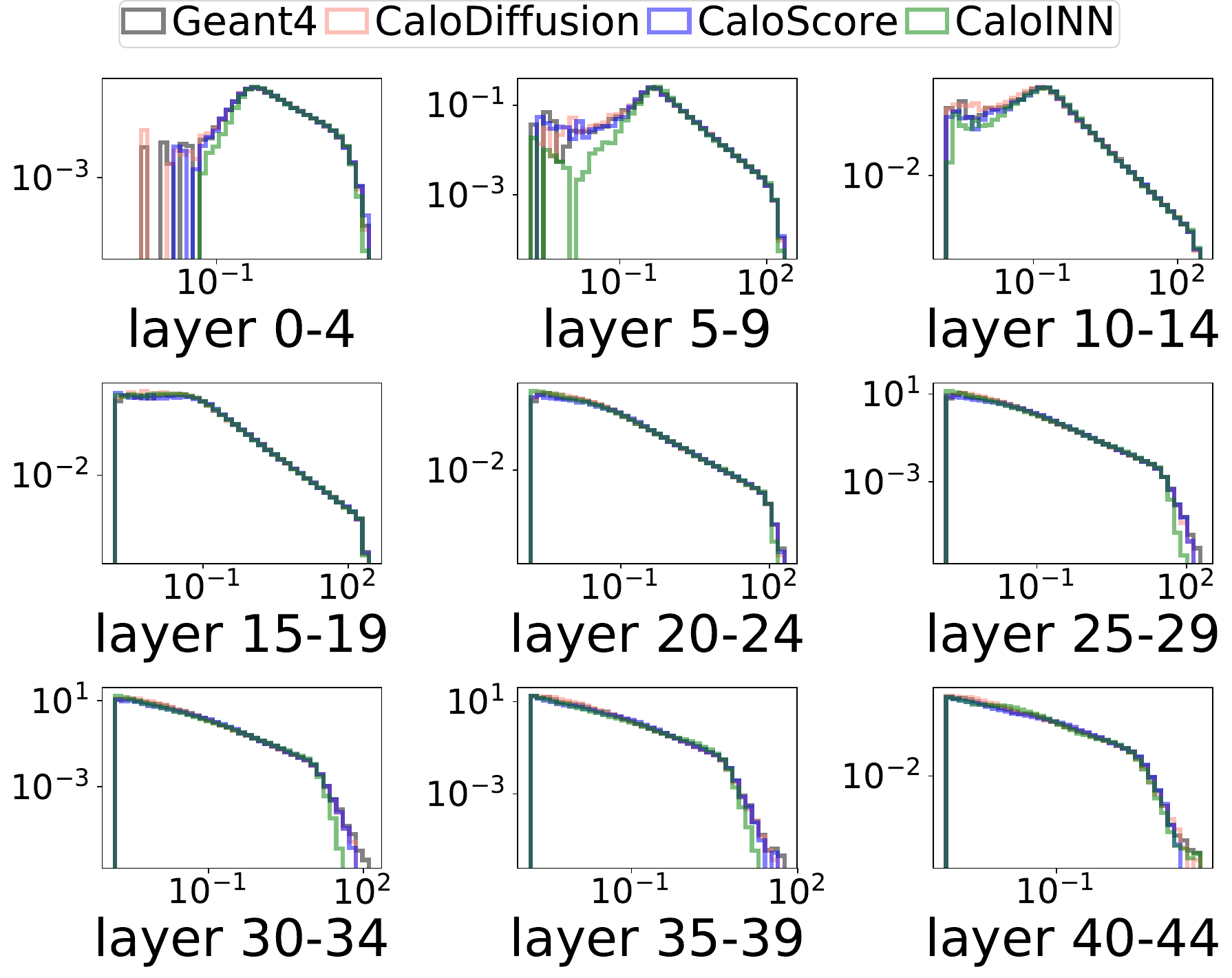}} 
\subfloat[Distribution of sparsity \label{fig:sparsity_ds3}]
{\includegraphics[width=0.5\textwidth]
{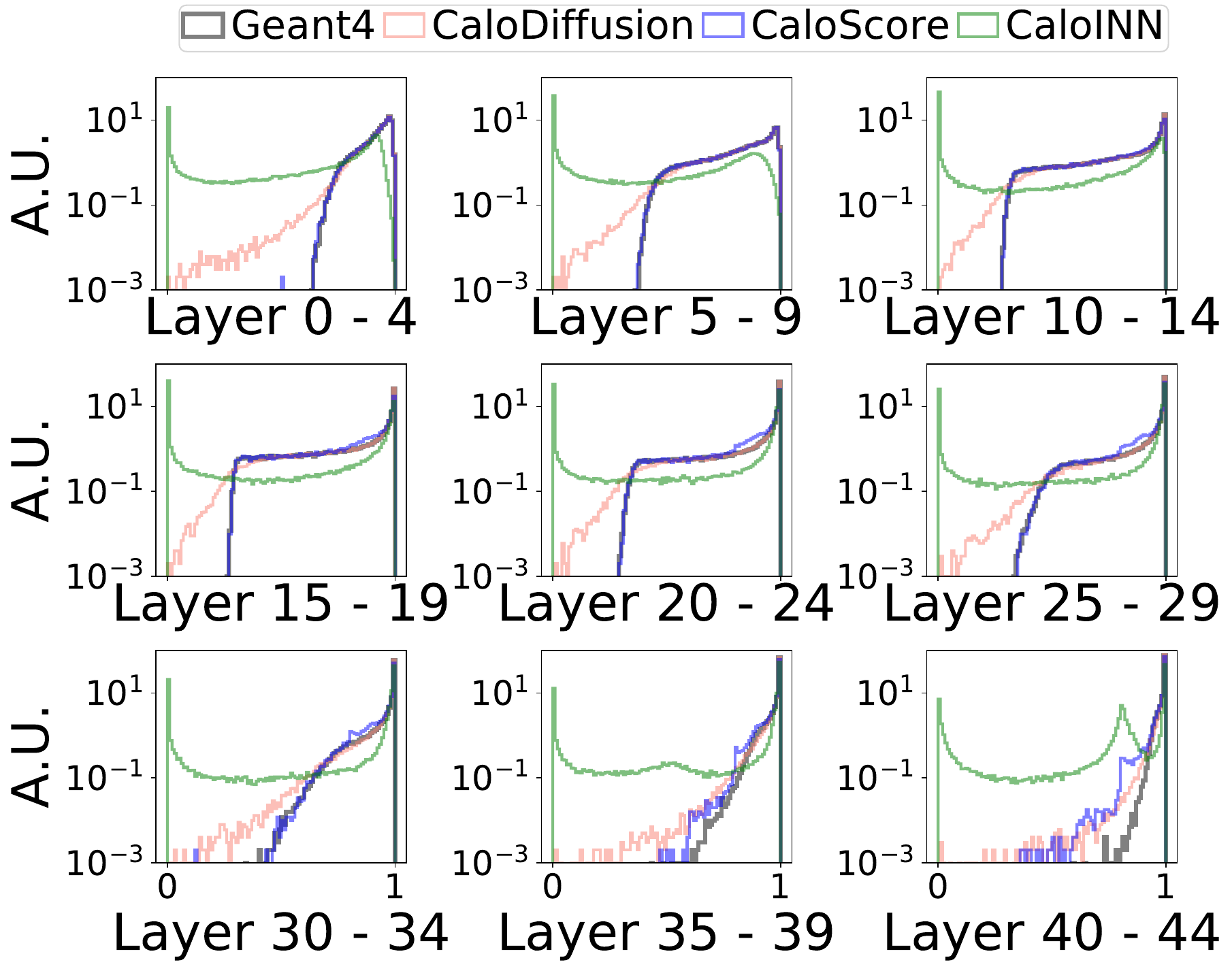}}
\caption{Histogram of two physics observables for dataset 3} \label{fig:histogram_observe_3}
\end{figure}

\subsection{EMD scores}

\begin{figure}[h]
\centering
\subfloat[Layer wise energy(in all ranges of incident energy)]
{\includegraphics[width=0.33\textwidth]{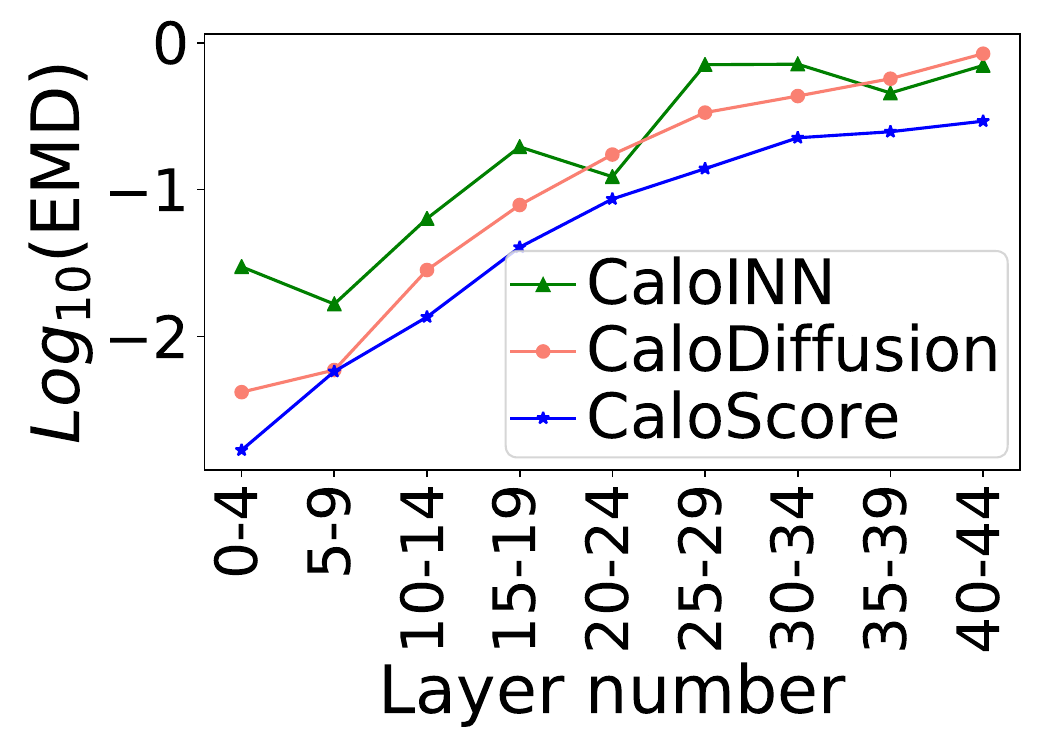}} 
\subfloat[Sparsity.]
{\includegraphics[width=0.33\textwidth]{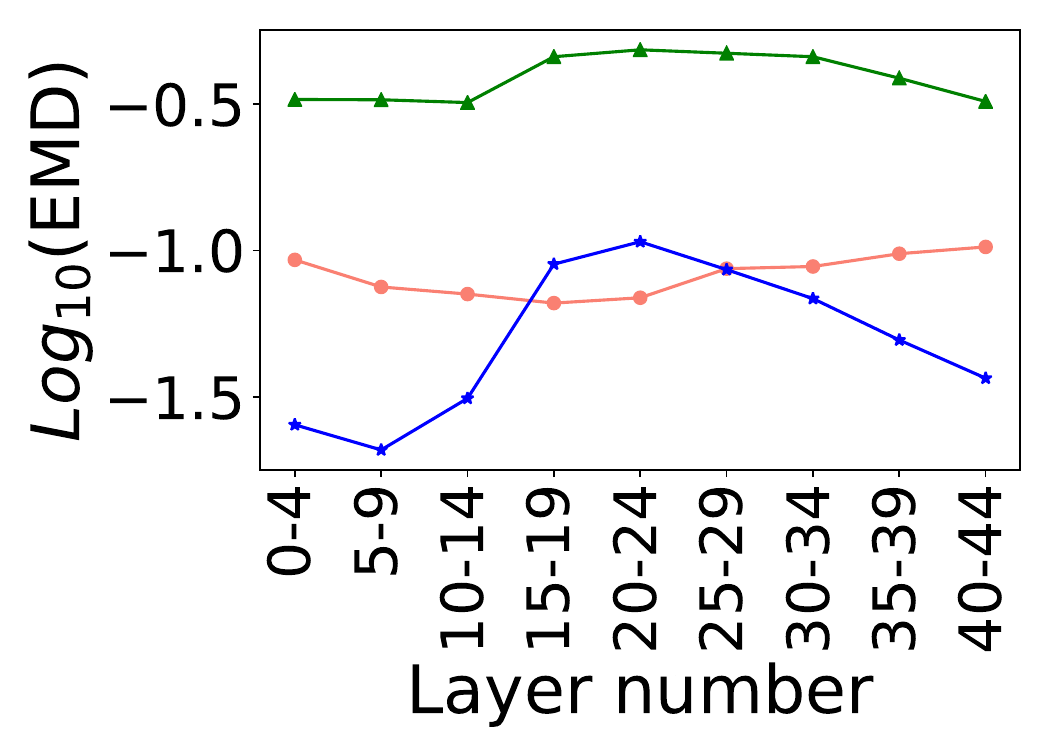}} 
\subfloat[Center of energy in $\eta$ direction.]
{\includegraphics[width=0.33\textwidth]{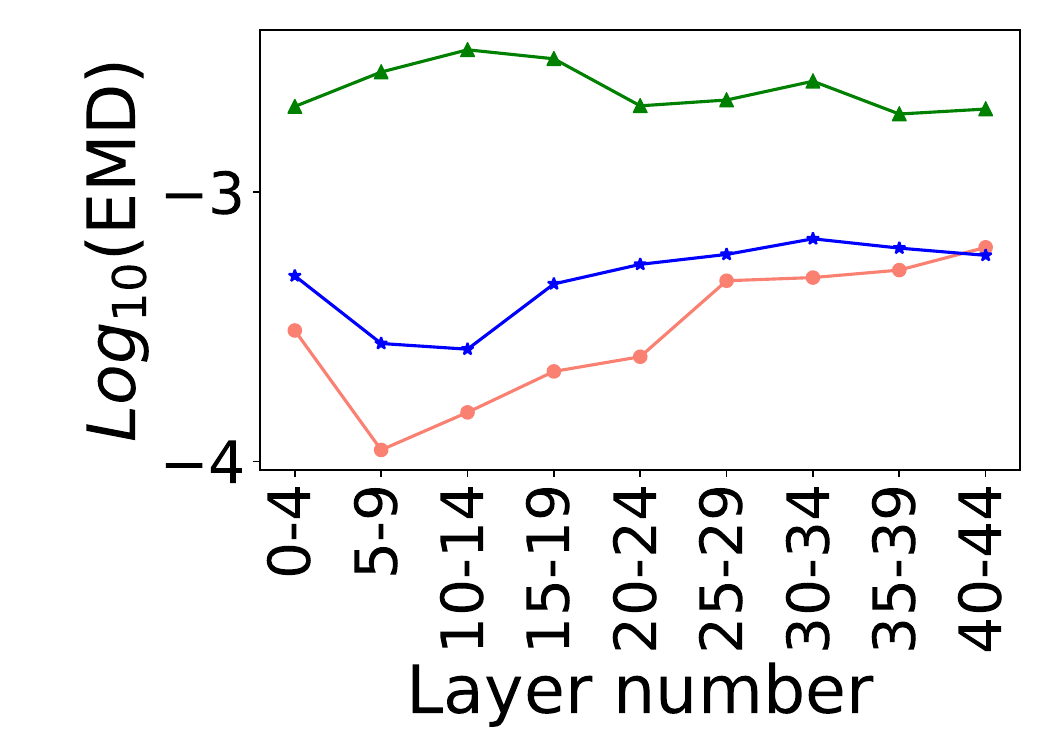}} 
\vspace{1mm}
\subfloat[Center of energy in $\phi$ direction]
{\includegraphics[width=0.33\textwidth]{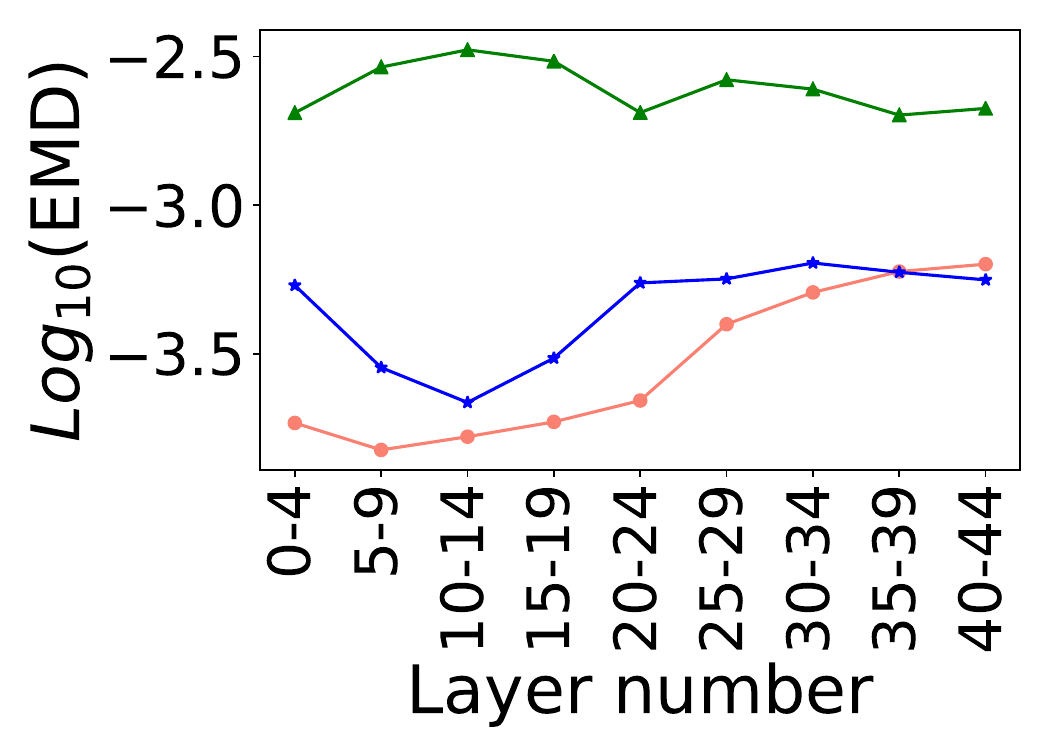}} 
\subfloat[Shower width in $\eta$ direction.]
{\includegraphics[width=0.33\textwidth]{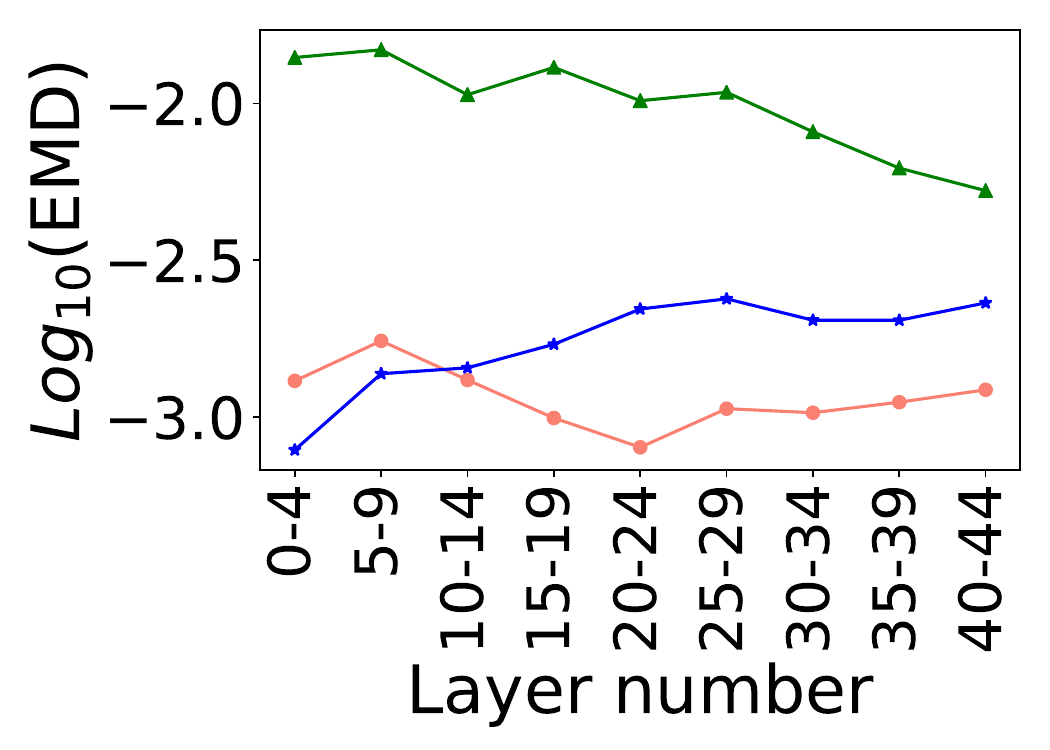}} 
\subfloat[Shower width in $\phi$ direction.]
{\includegraphics[width=0.33\textwidth]{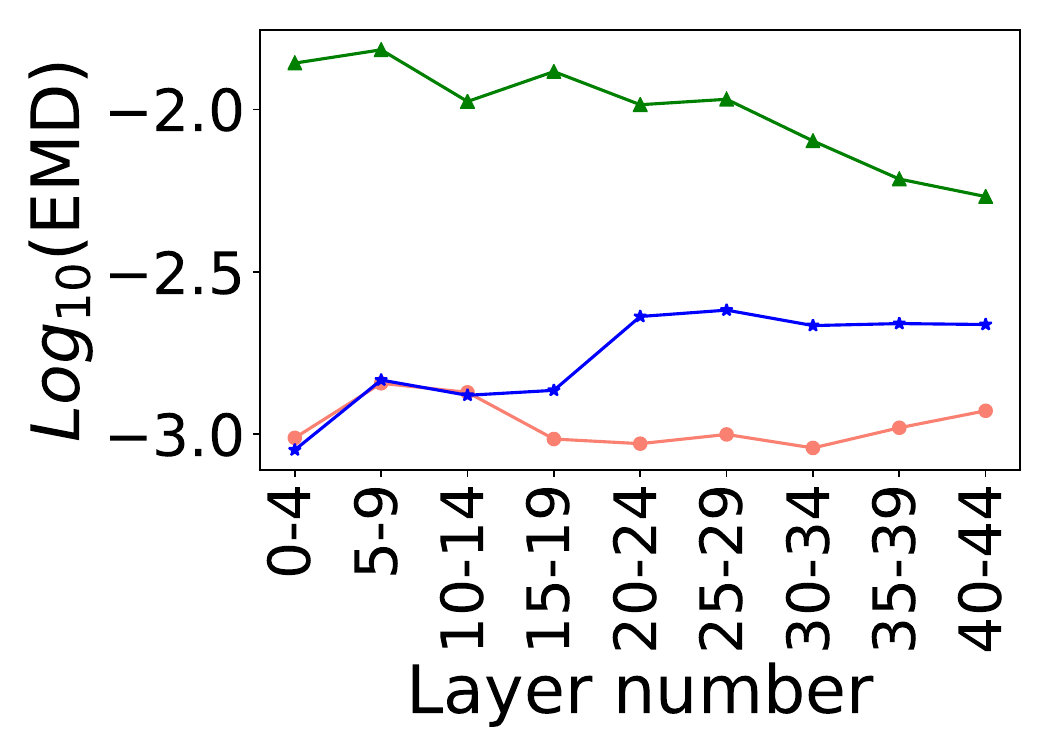}} 
\caption{EMD score for different high level features for dataset 2.} \label{fig:ds2_emd}
\end{figure}

\begin{figure}[h]
\centering
\subfloat[Layer wise energy(in all ranges of incident energy)]
{\includegraphics[width=0.33\textwidth]{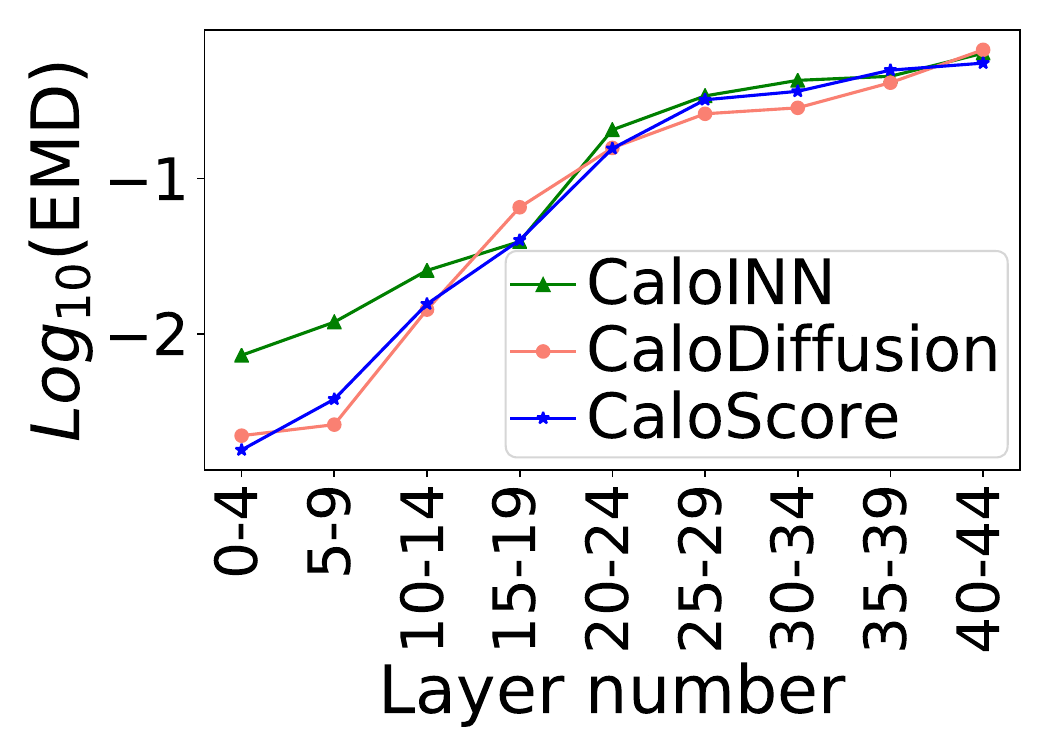}} 
\subfloat[Sparsity.]
{\includegraphics[width=0.33\textwidth]{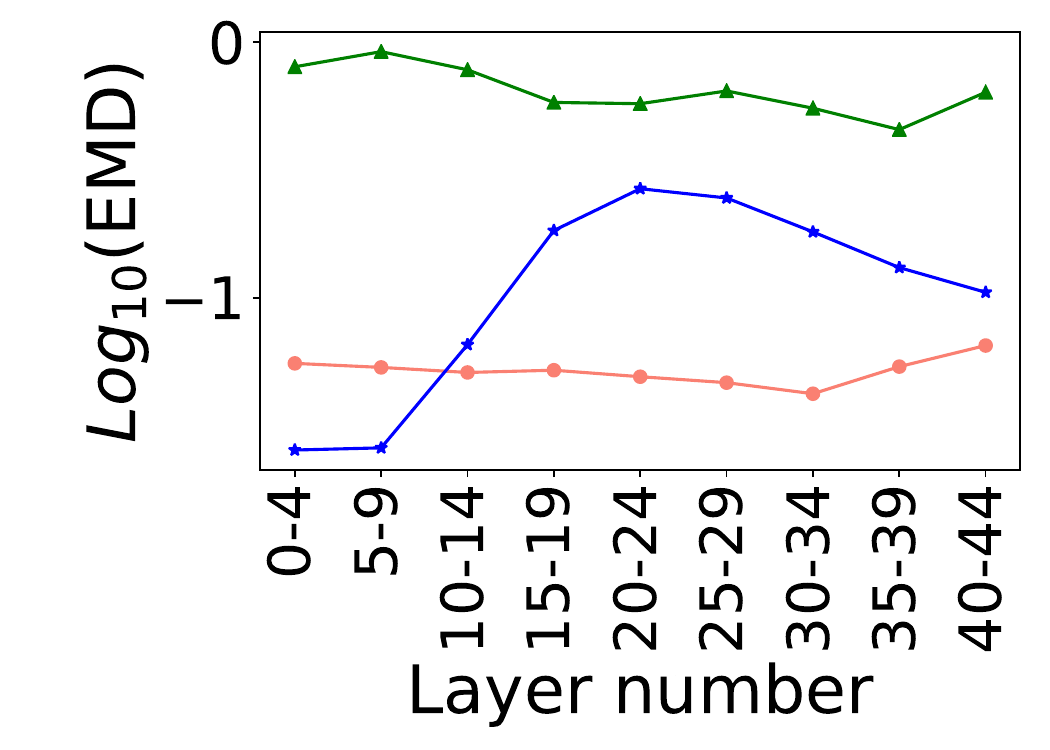}} \subfloat[Center of energy in $\eta$ direction.]
{\includegraphics[width=0.33\textwidth]{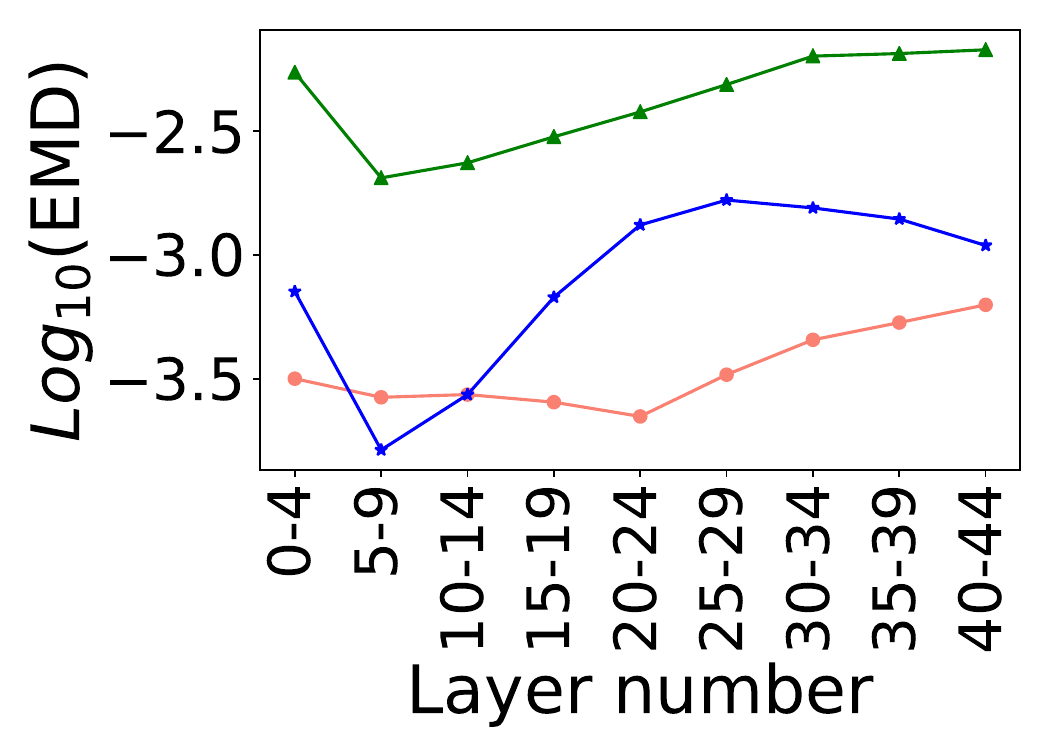}} 
\vspace{1mm}
\subfloat[Center of energy in $\phi$ direction]
{\includegraphics[width=0.33\textwidth]{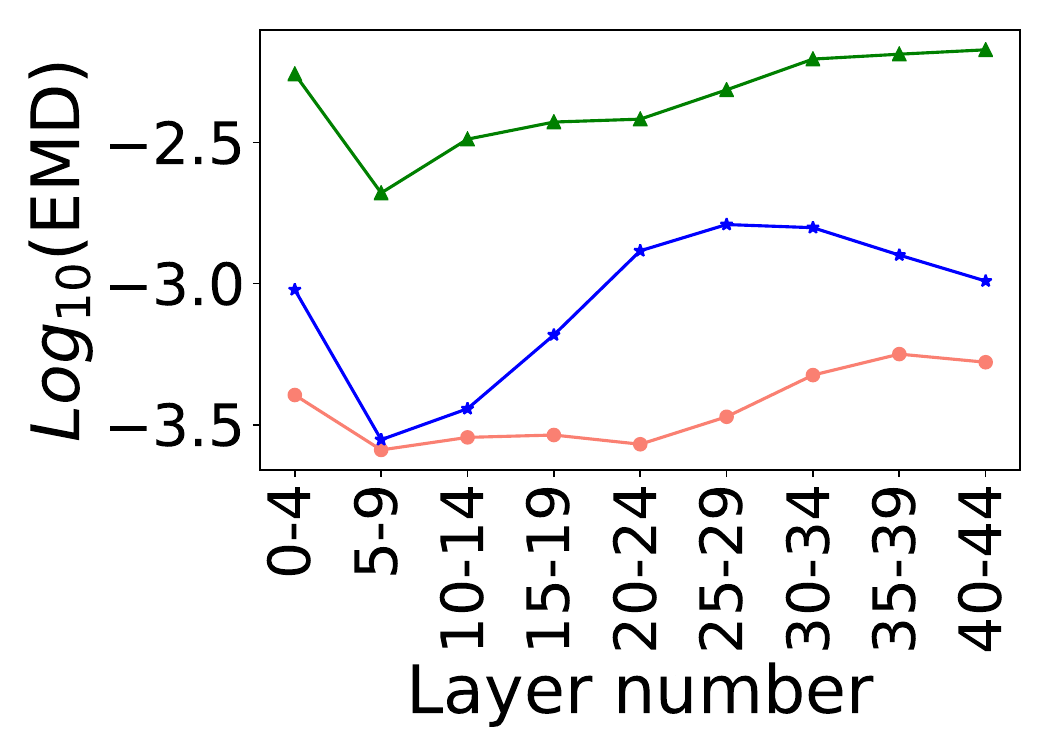}} 
\subfloat[Shower width in $\eta$ direction.]
{\includegraphics[width=0.33\textwidth]{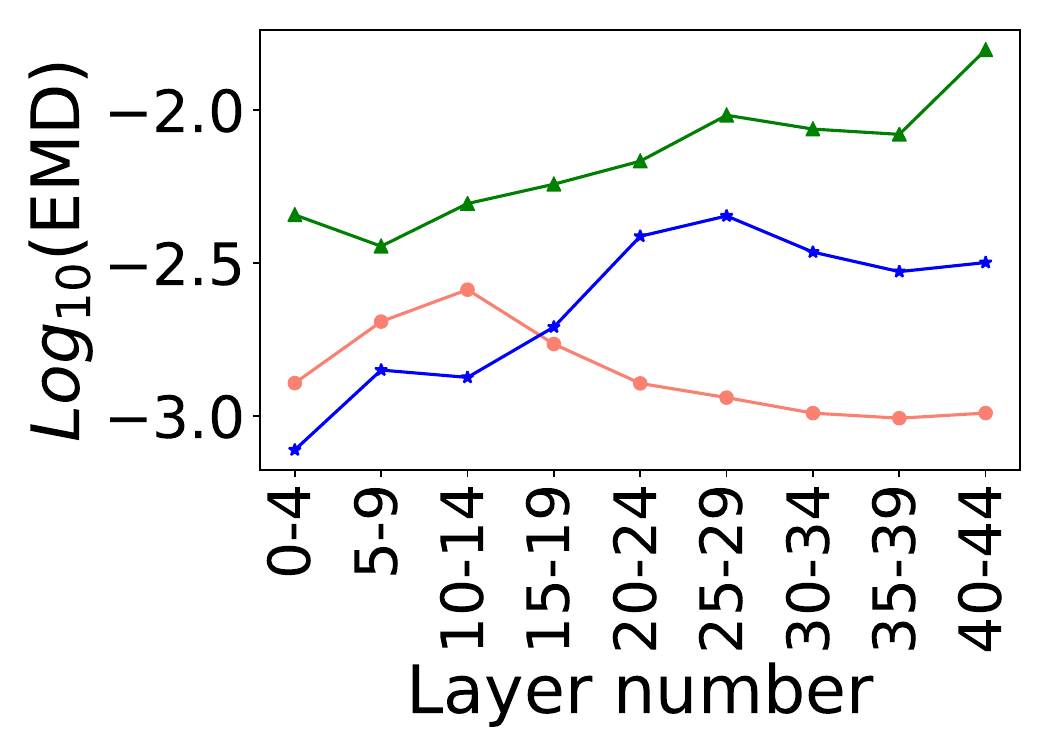}} 
\subfloat[Shower width in $\phi$ direction.]
{\includegraphics[width=0.33\textwidth]{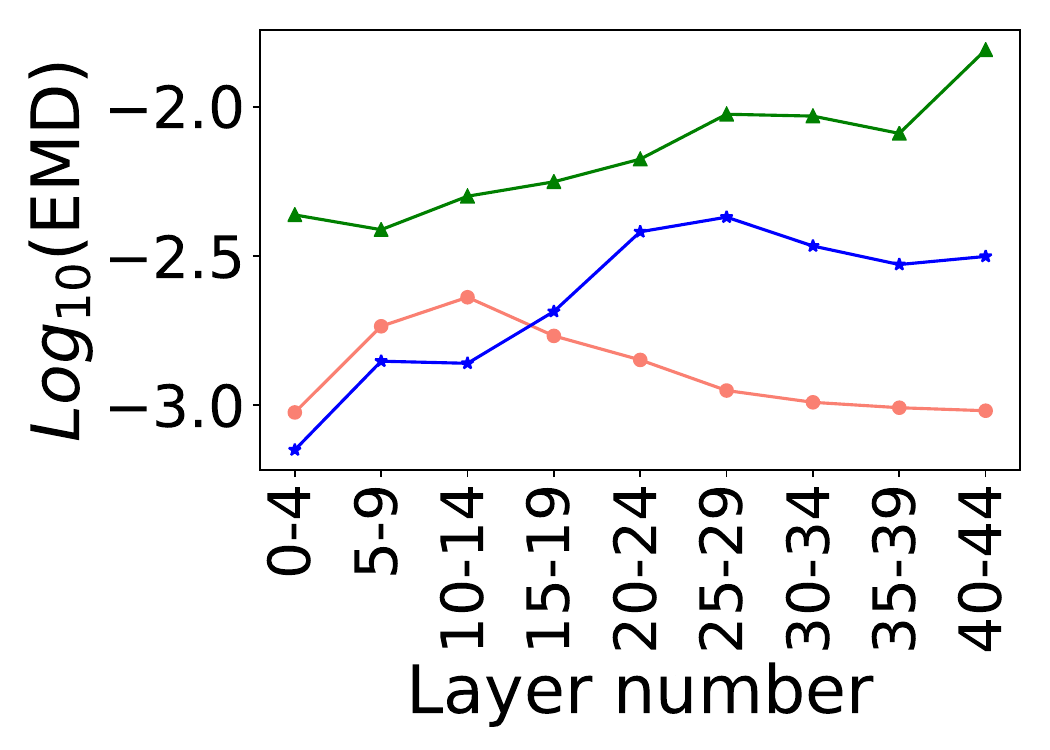}} 
\caption{EMD score for different high level features for dataset 3.} \label{fig:ds3_emd}
\end{figure}
In figure~\ref{fig:ds2_emd}~\ref{fig:ds3_emd},
we present the EMD scores using a logarithmic scale (base 10) to improve the clarity of differences across various physical features of the shower. The smaller the EMD score, the closer the two distributions are to each other. Here we compare each physical observable relative to Geant4. 

\textit{Analysis:} Discussed in main section.
% Table~\ref{tab:fpd_kpd} lists the FPD and KPD measures for each of the models. The FPD and KPD measure the similarity of physics observables based on multidimensional correlated features, both in terms of quality and diversity between two distributions. The smaller the difference, the more similar two distributons are.

% \textit{Analysis:} From table~\ref{tab:fpd_kpd}, we note that CaloScore generates showers that are significantly accurate for dataset 1 and 2, in terms of similarity of the generated showers w.r.t Geant4 compared to the other models. For dataset 3, CaloDiffusion generates more accurate showers whereas CaloINN does poorly both for FPD and KPD measures on all datasets. Additionally, CaloDiffusion with mixed-precision inference shows improved the FPD and KPD scores, evidence that lower precision does not necessarily compromise on the accuracy of the generated showers.

\subsection{Separation power}
\begin{figure}[h]
\centering
\subfloat[Layer wise energy(in all ranges of incident energy).]
{\includegraphics[width=0.33\textwidth]{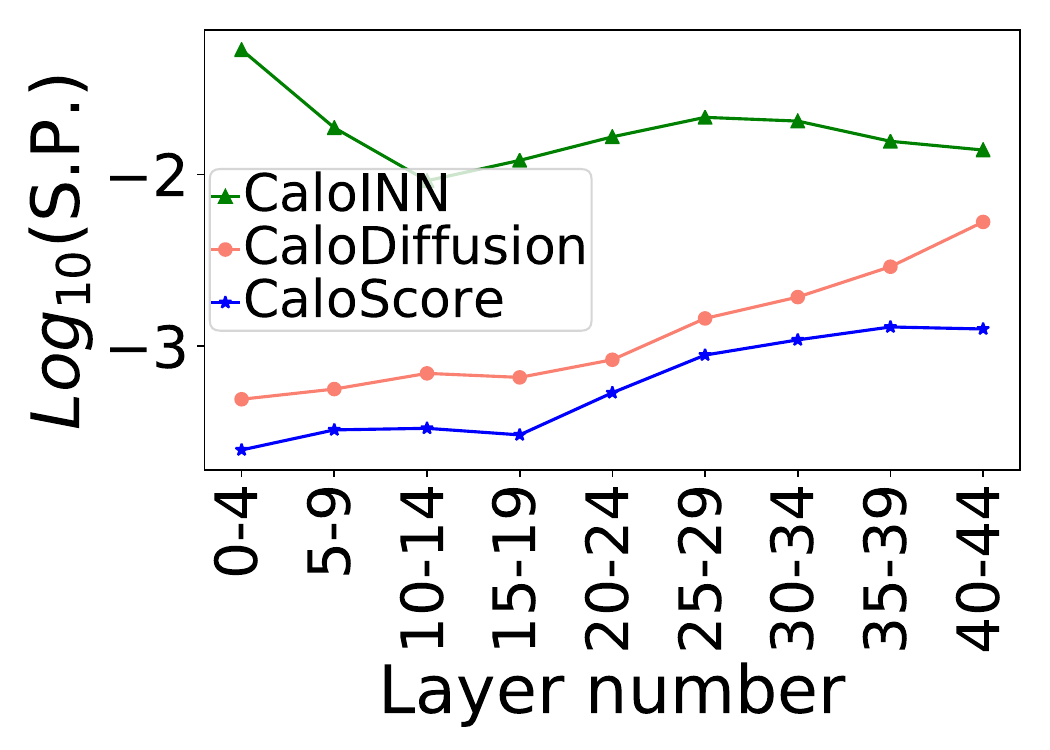}} 
\subfloat[Sparsity.]
{\includegraphics[width=0.33\textwidth]{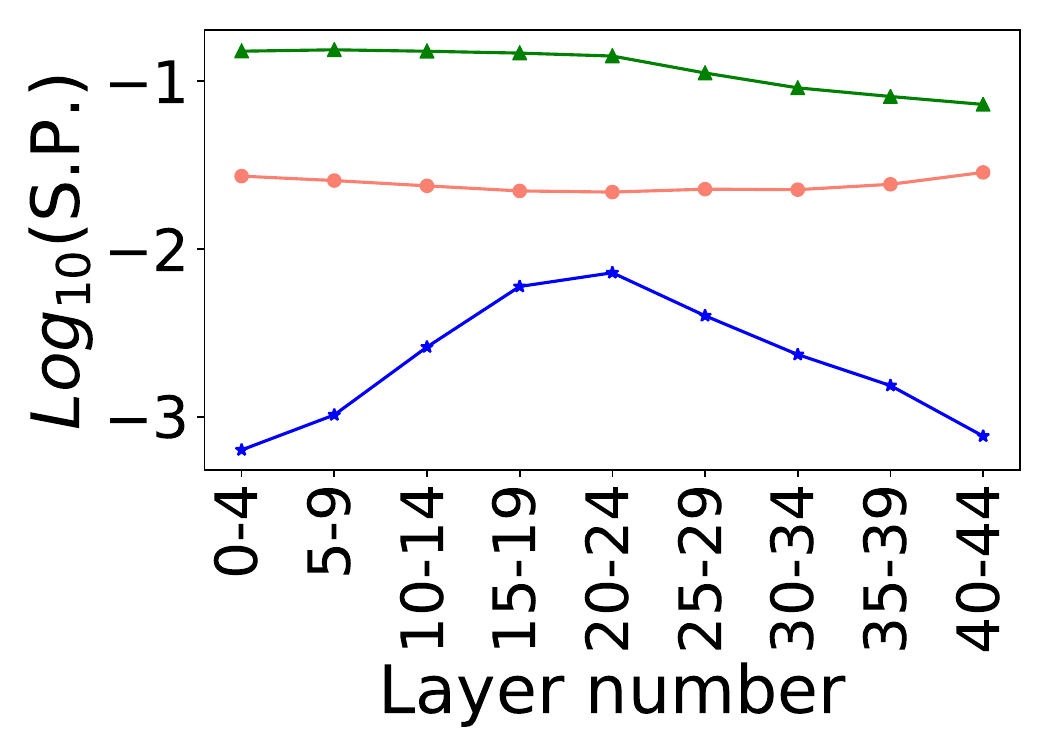}} 
\subfloat[Center of energy in $\eta$ direction.]
{\includegraphics[width=0.33\textwidth]{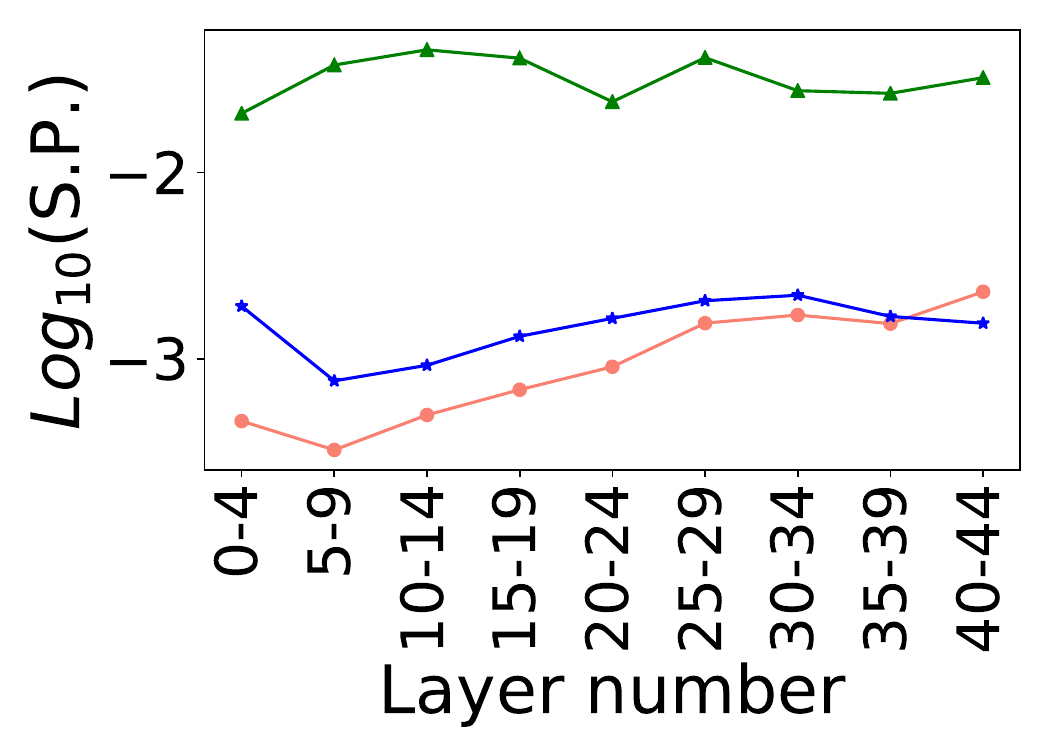}} 
\vspace{1mm}
\subfloat[Center of energy in $\phi$ direction.]
{\includegraphics[width=0.33\textwidth]{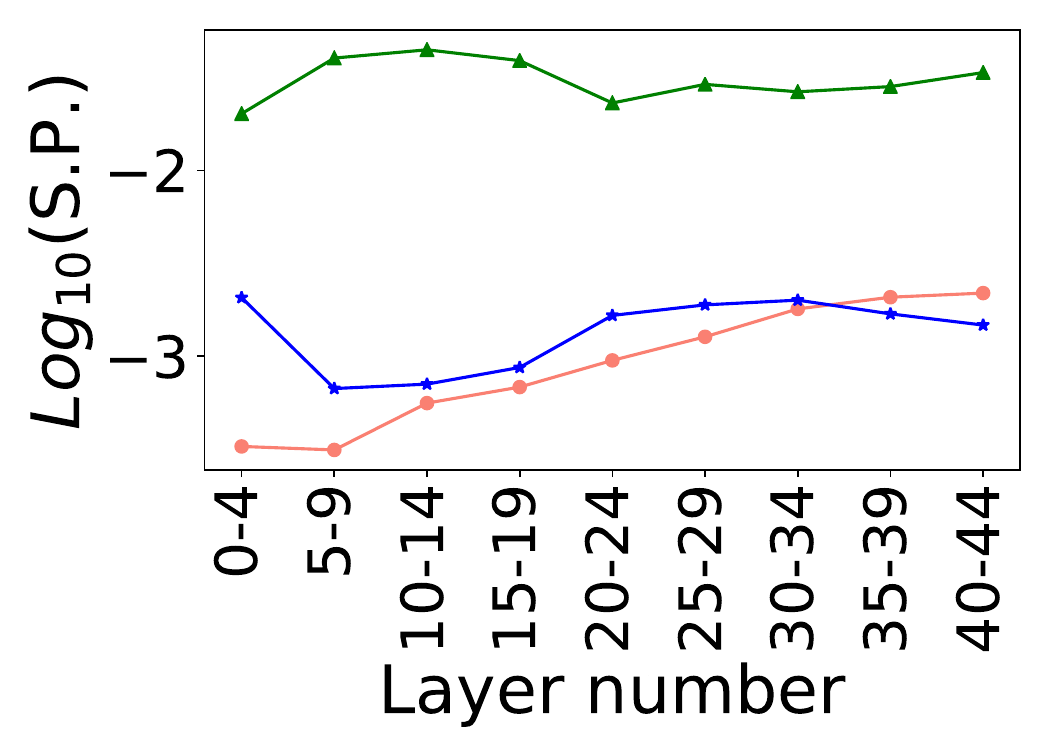}} 
\subfloat[Shower width in $\eta$ direction.]
{\includegraphics[width=0.33\textwidth]{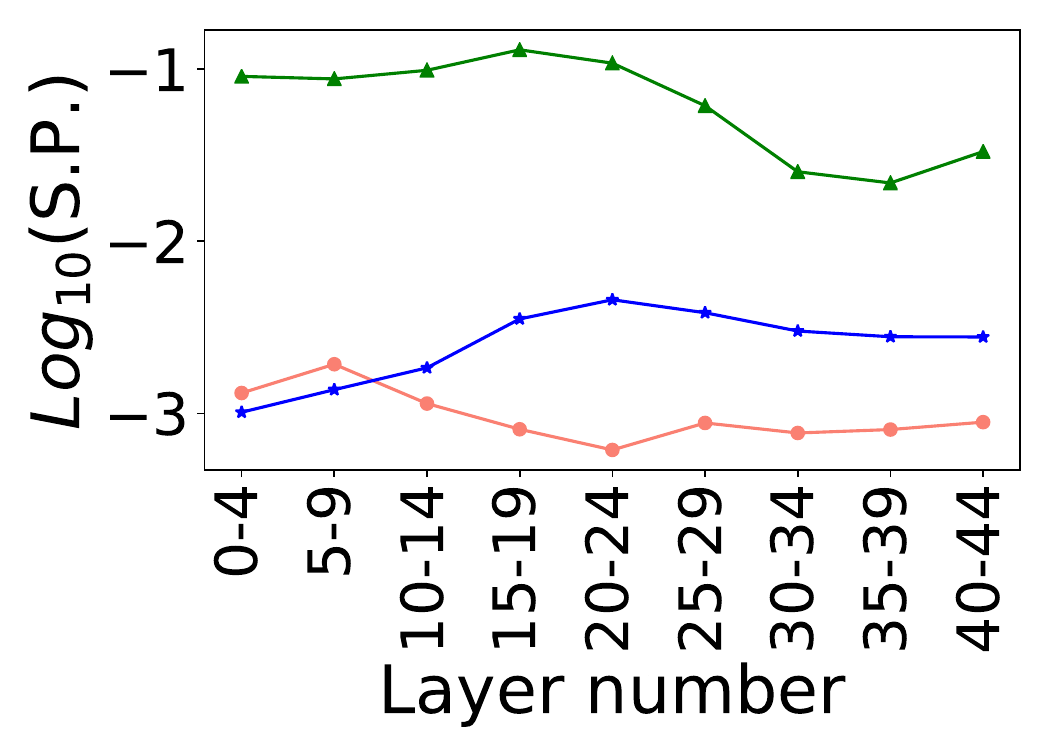}} 
\subfloat[Shower width in $\phi$ direction.]
{\includegraphics[width=0.33\textwidth]{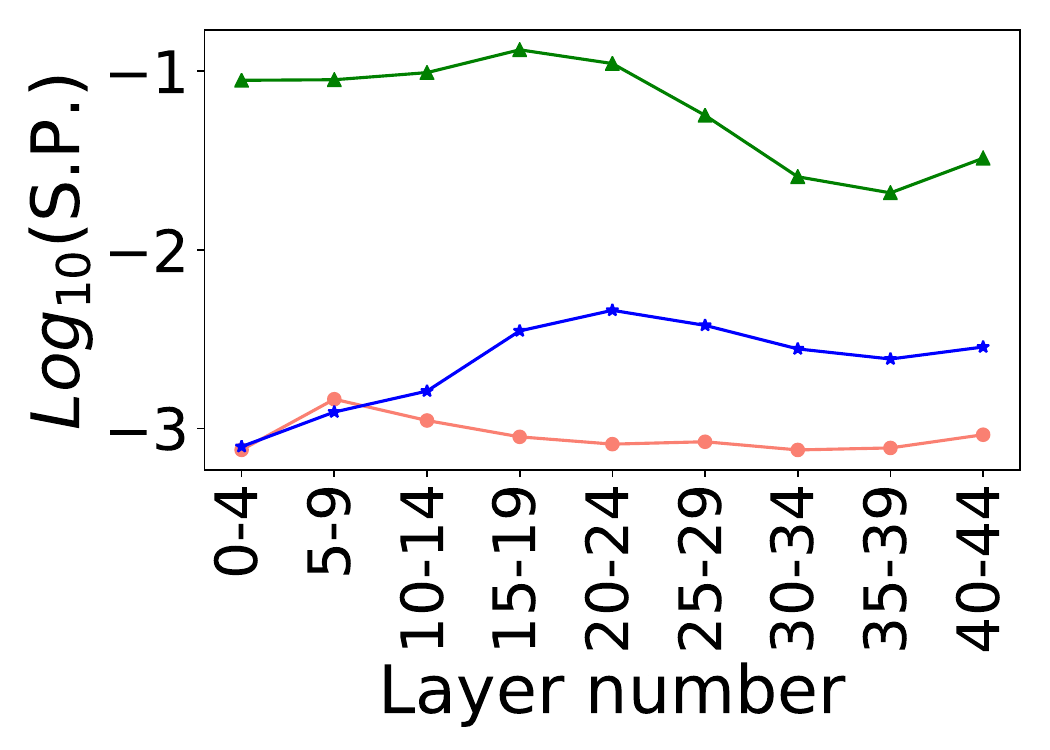}} 
\caption{Separation power(S.P.) of different high level features' histograms for dataset 2.} \label{fig:ds2_sep}
\end{figure}

\begin{figure}[h]
\centering
\subfloat[Layer wise energy(in all ranges of incident energy)]
{\includegraphics[width=0.33\textwidth]{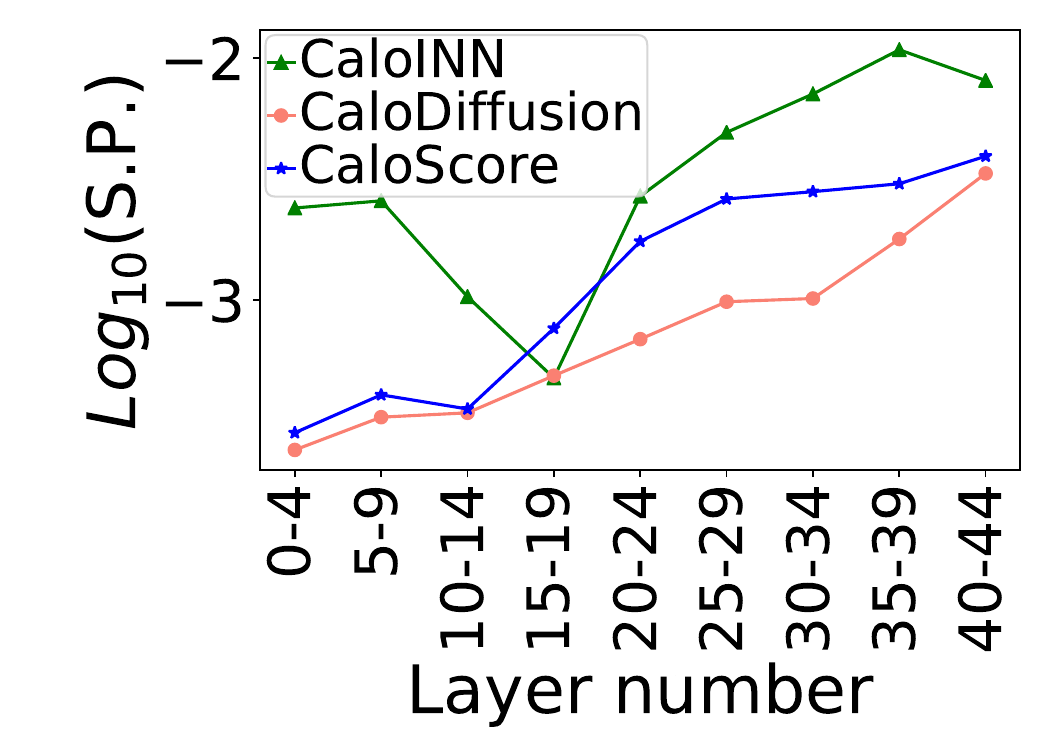}} 
\subfloat[Sparsity.]
{\includegraphics[width=0.33\textwidth]{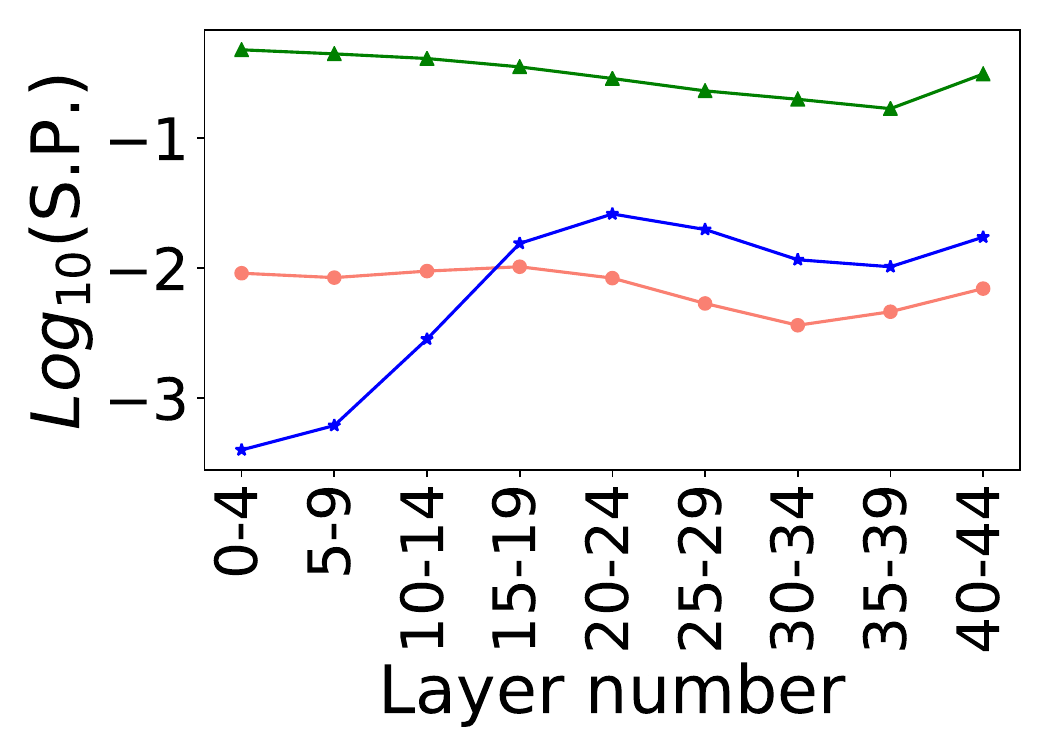}} 
\subfloat[Center of energy in $\eta$ direction.]
{\includegraphics[width=0.33\textwidth]{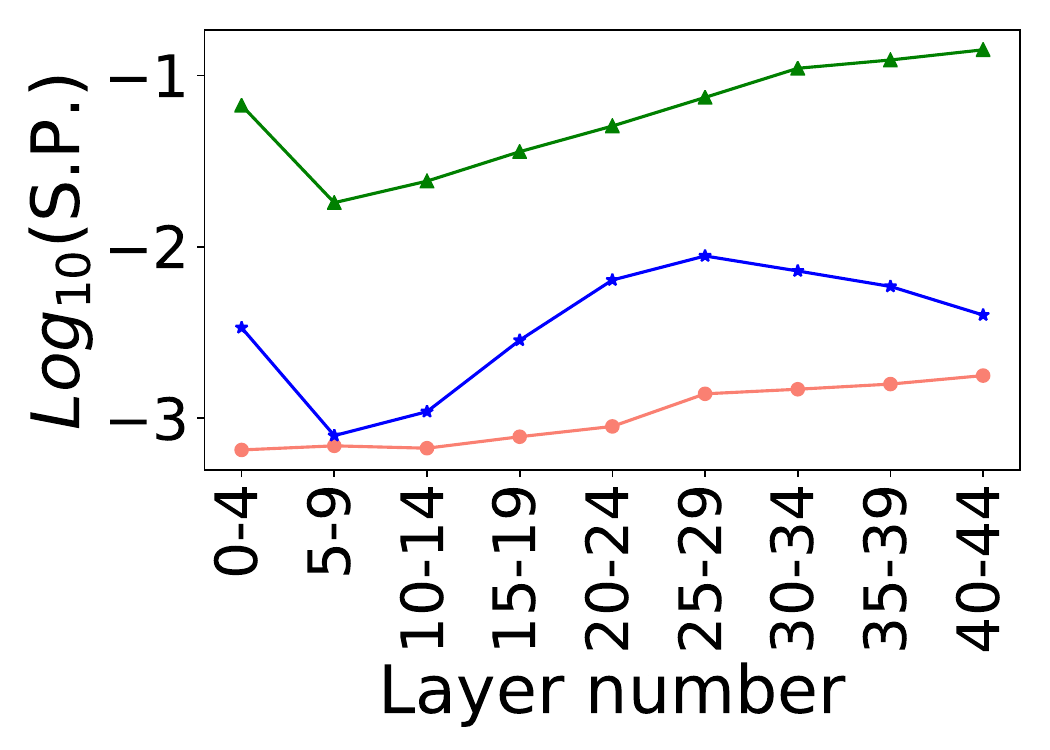}} 
\vspace{1mm}
\subfloat[Center of energy in $\phi$ direction]
{\includegraphics[width=0.33\textwidth]{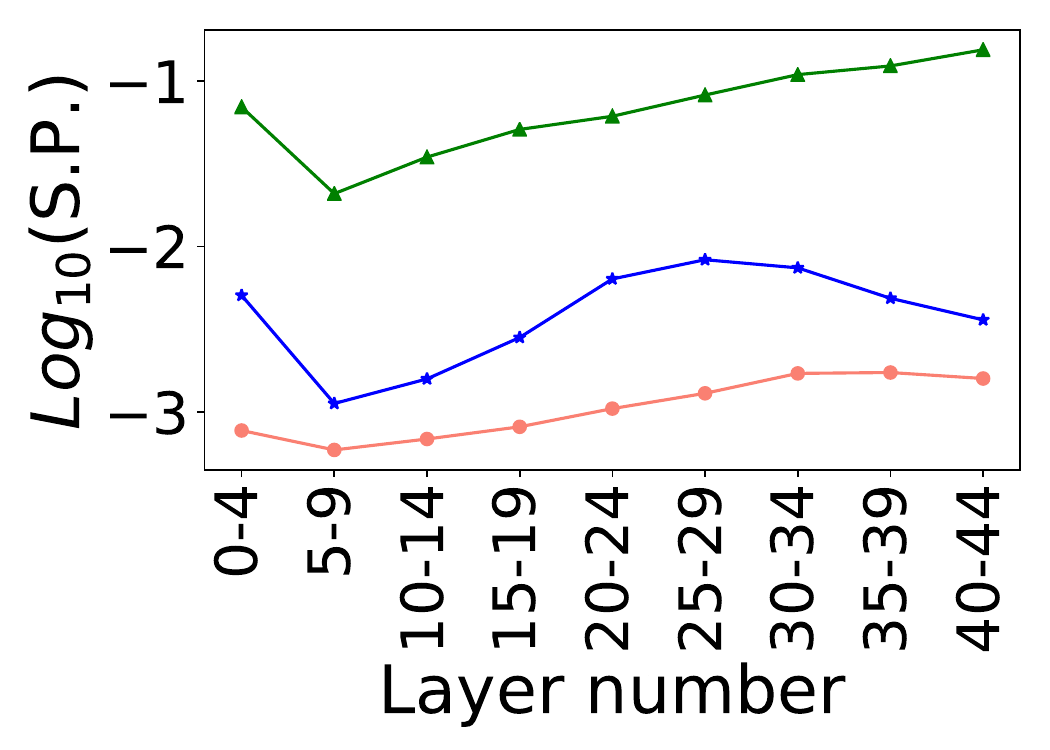}} 
\subfloat[Shower width in $\eta$ direction.]
{\includegraphics[width=0.33\textwidth]{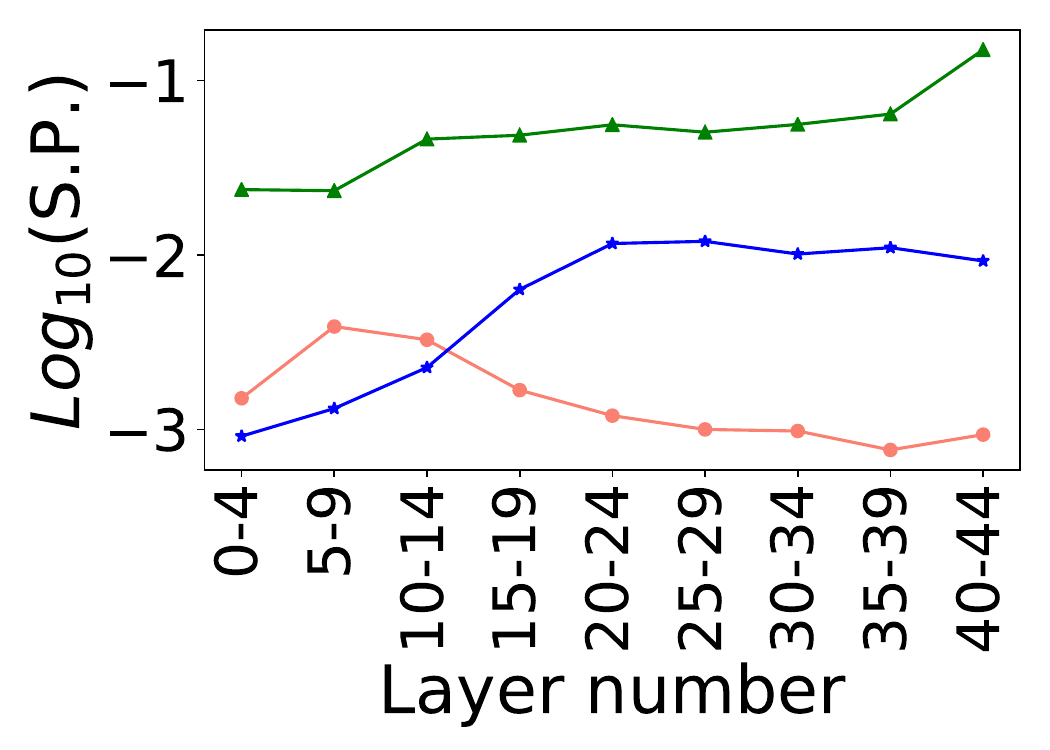}} 
\subfloat[Shower width in $\phi$ direction.]
{\includegraphics[width=0.33\textwidth]{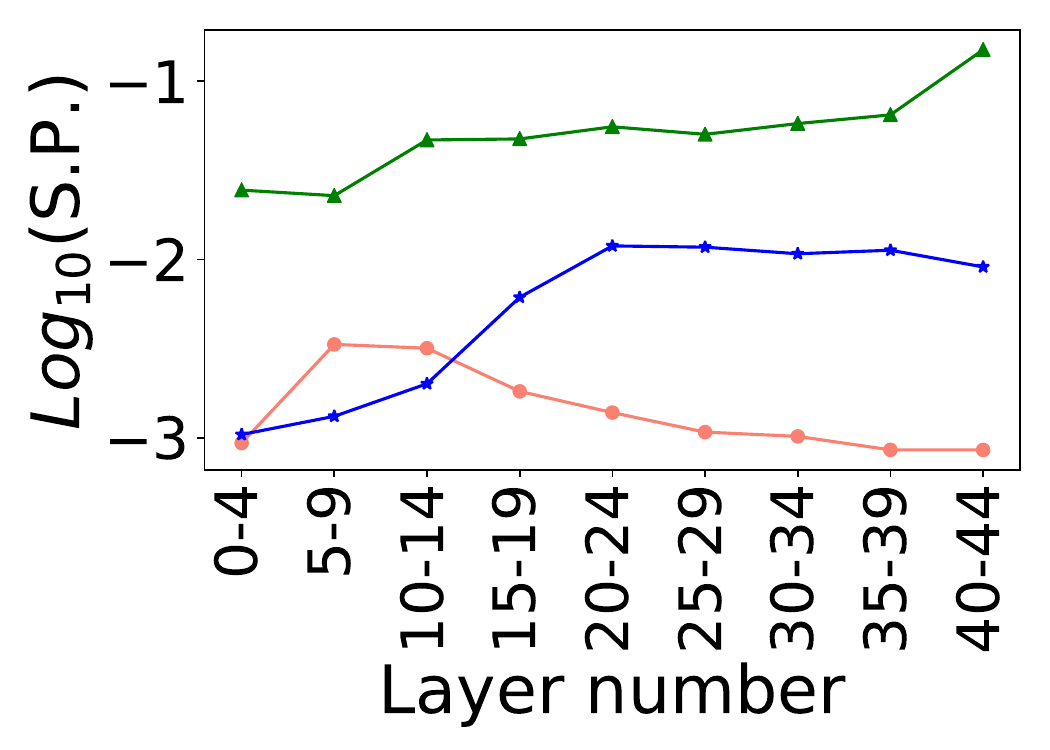}} 
\caption{Separation power(S.P.) of different high level features' histograms for dataset 3.} \label{fig:ds3_sep}
\end{figure}

Figure~\ref{fig:ds2_sep} and~\ref{fig:ds3_sep} show separation power between Geant4 and and other models for various high level features.

\textit{Analysis:} Discussed in main section.

\begin{figure}
    \centering
    \includegraphics[width=0.3\textwidth]{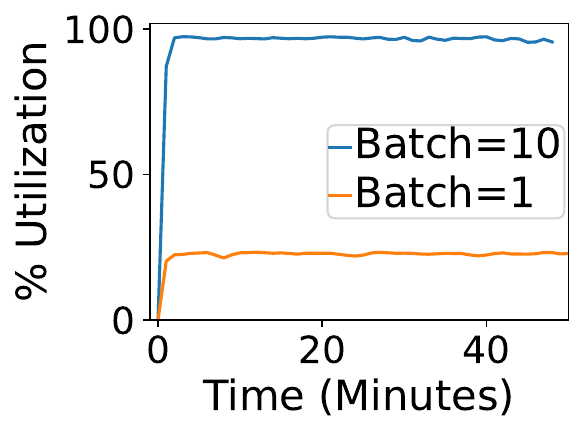}
    \caption{GPU usage percentage during mixed precision inference on the GPU. The blue plot represents dataset 3, while the orange plot corresponds to dataset 1 (pion).}
    \label{fig:gpu_uti}
\end{figure}

\subsection{Mixed precision inference evaluation}
\begin{figure}[h]
\centering
\subfloat[Layer wise energy(in all ranges of incident energy)]
{\includegraphics[width=0.33\textwidth]{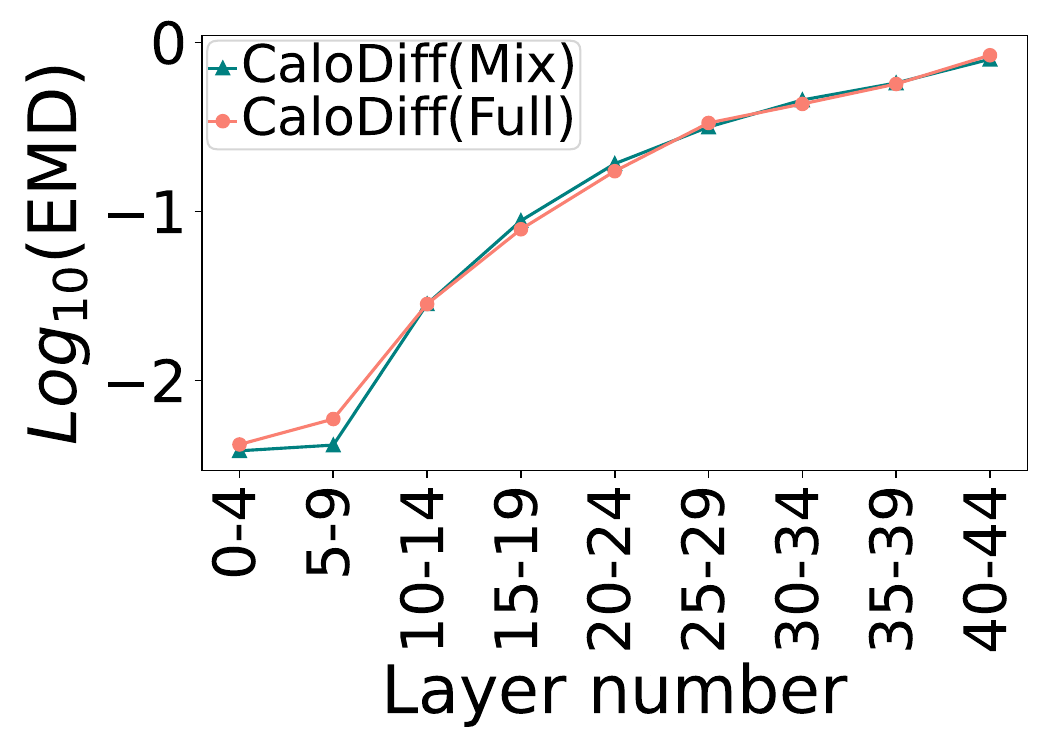}} 
\subfloat[Sparsity.]
{\includegraphics[width=0.33\textwidth]{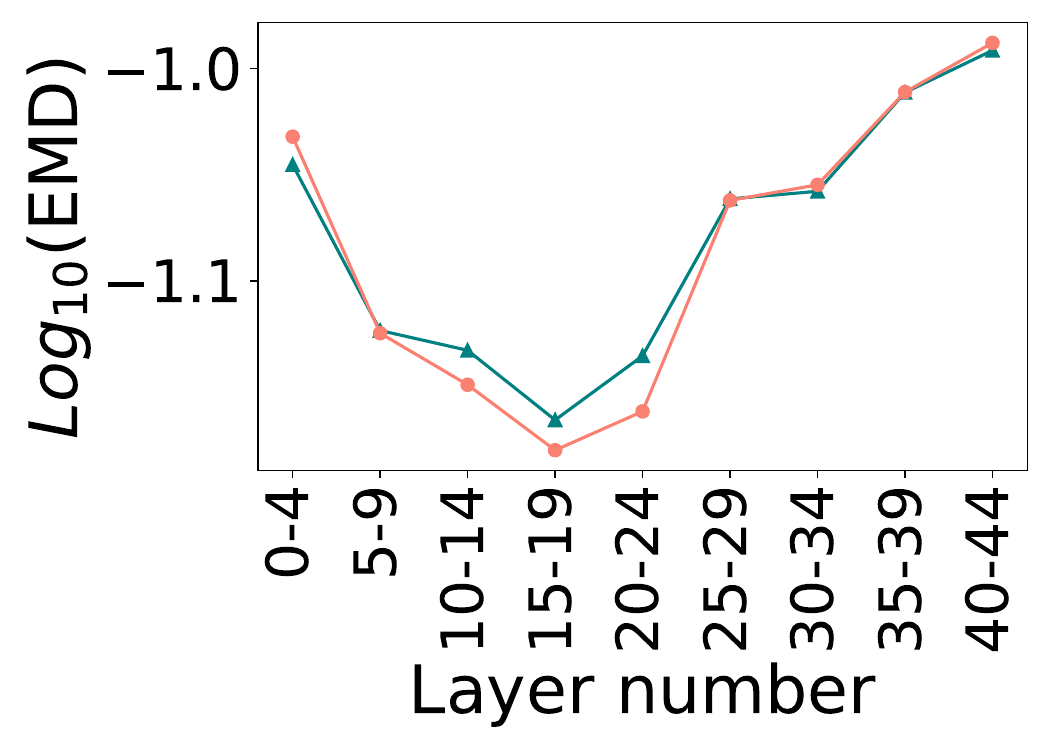}} \subfloat[Center of energy in $\eta$ direction.]
{\includegraphics[width=0.33\textwidth]{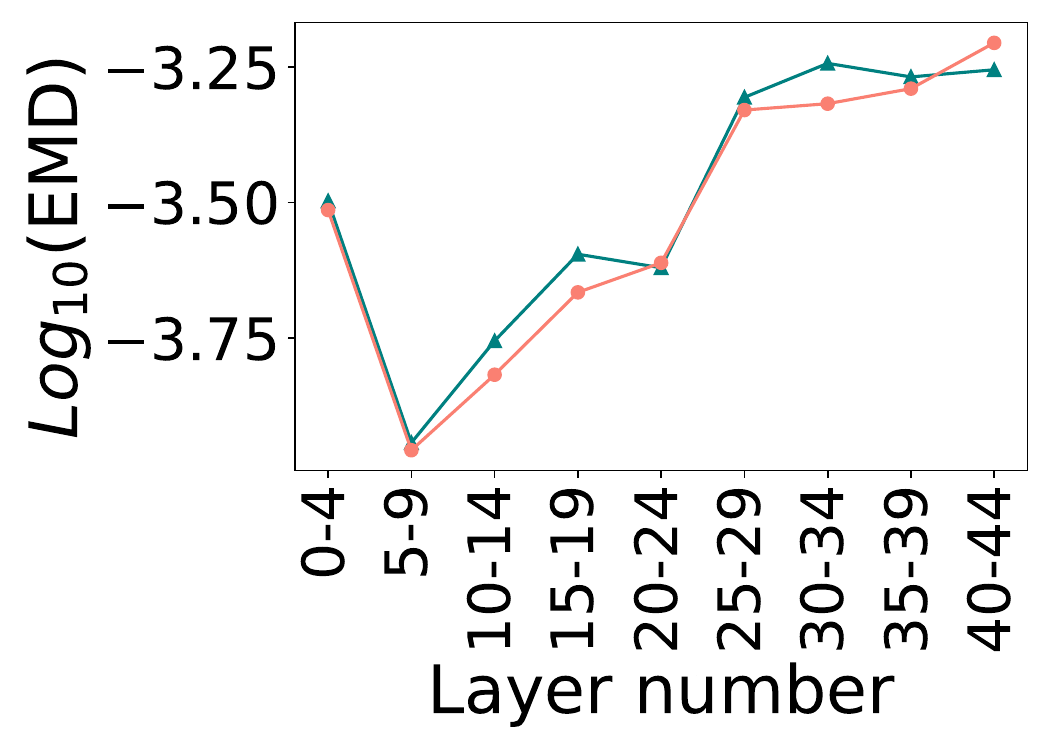}} 
\vspace{1mm}
\subfloat[Center of energy in $\phi$ direction]
{\includegraphics[width=0.33\textwidth]{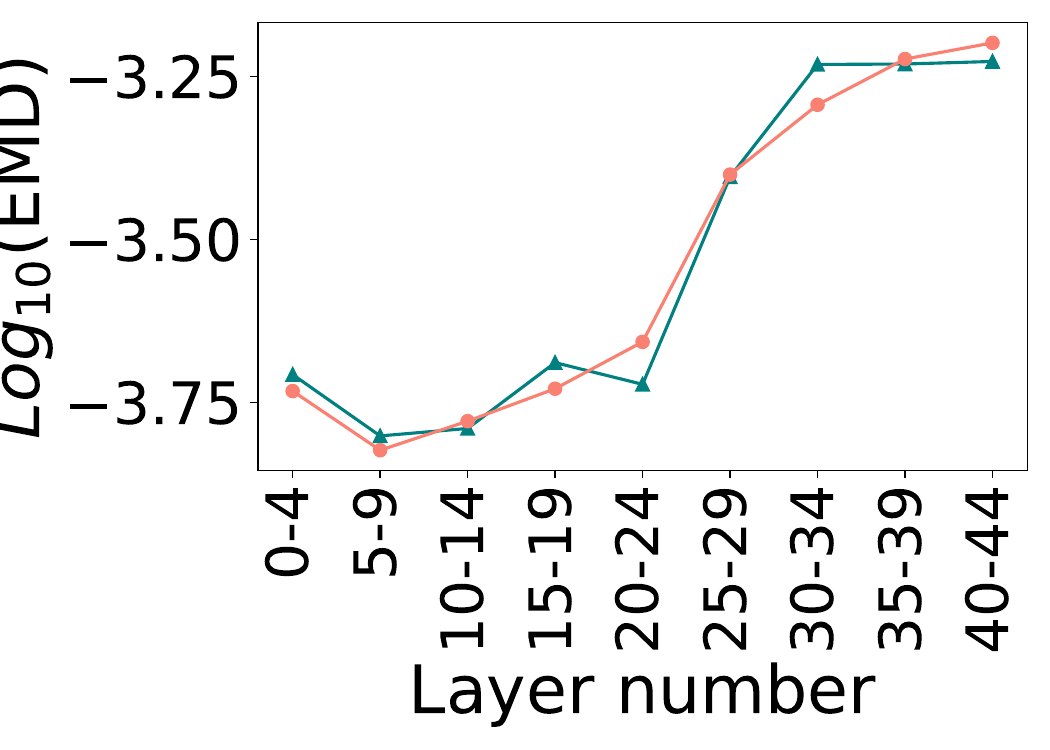}} 
\subfloat[Shower width in $\eta$ direction.]
{\includegraphics[width=0.33\textwidth]{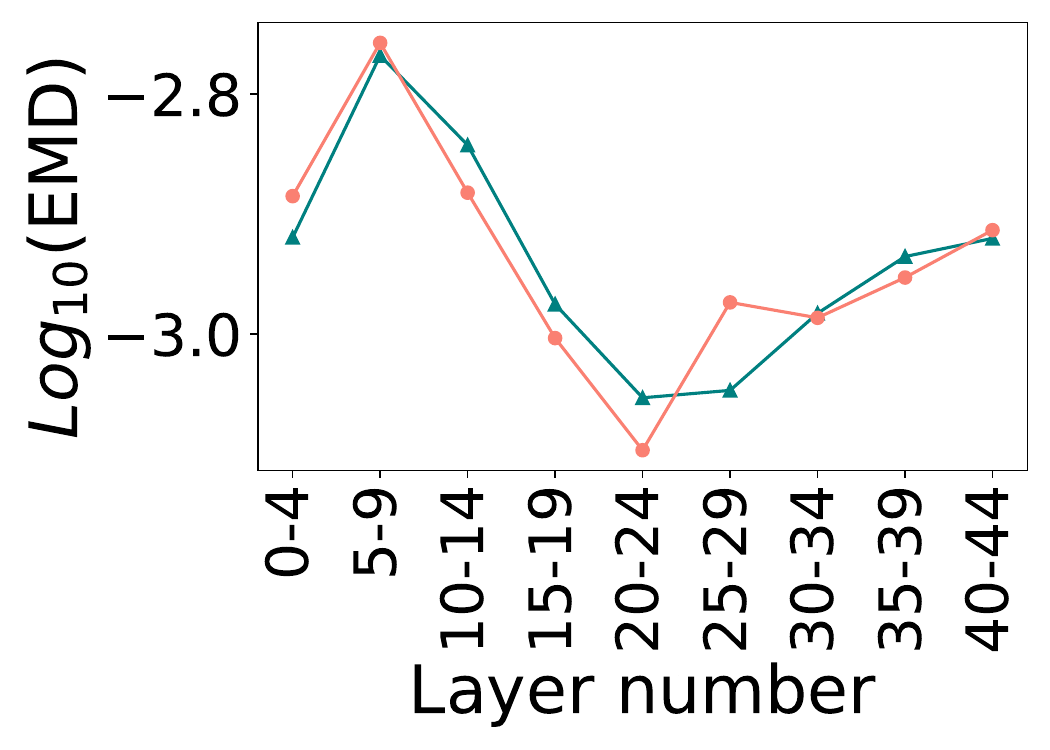}} 
\subfloat[Shower width in $\phi$ direction.]
{\includegraphics[width=0.33\textwidth]{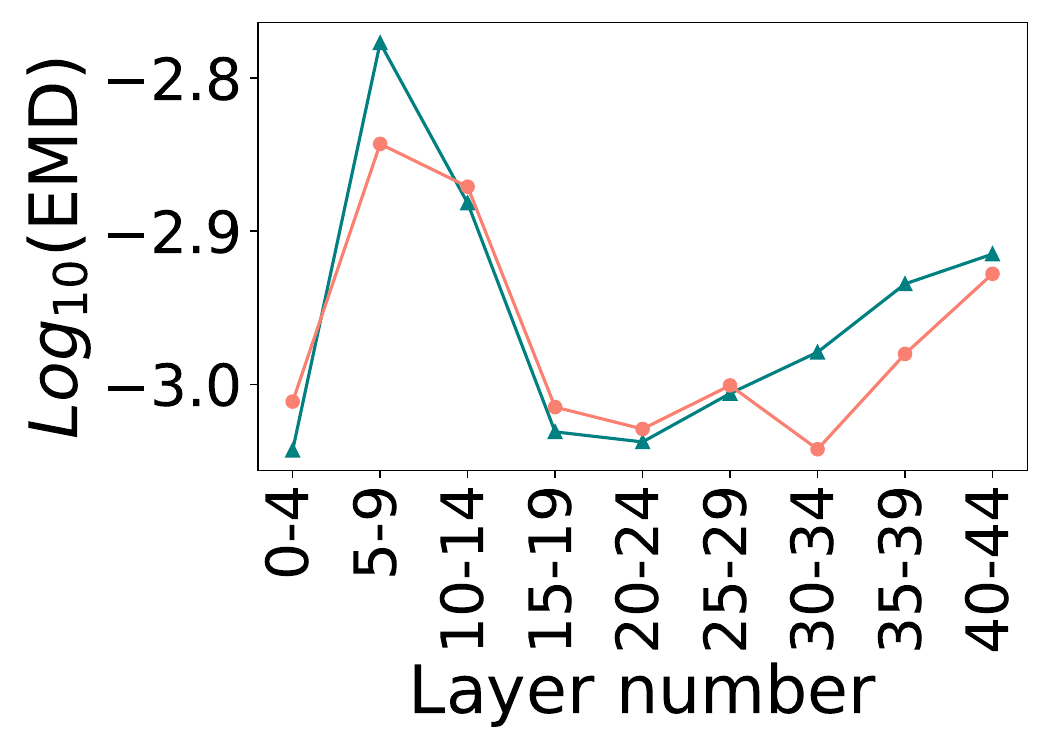}} 
\caption{EMD score for different high level features for dataset 2. Here CaloDiff(Full) denotes sample generated using full precision inference and CaloDiff(Mix) denotes sample generated using mix precision inference.} \label{fig:ds2_emd_mix}
\end{figure}

\begin{figure}[h]
\centering
\subfloat[Layer wise energy(in all ranges of incident energy)]
{\includegraphics[width=0.33\textwidth]{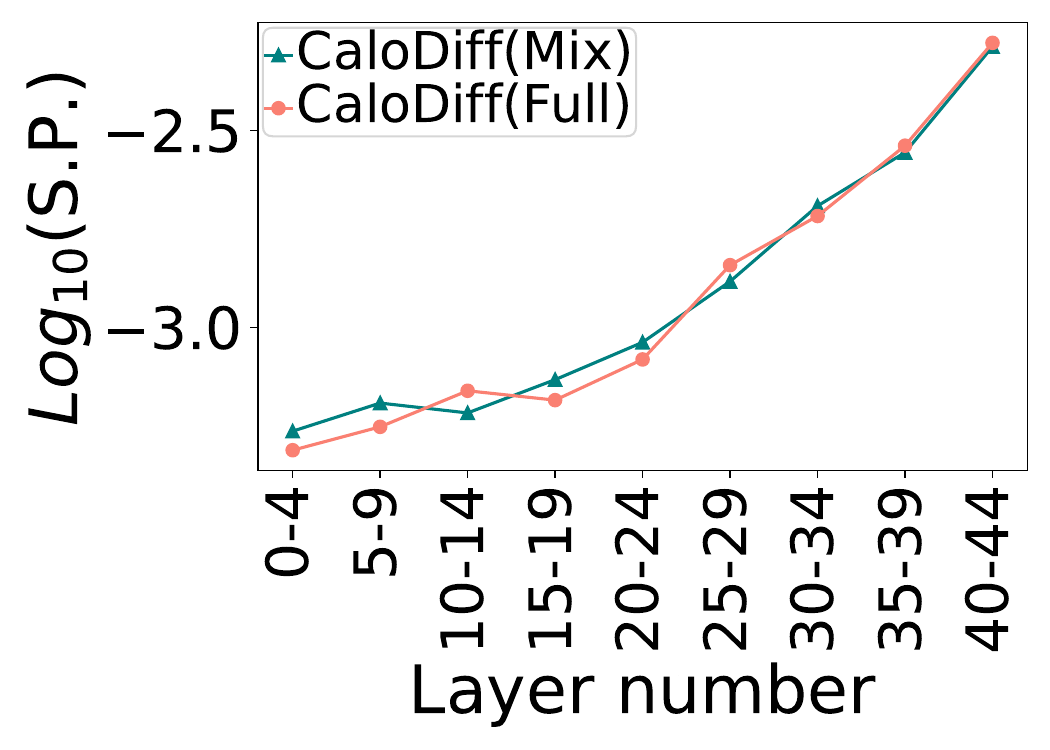}} 
\subfloat[Sparsity.]
{\includegraphics[width=0.33\textwidth]{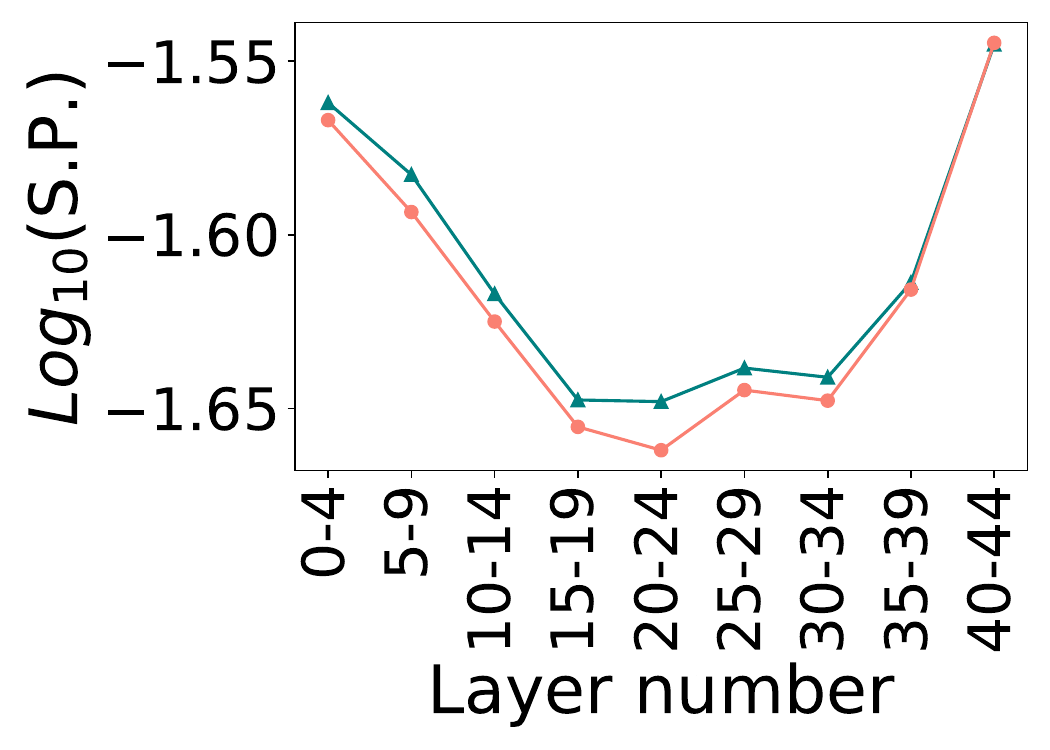}} \subfloat[Center of energy in $\eta$ direction.]
{\includegraphics[width=0.33\textwidth]{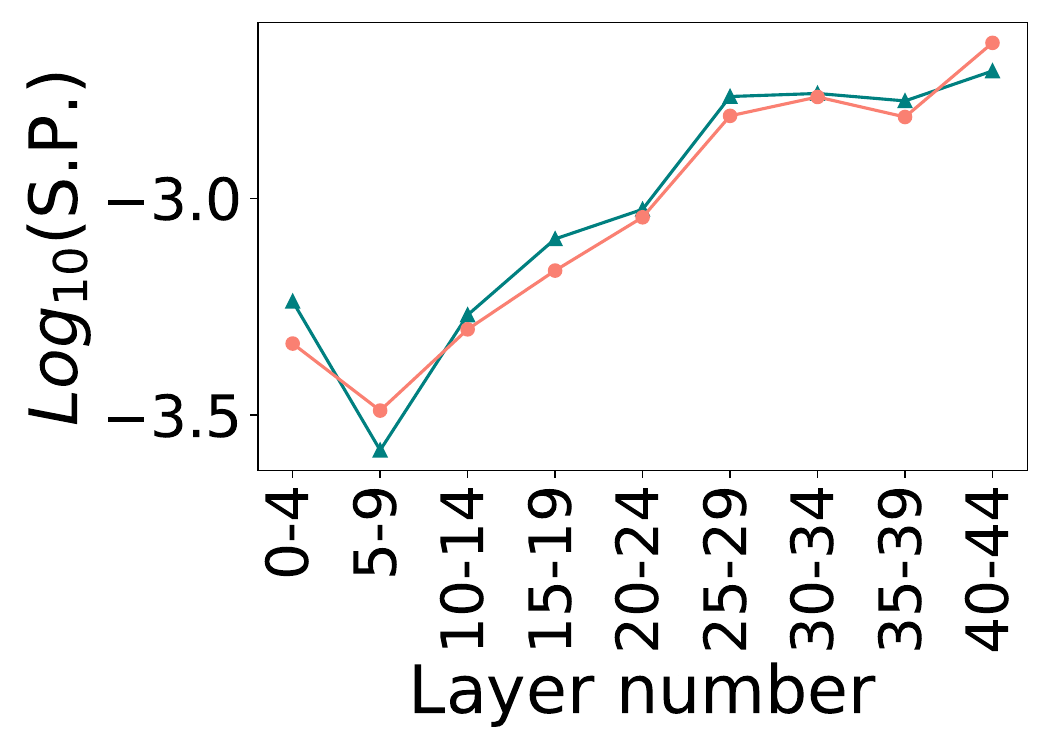}} 
\vspace{1mm}
\subfloat[Center of energy in $\phi$ direction]
{\includegraphics[width=0.33\textwidth]{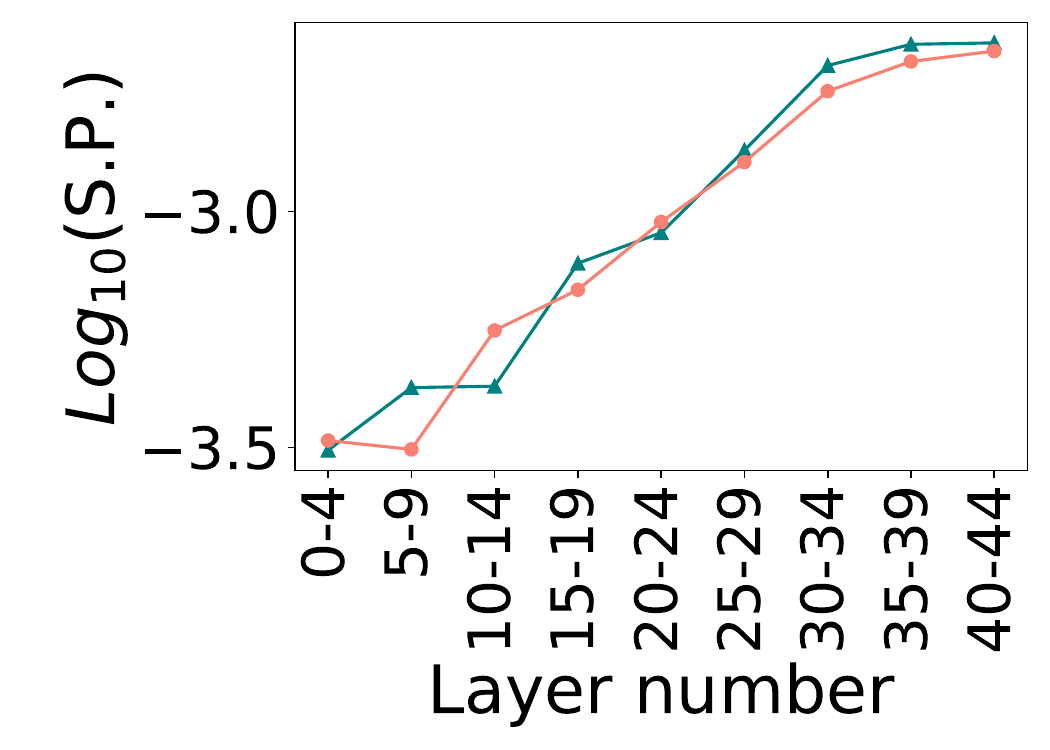}} 
\subfloat[Shower width in $\eta$ direction.]
{\includegraphics[width=0.33\textwidth]{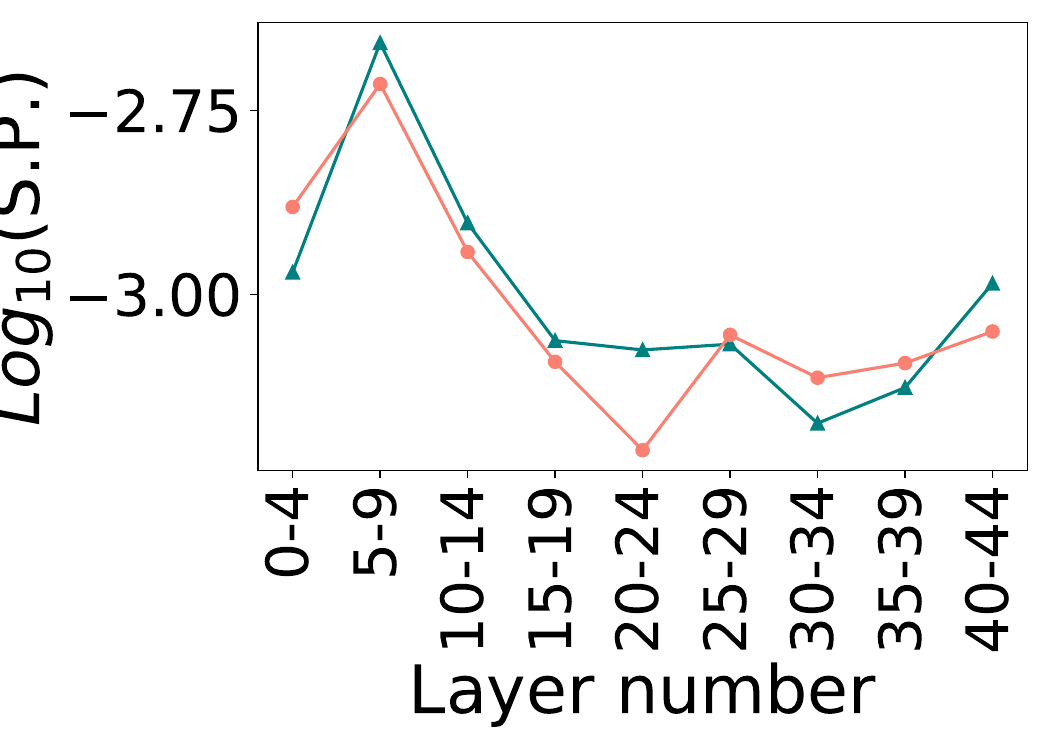}} 
\subfloat[Shower width in $\phi$ direction.]
{\includegraphics[width=0.33\textwidth]{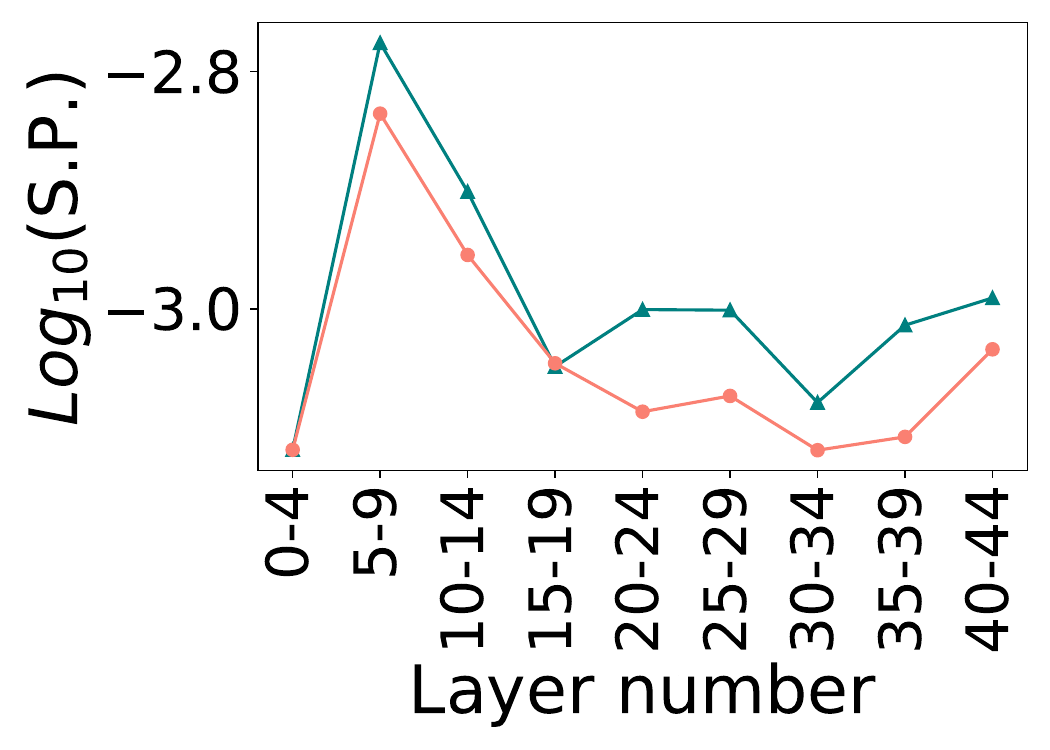}} 
\caption{Separation power for different high level features for dataset 2. Here CaloDiff(Full) denotes sample generated using full precision inference and CaloDiff(Mix) denotes sample generated using mix precision inference.} \label{fig:ds2_sep_mix}
\end{figure}

\begin{figure}[h]
\centering
\subfloat[Layer wise energy(in all ranges of incident energy)]
{\includegraphics[width=0.33\textwidth]{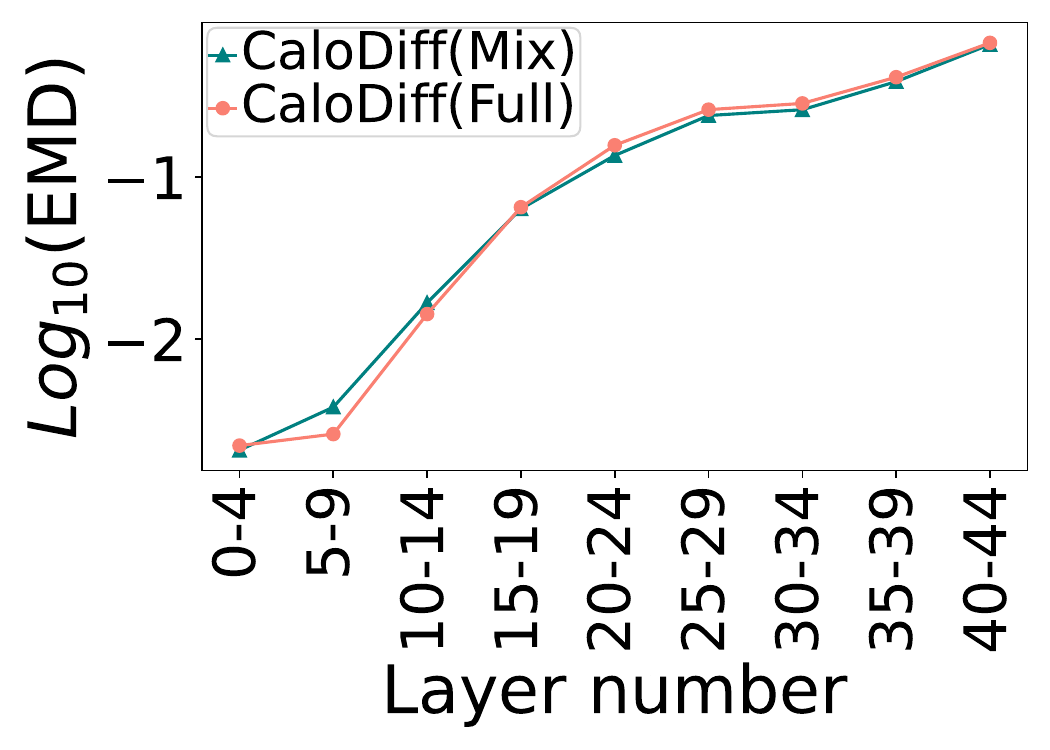}} 
\subfloat[Sparsity.]
{\includegraphics[width=0.33\textwidth]{images_subplot/final_dataset_3_mix/EMDs/emd_sparsity.pdf}} \subfloat[Center of energy in $\eta$ direction.]
{\includegraphics[width=0.33\textwidth]{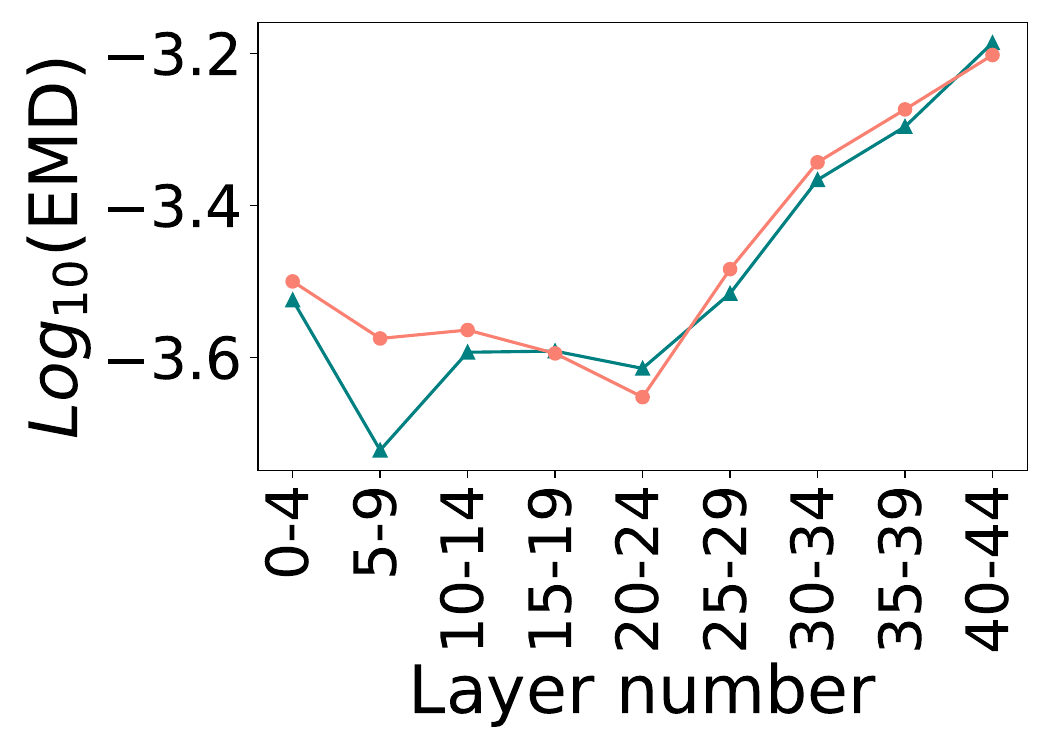}} 
\vspace{1mm}
\subfloat[Center of energy in $\phi$ direction]
{\includegraphics[width=0.33\textwidth]{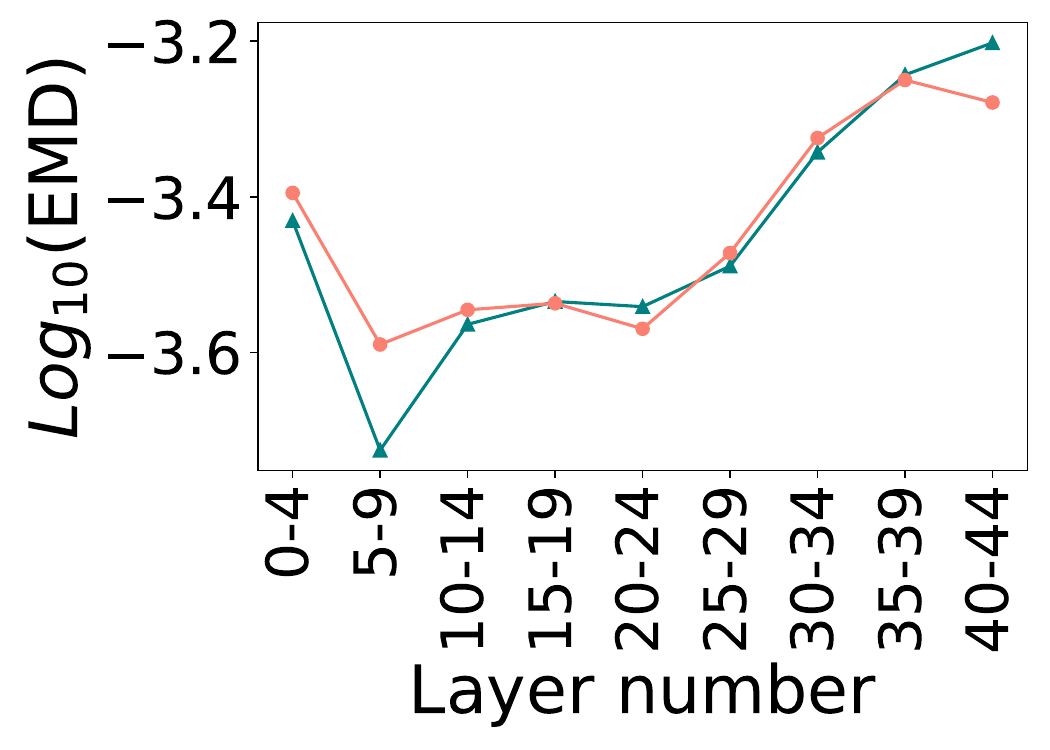}} 
\subfloat[Shower width in $\eta$ direction.]
{\includegraphics[width=0.33\textwidth]{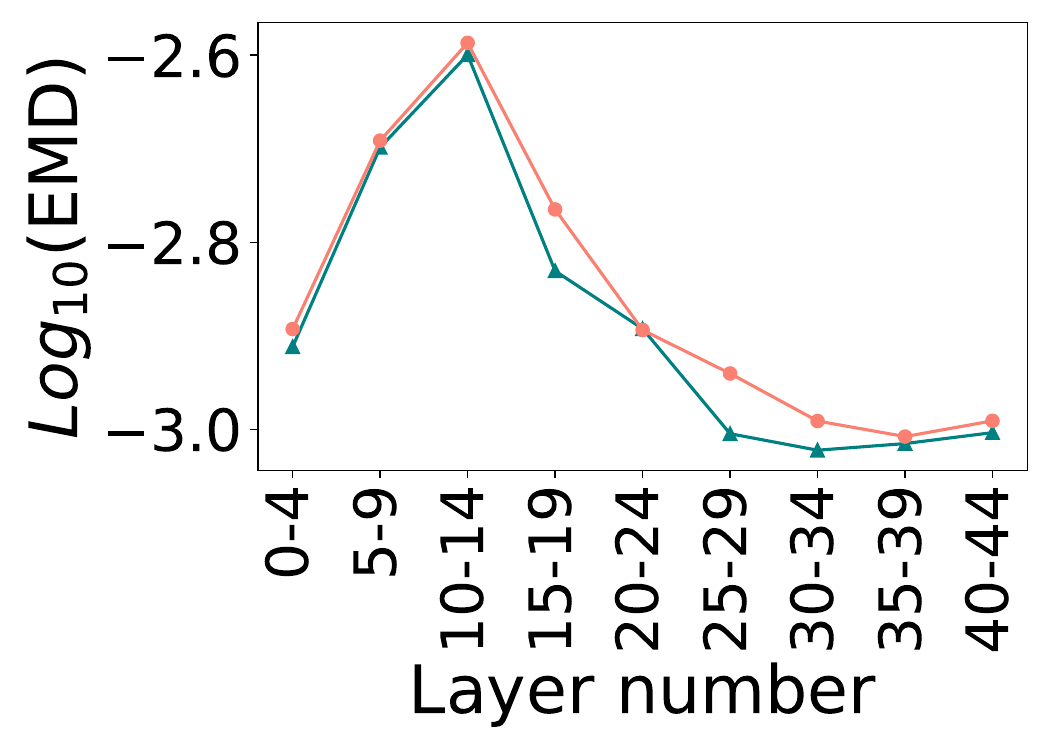}} 
\subfloat[Shower width in $\phi$ direction.]
{\includegraphics[width=0.33\textwidth]{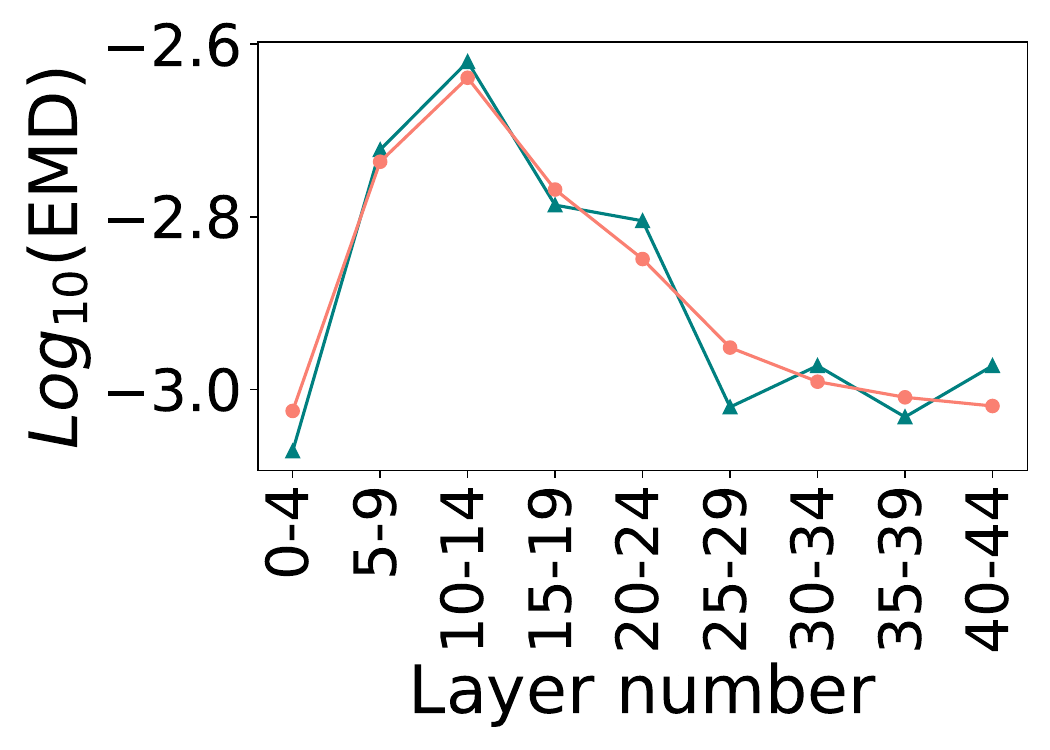}} 
\caption{EMD score for different high level features for dataset 3. Here CaloDiff(Full) denotes sample generated using full precision inference and CaloDiff(Mix) denotes sample generated using mix precision inference.} \label{fig:ds3_emd_mix}
\end{figure}

\begin{figure}[h]
\centering
\subfloat[Layer wise energy(in all ranges of incident energy)]
{\includegraphics[width=0.33\textwidth]{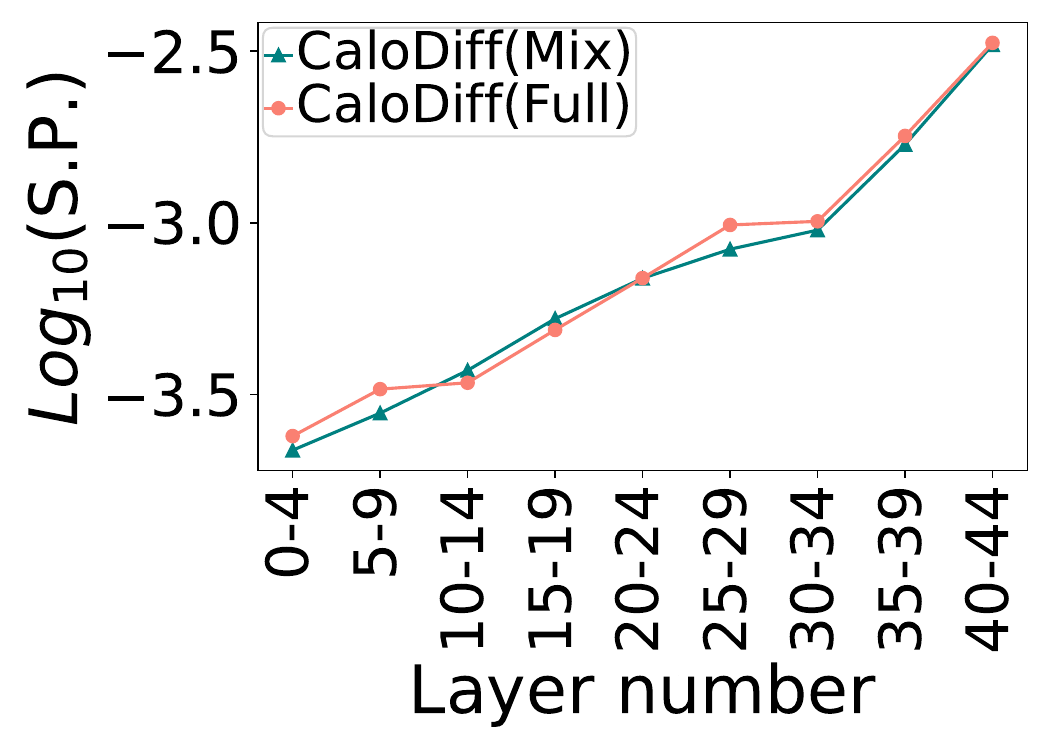}} 
\subfloat[Sparsity.]
{\includegraphics[width=0.33\textwidth]{images_subplot/final_dataset_3_mix/Seps/sep_sparsity.pdf}} \subfloat[Center of energy in $\eta$ direction.]
{\includegraphics[width=0.33\textwidth]{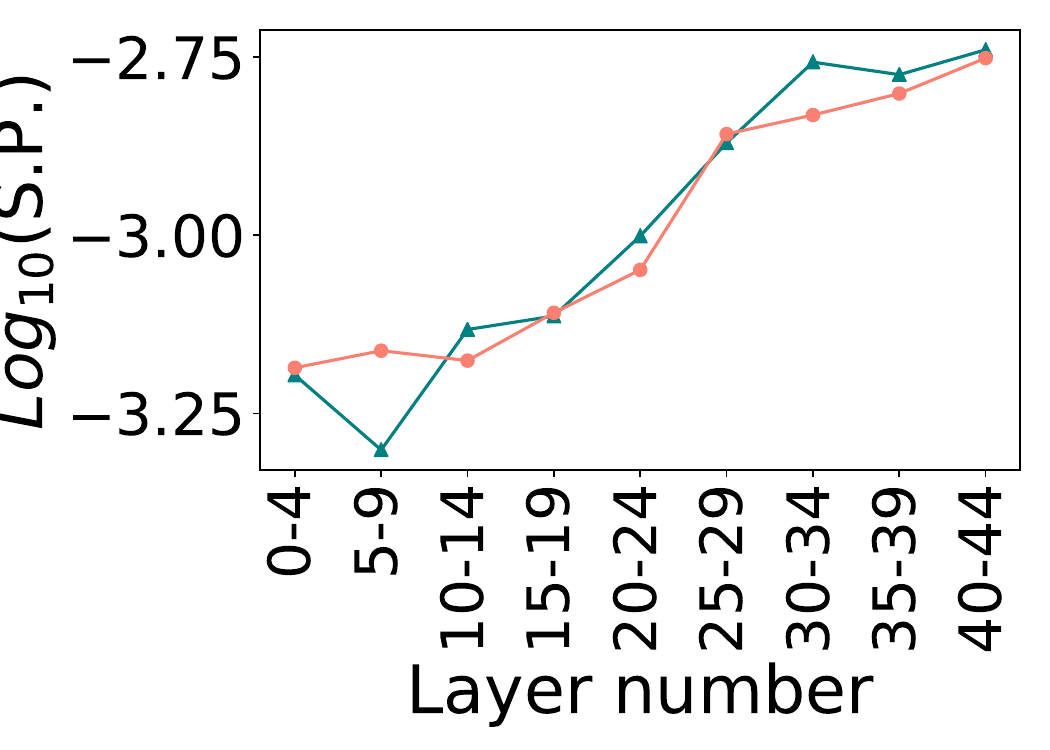}} 
\vspace{1mm}
\subfloat[Center of energy in $\phi$ direction]
{\includegraphics[width=0.33\textwidth]{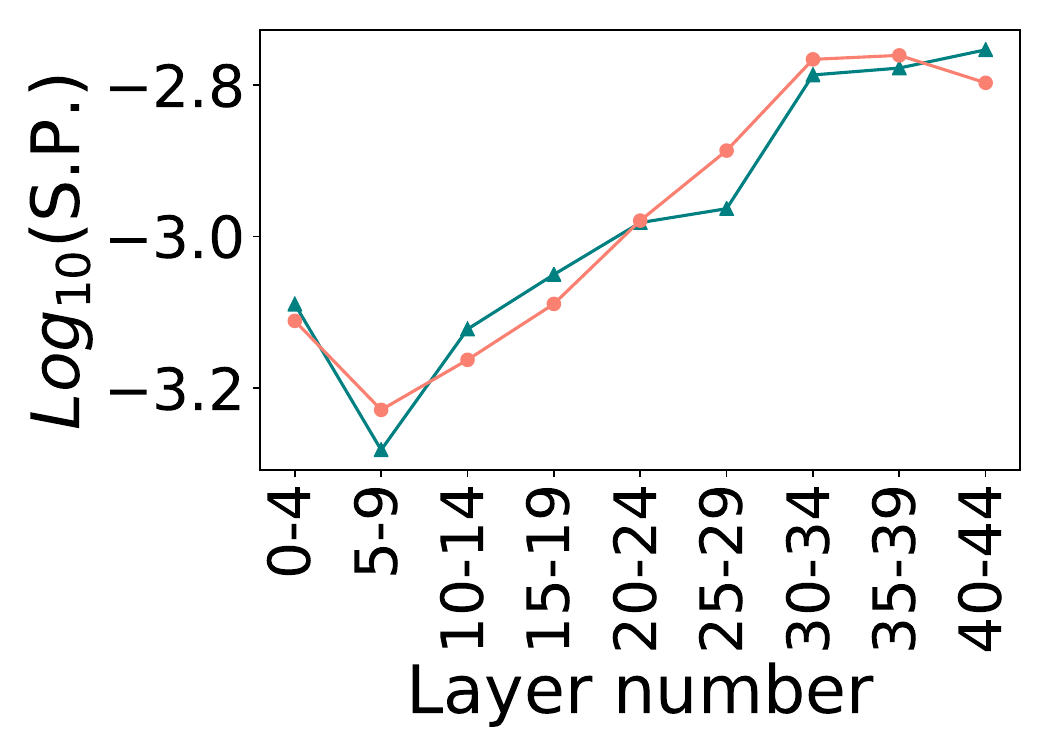}} 
\subfloat[Shower width in $\eta$ direction.]
{\includegraphics[width=0.33\textwidth]{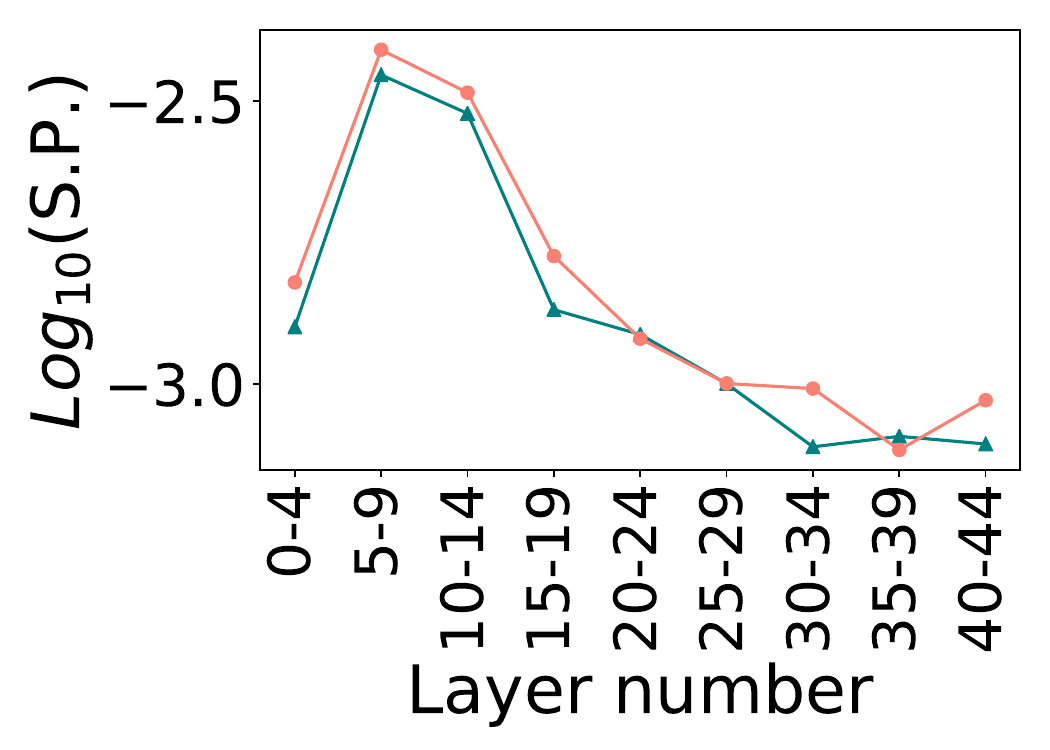}} 
\subfloat[Shower width in $\phi$ direction.]
{\includegraphics[width=0.33\textwidth]{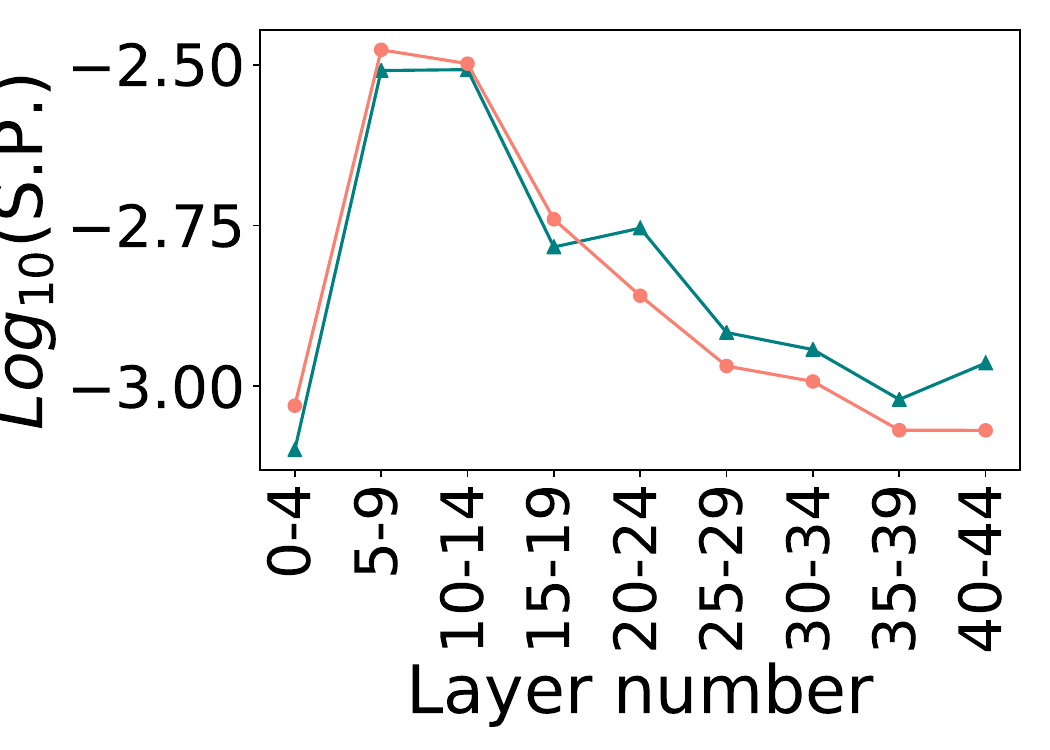}} 
\caption{Separation power for different high level features for dataset 3. Here CaloDiff(Full) denotes sample generated using full precision inference and CaloDiff(Mix) denotes sample generated using mix precision inference.} \label{fig:ds3_sep_mix}
\end{figure}

 Figures[~\ref{fig:ds2_emd_mix},~\ref{fig:ds2_sep_mix},~\ref{fig:ds3_emd_mix},~\ref{fig:ds3_sep_mix}] display the quantitative performance of mixed precision inference on the GPU for datasets 2 and 3.

\end{appendices}

\end{document}